\documentclass[11pt,a4paper]{article}
\pdfoutput=1
\usepackage{jheppub}
\usepackage{amsfonts, amssymb, amsmath, amsthm}
\usepackage{mathrsfs}
\usepackage{graphicx}
\usepackage[utf8]{inputenc}
\usepackage[english]{babel}
\usepackage{slashed}
\usepackage{tikz}
\usepackage{comment}

\usepackage{color}
\definecolor{dark-gray}{gray}{0.20}
\definecolor{gray}{gray}{0.30}
\definecolor{light-gray}{gray}{0.80}
\definecolor{dark-red}{rgb}{0.7,0,0}
\definecolor{dark-green}{rgb}{0.1,0.4,0}
\definecolor{dark-blue}{rgb}{0.3,0.3,0.7}
\definecolor{light-blue}{rgb}{0.8,0.8,1}
\definecolor{blue}{rgb}{0,0,1}
\definecolor{red}{rgb}{1,0,0}
\definecolor{green}{rgb}{0,1,0}

\usepackage{hyperref}
\hypersetup{
	colorlinks=true,
	linkcolor=dark-blue,
	citecolor=dark-red,
	urlcolor=dark-blue,
	linktoc=page
}

\theoremstyle{definition}

\theoremstyle{remark}

\newcommand{\be}{\begin{equation}}
\newcommand{\ee}{\end{equation}}
\newcommand{\ba}{\begin{aligned}}
\newcommand{\ea}{\end{aligned}}
\newcommand{\bea}{\begin{eqnarray}}
\newcommand{\eea}{\end{eqnarray}}

\title{Black hole thermodynamics at 4 derivatives, \\  natural variables and BPS limits}
\preprint{USTC-ICTS/PCFT-25-21}

\author[a,b]{Kiril Hristov}
\author[c]{, Peng-Ju Hu}
\author[c,d]{and Yi Pang}

\affiliation[a]{Faculty of Physics, Sofia University ``St.\ Kliment Ohridski'',\\ J. Bourchier Blvd. 5, 1164 Sofia, Bulgaria}
\affiliation[b]{INRNE, Bulgarian Academy of Sciences, Tsarigradsko Chaussee 72, 1784 Sofia, Bulgaria}
\affiliation[c]{Center for Joint Quantum Studies and Department of Physics, School of Science,\\ Tianjin University, 135 Yaguan Road, Tianjin 300350, China}
\affiliation[d]{Peng Huanwu Center for Fundamental Theory, Hefei, Anhui 230026, China}

\abstract{
\noindent We study Einstein-Maxwell theory in $D \geq 3$ spacetime dimensions including \emph{all} Lorentz-invariant parity-even four-derivative couplings. Building on the results of \cite{Hu:2023gru}, we consider static, charged, asymptotically flat black hole solutions to first order in the higher-derivative expansion. In $D=4$ and $D=5$, we compute the corrected black hole thermodynamics and compare with the Reall–Santos prescription based on the two-derivative background, highlighting a subtlety when both inner and outer horizons are involved. By introducing natural variables, as in \cite{Hristov:2023sxg}, we recast the on-shell actions in terms of left- and right-moving chemical potentials, which significantly simplifies the analysis.

We also compute first-order thermodynamic corrections for the most general rotating black holes in $D=4$ and $D=5$, without modifying the background solutions. We identify a novel BPS-like limit in $D=4$, extending known supergravity results beyond their traditional domain of validity. Finally, in $D=5$, the analysis of BPS and almost BPS limits enables an independent verification of the five-dimensional BPS thermodynamics. We clarify the origin of a discrepancy in the literature concerning higher-derivative supergravity localization, sharpening the tension between direct computations and predictions based on the $D=4$/$D=5$ connection.
}
\date{\today}
\begin{document}
\maketitle

\section{Introduction and discussion}
\label{sec:intro}

Even though General Relativity and its matter extensions—such as Einstein–Maxwell theory—remain non-renormalizable, it is increasingly clear that higher-derivative (HD) corrections capture key features of quantum gravity. These terms arise naturally in the low-energy effective actions of string theory, where they encode both perturbative and non-perturbative effects, including string loops and D-brane instantons, \cite{Zwiebach:2004tj,Polchinski:1998rq,Becker:2006dvp}. While gravity’s non-renormalizability limits its role as a quantum field theory, higher-derivative corrections—structured and constrained by string theory—offer a powerful window into quantum aspects of spacetime and black hole microphysics, \cite{Sen:2005wa,Sen:2007qy}.

Motivated by this, we adopt an agnostic perspective regarding the string-theoretic origin of these corrections and systematically analyze \emph{all} parity-even four-derivative terms consistent with diffeomorphism and gauge invariance that extend Einstein–Maxwell theory in arbitrary spacetime dimension $D \geq 3$. We treat them as small perturbations to the two-derivative action, thus focusing on black hole solutions that already exist in Einstein-Maxwell theory. Looking first at static backgrounds, we demonstrate that explicit black hole solutions incorporating these corrections can still be constructed, and their thermodynamics reliably computed, see \cite{Campanelli:1994sj,LopesCardoso:1998tkj,Kats:2006xp,Charmousis:2008kc,Charles:2015eha,Cheung:2018cwt,Cano:2018aod,Cano:2019ore,Cremonini:2019wdk,Chen:2020hjm,Agurto-Sepulveda:2022vvf,Ma:2023qqj} and references thereof. Technically, the construction of such solutions becomes feasible by taking the zero cosmological constant limit of the recently discovered static AdS black holes in the presence of arbitrary four-derivative corrections \cite{Hu:2023gru}. The thermodynamics of these configurations can be determined in two complementary ways: either directly from the corrected solutions or by applying the Reall-Santos (RS) method to the original two-derivative solutions, as described in \cite{Reall:2019sah}.

In the rotating case, obtaining fully back-reacted solutions becomes significantly more challenging from a technical standpoint. Nevertheless, we can still compute the thermodynamic corrections by evaluating the Euclidean on-shell action and conserved charges via the RS method. We pay special attention to a subset of four-derivative terms inspired by supergravity constructions in $D = 4$ and $D = 5$, see \cite{deWit:1980lyi,Bergshoeff:1980is,deWit:1982na,deWit:2006gn,Hanaki:2006pj,Butter:2011sr,Ozkan:2013nwa,Butter:2013lta,Kuzenko:2015jxa,Hegde:2019ioy,Gold:2023dfe,Gold:2023ymc,Gold:2023ykx, Gold:2025ttt}, and the associated BPS limits that relate to a number of recent developments on supergravity localization, \cite{BenettiGenolini:2019jdz,Hosseini:2019iad,Hristov:2021qsw,BenettiGenolini:2023ndb,Hristov:2024cgj,BenettiGenolini:2024lbj,Cassani:2024kjn,Colombo:2025ihp}. This approach allows us both to test the broader framework and to establish connections with explicit effective actions arising from string compactifications, see \cite{Eloy:2020dko,Garousi:2020gio,Codina:2020kvj,Lescano:2021guc,Liu:2023fqq,Jayaprakash:2024xlr,Ozkan:2024euj} for recent developments and review. As a result, the present work yields a range of new technical findings that are difficult to summarize succinctly. In what follows in this section, we instead highlight several conceptual advances that emerge from our analysis.

In this work, our thermodynamic analysis extends beyond the conventional focus on the outer black hole horizon to include the thermodynamics associated with the inner horizon. This broader perspective enables a reformulation of the thermodynamic potentials as specific linear combinations of their values on the two horizons—referred to as left- and right-moving, or natural variables, in \cite{Hristov:2023sxg}. This reformulation is motivated by the observation that the first law of black hole thermodynamics holds independently at each horizon, specifically at the inner ($r = r_-$) and outer ($r = r_+$) horizons,~\footnote{For simplicity, we consider a setting with a single electric charge and angular momentum, but the principle extends straightforwardly to cases with additional charges and chemical potentials.}
\be 
 \beta_\pm\, \delta M - \delta S_\pm - \beta_\pm \Phi_\pm\, \delta Q - \beta_\pm \Omega_\pm\, \delta J  = 0\ , \quad \Rightarrow \quad  I_\pm=I_\pm (\beta_\pm, \Phi_\pm, \Omega_\pm)\ ,
\ee 
where $M, Q, J$ denote the conserved asymptotic charges (mass, electric charge, and angular momentum, respectively); $\beta, \Phi, \Omega$ are their conjugate chemical potentials (inverse temperature, electric potential, and angular velocity); $S$ is the entropy, and $I$ the on-shell action, with the subscripts $\pm$ indicating the horizon to which each quantity pertains.

Natural variables are then defined by taking linear combinations of the chemical potentials $\beta_{l,r},\ \varphi_{l,r}, \ \omega_{l,r}$ and corresponding on-shell actions $I_{l,r}$ evaluated at both horizons. As shown in a series of works \cite{Hristov:2023sxg,Hristov:2023cuo,Hristov:2024lzr,Hristov:2024cjp}, these variables possess several remarkable properties: the resulting on-shell actions become considerably simpler and can be expressed explicitly in terms of the natural potentials. Furthermore, the formalism admits a smooth BPS limit, in which the right-moving on-shell action vanishes, while the left-moving one reproduces the expected result from supersymmetric considerations. This combination of features appears to be quite unique, but the fundamental importance of these variables is still to be understood microscopically.~\footnote{It is tempting to speculate that the two sectors relate to different fermionic boundary conditions in supersymmetric settings, which microscopically define different partition functions.} In contrast, other frameworks for deriving BPS thermodynamics \cite{Cassani:2019mms,Cassani:2021dwa,Bobev:2022bjm,Cassani:2022lrk,Cassani:2024tvk} do not straightforwardly extend to general thermal backgrounds. Additionally, one might be interested in the near-BPS limit, see \cite{Heydeman:2020hhw,Boruch:2022tno,Heydeman:2024ezi}, which we do not discuss here.

In connection with this discussion, it is important to emphasize that HD corrections typically modify the causal structure of the original two-derivative solutions, \cite{Cano:2018aod}. In particular, the number of black hole horizons may change: the standard pair of inner and outer horizons is generally supplemented by additional, \emph{small} inner horizons that are absent in the two-derivative theory. We interpret the emergence of these additional horizons as a distinctive feature of the HD expansion of the effective gravitational theory. In the present analysis, we do not attempt to incorporate the small horizons into the definition of the natural variables, as we view four-derivative black hole thermodynamics primarily as a subleading correction rather than an autonomous thermodynamic framework. Nonetheless, it would be worthwhile to explore whether similar structures persist at the level of six- and higher-derivative corrections, which may further refine the features observed at four-derivative order.

Given the generality of the HD terms considered, we once again demonstrate empirically that the natural variables approach yields a substantial simplification and proves particularly effective in analyzing the BPS limit. This framework, however, naturally prompts the question of how the RS method should be applied when both the inner and outer horizons are treated simultaneously. At first glance, this appears problematic for a simple reason. The RS trick relies on the uncorrected two-derivative solutions, keeping the intrinsic quantities $\beta$, $\Phi$, and $\Omega$ fixed, and computes the corrected on-shell action in the presence of HD terms. While this approach is a priori valid at each individual horizon—where a first law of black hole thermodynamics can still be formulated—it is specifically designed to evaluate the corrected on-shell action associated with a \emph{single} horizon, rather than the linear combinations of quantities required in the natural variables formalism.

The root of the difficulty is that in the RS method the asymptotic charges are derived from the variation of the on-shell action, which in turn depends on the choice of coordinate periodicity dictated by the choice of horizon. As a result, the naive application of the first law at each horizon leads to: 
\be 
 \beta_i\delta M_i - \delta S_i - \beta_i \Phi_i\, \delta Q_i - \beta_i \Omega_i\, \delta J_i = 0\ , 
\ee 
where the index $i$ runs over the distinct horizons. Crucially, this expression treats the charges $M_i$, $Q_i$, and $J_i$ as horizon-dependent quantities. This leads to an inconsistency, since the asymptotic charges are global, uniquely defined quantities characterizing the spacetime. The result is an unphysical outcome: $M_+ \neq M_-$, $Q_+ \neq Q_-$, $J_+ \neq J_-$, and so on, across the different horizons. This contradiction makes it clear that the results of applying the RS trick separately at each horizon cannot simply be superimposed within the natural variables framework.

Fortunately, this apparent obstacle also points to its own resolution, which becomes evident when examining the fully corrected backgrounds. The key is to impose the equality of asymptotic charges by hand across the horizons, $M_+ \stackrel{!}{=} M_-, \, Q_+ \stackrel{!}{=} Q_-,\,  J_+ \stackrel{!}{=} J_-$, effectively interpreting the resulting constraints among chemical potentials as additional information about the structure of HD corrections. We show that this augmented version of the RS trick~\footnote{Note that this is a different type of extension of the RS trick, as compared to the improvement needed for asymptotically AdS black holes, see \cite{Hu:2023gru}.}  suffices for the purpose of defining the natural variables.

In the second part of our work, focusing on general rotating black holes, the natural variables formulation enables us to rederive the BPS limits in the $D = 4$ and $D = 5$ cases from a new perspective, thereby extending several earlier results on the topic. In the four-dimensional case, our findings are consistent with previous results on BPS black holes in supergravity, see \cite{LopesCardoso:1998tkj,Charles:2015eha,Bobev:2020egg,Hristov:2023cuo}. Remarkably, however, we observe that the same results hold across an even broader class of theories that do not correspond to combinations arising from off-shell four-derivative super-invariants exactly. This additional freedom in the couplings can be understood through field redefinitions, but it nevertheless represents a meaningful extension.

The five-dimensional case, which is technically more involved, allows us to rederive — from a different viewpoint—the recent result of \cite{Cassani:2024tvk} for the BPS limit presented first in \cite{deWit:2009de} in a canonical ensemble. We also explore the so-called almost BPS limit, which is closely related but distinct in that it does not preserve supersymmetry of the full background, see \cite{Hristov:2023cuo}. A conceptually striking feature is that supergravity, through the $D=4$/$D=5$ connection, \cite{Andrianopoli:2004im,Behrndt:2005he,Banerjee:2011ts}, provides two complementary perspectives on both the BPS and almost BPS limits. The surprising outcome, already noted in \cite{deWit:2009de}, is that the BPS limit appears to be \emph{incorrectly} reproduced from the four-dimensional viewpoint, see the work of \cite{Castro:2007ci,Castro:2008ne,Gupta:2021roy} and the four-dimensional OSV formula, \cite{Ooguri:2004zv}. We discuss this discrepancy in detail in Section~\ref{sec:5dBPS}, though we do not fully resolve it. Nevertheless, our findings strongly indicate the need for a careful examination of   the application of fixed-point formulas/supergravity localization, see \cite{Ooguri:2004zv,BenettiGenolini:2019jdz,Hosseini:2019iad,Hristov:2021qsw,BenettiGenolini:2023ndb,Hristov:2024cgj,BenettiGenolini:2024lbj,Cassani:2024kjn,Colombo:2025ihp}, simultaneously on both sides of the $D=4$/$D=5$ connection in presence of higher-derivative corrections.

The rest of the paper is organized as follows. Part I broadly deals with static charged black holes in Einstein-Maxwell theory and its four-derivative deformations. In section \ref{sec:gensol} we write down the most general parity-even four-derivative deformations of Einstein-Maxwell theory in $D \geq 3$ and give the back-reacted black hole solutions to first order in the higher-derivative corrections, elaborating on the standard thermodynamic quantities. In section \ref{sec:4dstatic} we focus on the $D=4$ case taking the natural variables approach described above. We find the left- and right-moving on-shell actions and observe some further simplifications in the case of supersymmetry. This also allows us to demonstrate how these results can be alternatively derived using the RS method.  In section \ref{sec:5dstatic} we repeat the same steps in $D=5$, emphasizing the differences with $D=4$. In part II we include rotation in our four-derivative analysis, discussing the $D=4$ case in section \ref{sec:4drotating} and the $D=5$ case in \ref{sec:5drotating}. Apart from general observations about the left- and right-moving thermodynamics, we focus much of the discussion on the BPS limits in supergravity and comment on the relation with previous literature and the discrepancy we described above.

\part{Static black holes}

\section{General 4-derivative solutions}
\label{sec:gensol}

We start from Einstein-Maxwell theory extended by 4-derivative interactions
of the form
\begin{align}
I & =\frac{1}{16\pi G_{N}}\int d^{D}x\sqrt{-g}\mathcal{L},\ \ \mathcal{L}=R-\frac{1}{4\sigma^{2}}F^{\mu\nu}F_{\mu\nu}+\Delta\mathcal{L}\ ,\label{Action}
\end{align}
where 
\begin{align}
\Delta\mathcal{L}= & c_{1}R^{2}+c_{2}R_{\mu\nu}R^{\mu\nu}+c_{3}R_{\mu\nu\rho\sigma}R^{\mu\nu\rho\sigma}\nonumber \\
+ & \frac{c_{4}}{\sigma^{2}}RF_{\mu\nu}F^{\mu\nu}+\frac{c_{5}}{\sigma^{2}}R_{\mu\nu}F^{\mu\rho}F^{\nu}{}_{\rho}+\frac{c_{6}}{\sigma^{2}}R_{\mu\nu\rho\sigma}F^{\mu\nu}F^{\rho\sigma}\nonumber \\
+ & \frac{c_{7}}{\sigma^{4}}F_{\mu\nu}F^{\mu\nu}F_{\rho\sigma}F^{\rho\sigma}+\frac{c_{8}}{\sigma^{4}}F_{\mu\nu}F^{\nu\rho}F_{\rho\sigma}F^{\sigma\mu}+\frac{c_{9}}{\sigma^{2}}\nabla^{\mu}F_{\mu\nu}\nabla_{\rho}F^{\rho\nu}\ .\label{dL}
\end{align}
Note that we have included the parameter $\sigma$ above for convenience in relating to different conventions in the literature, but it can be freely fixed from the beginning, e.g.\ to $1$, without loss of generality. The coefficients $c_1, ..., c_9$ instead parametrize the nine different 4-derivative couplings of the metric and gauge fields that preserve Lorentz covariance. We consider the coefficients to be small, such that we solve the resulting equations of motion at first order. This technically simplifies our analysis, which is analogous to the approach in \cite{Campanelli:1994sj,Hu:2023qhs}. The field equations are given by
\begin{align}
g_{\mu\nu} & :\ P_{(\mu}^{\ \alpha\beta\gamma}R_{\nu)\alpha\beta\gamma}-2\nabla^{\alpha}\nabla^{\beta}P_{\alpha(\mu\nu)\beta}-\frac{\mathcal{L}}{2}g_{\mu\nu}-\frac{1}{2}\mathcal{E}_{(\mu}^{\ \ \rho}F_{\nu)\rho}=0\,,\nonumber \\
A_{\mu} & :\ \nabla_{\mu}\mathcal{E}^{\mu\nu}=0\,,
\end{align}
where we have defined $P^{\mu\nu\rho\sigma}:=\partial\mathcal{L}/\partial R_{\mu\nu\rho\sigma},\ \mathcal{E}^{\mu\nu}:=-2\partial\mathcal{L}/\partial F_{\mu\nu}$:
\begin{align}
P^{\mu\nu\rho\sigma}= & Rg^{\mu[\rho}g^{\sigma]v}+2c_{1}Rg^{\mu[\rho}g^{\sigma]v}+c_{2}(R^{\mu[\rho}g^{\sigma]\nu}-R^{\nu[\rho}g^{\sigma]\mu})+2c_{3}R^{\mu\nu\rho\sigma}\nonumber \\
 & +\frac{c_{4}}{\sigma^{2}}g^{\mu[\rho}g^{\sigma]v}F_{\alpha\beta}F^{\alpha\beta}+\frac{c_{5}}{\sigma^{2}}\frac{1}{2}(S^{\mu[\rho}g^{\sigma]\nu}-S^{\nu[\rho}g^{\sigma]\mu})+\frac{c_{6}}{\sigma^{2}}F^{\mu\nu}F^{\rho\sigma}\ ,\nonumber \\
\mathcal{E}^{\mu\nu}= & \frac{1}{\sigma^{2}}F^{\mu\nu}-4(\frac{c_{4}}{\sigma^{2}}RF^{\mu\nu}-\frac{c_{5}}{\sigma^{2}}R_{\ \rho}^{[\mu}F^{\nu]\rho}+\frac{c_{6}}{\sigma^{2}}R^{\mu\nu\rho\sigma}F_{\rho\sigma}+2\frac{c_{7}}{\sigma^{4}}F^{\mu\nu}F^{\rho\sigma}F_{\rho\sigma})\nonumber \\
 & -8\frac{c_{7}}{\sigma^{4}}F^{\mu\rho}F^{\nu\sigma}F_{\rho\sigma}-\frac{c_{9}}{\sigma^{2}}\nabla^{[\mu}\nabla_{\rho}F^{\nu]\rho}\ ,\qquad S^{\mu\nu}:=F_{\ \rho}^{\mu}F^{\nu\rho}\ .
\end{align}
Notice that the last term, proportional to $c_9$, is proportional to the original Maxwell equation. In the perturbative expansion around two-derivative solutions it therefore drops out of all our calculations, effectively leaving us with eight different corrections.

 This theory, \eqref{Action}, admits the static charged black hole solution \cite{Cheung:2018cwt,Ma:2023qqj,Hu:2023gru}
\begin{align}
ds_{D}^{2}= & -h(r)dt^{2}+\frac{dr^{2}}{f(r)}+r^{2}d\Omega_{D-2}^{2}\,,\ h_{0}=f_{0}=1-\frac{2m}{r^{D-3}}+\frac{q^{2}}{r^{2(D-3)}}\,,\nonumber \\
h= & h_{0}+\Delta h+\mathcal{O}(c_{i}^{2})\,,\ \ \ \ f=f_{0}+\Delta f+\mathcal{O}(c_{i}^{2})\,,\nonumber \\
A= & \psi(r)dt\ ,\ \ \ \ \psi=\psi_{0}+\Delta\psi+\mathcal{O}(c_{i}^{2})\,,\ \ \ \psi_{0}=-\sqrt{\frac{2(D-2)}{(D-3)}}\frac{\sigma q}{r^{D-3}}\ ,
\label{RNsolution}
\end{align}
where $m,\,q$ denote the ``mass'' and ``charge'' parameters, and $d\Omega_{D-2}^{2}$ the metric of the round $(D-2)$-sphere. The $D$-dimensional corrected black hole solution $\Delta h,\,\Delta f,\,\Delta\psi$ is listed in appendix \ref{AppendixA}. 

In this paper, all thermodynamic quantities are calculated up to $\mathcal{O}(c_{i})$. The black hole horizon is located
at
\begin{equation}
r_{h}=r_{0}+\Delta r+\mathcal{O}(c_{i}^{2})\,,\qquad h_{0}(r_{0})=h(r_{h})=0\, .
\end{equation}
where $\Delta r$ is listed in appendix \ref{AppendixA}, and $r_{0}$
can be either an inner horizon $r_{-}$ or an outer horizon $r_{+}$
determined by
\begin{equation}
m=\frac{q^{2}}{2r_{0}^{D-3}}+\frac1{2}r_{0}^{D-3}\,,\label{m}
\end{equation}
For simplicity, the parameter $m$ has been replaced with \eqref{m}
in the following, effectively replacing the parameter $m$ with the parameter $r_0$, which coincides with the position of the horizon only in the two-derivative case and can otherwise be considered as a parametrization choice. This choice is particularly suitable for our purposes, given that we want to look at the thermodynamics on the new inner and outer horizons, which are presicely the corrected values of $r_0 = r_+$ and $r_0 = r_-$. Thus, using the $r_0$ parametrization we can simply describe both of the corrected original horizons by picking the two different solutions of \eqref{m},
\be
	r_\pm^{(D-3)} = m \pm \sqrt{m^2 - q^2}\ .
\ee

The temperature and inverse temperature on a given horizon are then found to be
\begin{align}
T & =\frac{\kappa_{{\rm surf}}}{2\pi}\,,\ \ \beta=\frac{1}{T}=\beta_{0}+\Delta\beta+\mathcal{O}(c_{i}^{2})\,,\ \ \beta_{0}=\frac{4\pi r_{0}}{(D-3)(1-q^{2}r_{0}^{-2(D-3)})}\ ,\label{T}
\end{align}
where the $D$-dimensional corrected thermodynamics $\Delta\beta,\,\Delta S,\,\Delta\Phi,\,\Delta I$
are listed in appendix \ref{AppendixA}. The formula for surface gravity
is
\begin{equation}
\kappa_{{\rm surf}}=-\frac{g^{\mu\nu}\partial_{\mu}\xi^{2}\partial_{\nu}\xi^{2}}{4\xi^{2}}|_{r=r_{h}},\ \xi^{2}=\xi^{\mu}\xi_{\mu}\,.
\end{equation}
where $\xi=\partial_{t}$ is a Killing vector. The black hole entropy is given by Wald entropy
\begin{equation}
S=-\frac1{8 G_{N}}\int d^{D-2}x\sqrt{-\gamma}P^{\mu\nu\rho\sigma}\epsilon_{\mu\nu}\epsilon_{\rho\sigma}=S_{0}+\Delta S+\mathcal{O}(c_{i}^{2})\,,\ \ S_{0}=\frac{r_{0}^{D-2}\, \omega_{D-2}}{4G_{N}}\ ,\label{S}
\end{equation}
where $\omega_{D-2}$ denotes the dimensionless volume of the $(D-2)$-sphere. $\gamma$ denotes the determinant
of the induced metric on the two-dimensional horizon. $\epsilon_{\mu\nu}$
denotes the unit binormal to the horizon, and normalized such that
$\epsilon_{\mu\nu}\epsilon^{\mu\nu}=-2$. The electrostatic potential
$\Phi$ is given by
\begin{equation}
\Phi=A_{t}|_{r_{h}}^{\infty}=\Phi_{0}+\Delta\Phi+\mathcal{O}(c_{i}^{2})\,,\qquad \Phi_{0}=\sqrt{\frac{2(D-2)}{(D-3)}}\frac{\sigma q}{r_{0}^{D-3}}\ ,\label{Phi}
\end{equation}
The electric charge $Q$ is
\begin{equation}
Q=\frac{1}{16\pi G_{N}}\int\star\mathcal{E}_{(2)}=Q_{0}+\mathcal{O}(c_{i}^{2})=\frac{\sqrt{2(D-2)(D-3)}\, \omega_{D-2}}{16\pi G_{N}\sigma}\, q\ .\label{Q}
\end{equation}
where $\mathcal{E}_{(2)}=\frac{1}{2}\mathcal{E}_{\mu\nu}dx^{\mu}\land dx^{\nu}$. $\star$  denotes the Hodge dual operator.  The conserved charge $M$ associated with a Killing vector $\xi_{t}$
can be computed by the Komar integral
\begin{equation}
M=-\frac{1}{16\pi G_{N}}\int\star d\xi_{t}=M_{0}+\mathcal{O}(c_{i}^{2})=\frac{(D-2)\, \omega_{D-2}}{8\pi G_{N}}\, m\,.\label{M}
\end{equation}
We find that the charge $Q$ and mass $M$ are not corrected at first subleading order, when wirtten in terms of the parameters $m$ and $q$. Note that this does not mean the physical mass and charge are uncorrected, but only that the parameters $m$ and $q$ are particularly suitable for the description of the asymptotic charges. they however appear in a much more complicated way in the corrected solutions, as clearly seen in appendix \ref{AppendixA}.

The full action should include the Gibbons-Hawking-York boundary term
\begin{equation}
I=\frac{1}{16\pi G_{N}}\int_{\mathcal{M}}d^{D}x\sqrt{-g}\mathcal{L}+\frac{1}{8\pi G_{N}}\int_{\partial\mathcal{M}}d^{D-1}x\sqrt{-h}(K[h]-K[h_{b}])\ ,
\end{equation}
where $h_{ij}$ denotes the induced metric on $\partial\mathcal{M}$, $h_{b,ij}$
is the background metric, $K[h]$ is the trace of the extrinsic curvature.
To obtain a finite action, we can also subtract the contribution of background
spacetime as \cite{Gibbons:1976ue}\footnote{A generalization of the Reall-Santos method based on the background subtraction  in asymptotically AdS spacetime can be found in \cite{Guo:2025muo,Guo:2025ohn}}. After some calculations, we obtain the Euclidean on-shell action
\begin{equation}
I=I_{0}+\Delta I+\mathcal{O}(c_{i}^{2})\,,\qquad I_{0}=\frac{r_{0}^{D-2}\, \omega_{D-2}}{4G_{N}(D-3)}\,.
\end{equation}

The above quantities are all explicitly dependent on the position of the position of the two-derivative horizon, $r_0$, which can be freely taken to be any of the roots of the function $h_0 (r_0)$. Thus $r_0$ can take two different values, corresponding to the outer and inner horizon,  $r_0 = r_+$ and $r_0 = r_-$.
We can check the first law of thermodynamics and quantum statistical
relation hold on both horizons:
\begin{equation}
\label{eq:firstlaw}
\delta M=\beta_\pm^{-1}\,\delta S_\pm+\Phi_\pm\, \delta Q+\mathcal{O}(c_{i}^{2})\,,\qquad I=\beta_\pm\, (M-\beta_\pm^{-1}S_\pm-\Phi_\pm Q)+\mathcal{O}(c_{i}^{2})\,.
\end{equation}
Note that in the static case at 2-derivative order, we can actually
compute fully explicitly the on-shell action and thermodynamics in
terms of the chemical potential at each horizon $r_{\pm}$ \cite{Hristov:2023sxg}.
It can be generalized to 4-derivative order
\begin{align}
\label{eq:correctedthermo}
I_{\pm} & =I_{0,\pm}+\Delta I_{\pm}\,,\ \ I_{0,\pm}=\frac{(4\pi \sigma^{2})^{2-D}\, \omega_{D-2}}{(D-3)2^{D}G_{N}}\big((D-3)\beta_{\pm}(2\sigma^2-\frac{D-3}{D-2}\Phi_{\pm}^{2})\big)^{D-2}\,,\nonumber \\
S_{\pm} & =S_{0,\pm}+\Delta S_{\pm}\,,\ \ S_{0,\pm}=(D-3)I_{0,\pm}\,,\ \ \ \Delta S_{\pm}=(D-5)\Delta I_{\pm}\,,
\end{align}
where
\begin{align}
\label{eq:onshellinPhibeta}
\Delta I_{\pm} & =-\frac{\pi\big((D-3)\beta_{\pm}\big)^{D-4}\big(2(D-2)\sigma^2-(D-3)\Phi_{\pm}^{2}\big)^{D-5}\omega_{D-2}}{(8(D-2)\pi \sigma^{2})^{D-3}(3D-7)G_{N}}\Big(4c_{3}(D-2)^{4}(3D-7)\sigma^{4}\nonumber \\
 & -4(c_{3}+c_{6})(3D-7)(D^{2}-5D+6)^{2}\sigma^{2} \Phi_{\pm}^{2}+(D-3)^{3}\big((c_{1}+2c_{2}+7c_{3}+4c_{4}+4c_{5}\nonumber \\
 & +10c_{6}+16c_{7}+8c_{8})D^{2}-(8c_{1}+11c_{2}+36c_{3}+24c_{4}+20c_{5}+46c_{6}+64c_{7}+32c_{8})D\nonumber \\
 & +4(4c_{1}+4c_{2}+12c_{3}+8c_{4}+6c_{5}+13c_{6}+16c_{7}+8c_{8})\big)\Phi_{\pm}^{4}\Big)\,,
\end{align}
with $\beta_{\pm},\ \Phi_{\pm},\ I_{\pm},\ S_{\pm}$ computed
at each horizon $r_{\pm}$ independently. $I_{0,\pm},\ S_{0,\pm}$
denote the 2-derivative results.

\section{Thermodynamics in $D=4$}
\label{sec:4dstatic}
\subsection{Natural variables}
\subsubsection{The ordinary method \label{subsec:The-ordinary-method}}
In this section, we will compute the thermodynamics in natural variables
using the corrected black hole solution. It is obtained by setting $D=4$ in appendix
\ref{AppendixA}

\begin{align}
\Delta h & =\frac{2q^{2}}{5r^{6}}\big[5\big(c_{2}+2(2c_{3}-6c_{4}-c_{5}+c_{6})\big)mr-5(c_{2}+4c_{3}-4c_{4}+2c_{6})r^{2}\nonumber \\
 & -\big(c_{2}+2(2c_{3}-20c_{4}-4c_{5}+c_{6}+8c_{7}+4c_{8})\big)q^{2}\big]\ ,\nonumber \\
\Delta f & =\Delta h-\frac{2q^{2}}{r^{6}}(c_{2}+4c_{3}+20c_{4}+6c_{5}+6c_{6})\big(r(r-2m)+q^{2}\big)\ ,\nonumber \\
\Delta\psi & =-\frac{2\sigma q}{r^{9}}\big[4c_{6}mr^{5}+\frac{1}{5}\big(c_{2}+4c_{3}+20c_{4}+2c_{5}-18c_{6}-32(2c_{7}+c_{8})\big)q^{2}r^{4}\big]\ .
\end{align}
  Up to $\mathcal{O}(c_{i})$, the corrected thermodynamics is obtained by setting $D=4$ in appendix \ref{AppendixA},
\begin{align}
\beta & =\frac{4\pi r_{0}}{1-\frac{q^{2}}{r_{0}^{2}}}+\frac{16\pi q^{2}\big(5c_{6}r_{0}^{4}+q^{2}f_{4D}(q^{2}-2r_{0}^{2})\big)}{5r_{0}(r_{0}^{2}-q^{2})^{3}}\,,\nonumber \\
\Phi & =\frac{2\sigma q}{r_{0}}+\frac{4g\left(5c_{6}qr_{0}^{4}+q^{3}f_{4D}(q^{2}-2r_{0}^{2})\right)}{5r_{0}^{5}(r_{0}^{2}-q^{2})}\,,\nonumber \\
S & =\frac{\pi r_{0}^{2}}{G_{N}}+\frac{4\pi(-5(c_{3}+c_{6})q^{2}r_{0}^{2}+5c_{3}r_{0}^{4}+q^{4}f_{4D})}{5G_{N}r_{0}^{2}(r_{0}^{2}-q^{2})}\,,\nonumber \\
I & =\frac{\pi r_{0}^{2}}{G_{N}}-\frac{4\pi(-5(c_{3}+c_{6})q^{2}r_{0}^{2}+5c_{3}r_{0}^{4}+q^{4}f_{4D})}{5G_{N}r_{0}^{2}(r_{0}^{2}-q^{2})}\,,\nonumber \\
M & =\frac{q^{2}+r_{0}^{2}}{2G_{N}r_{0}}\,,\qquad Q=\frac{q}{2\sigma G_{N}}\,.\label{Thermo4D}
\end{align}
where $\omega_{D-2}= \omega_2=4\pi$, and we defined for short $f_{4D}:=c_{2}+4c_{3}+2c_{5}+7c_{6}+8(2c_{7}+c_{8})$. From $f_{0}(r_{\pm})=0$, we find
\begin{equation}
m=\frac{1}{2}(r_{-}+r_{+})\ ,\qquad q=\sqrt{r_{-}r_{+}}\ ,\label{mq4D}
\end{equation}
where $r_{+},\ r_{-}$ represent the outer and inner horizon radius
of the leading order, respectively. Substituting eq.(\ref{mq4D})
into the corrected thermodynamics \eqref{Thermo4D}, we can write
the thermodynamics in terms of the horizon radius $r_{\pm}$ \cite{Hristov:2023sxg}.
We find the first law of thermodynamics and quantum statistical relation
hold, \eqref{eq:firstlaw}, with the same asymptotic charges on each horizon, $M=M_{+}=M_{-},\ Q=Q_{+}=Q_{-}$. The on-shell actions at the two horizons, given generically in \eqref{eq:correctedthermo}, with the substitution $D=4$ not resulting in any notable simplification.

Instead, we can now compute the left- and right-moving variables as defined in \cite{Hristov:2023sxg,Hristov:2023cuo,Hristov:2024lzr},
\begin{align}
\beta_{l,r} :=\frac{1}{2}(\beta_{+}\pm\beta_{-})\,,\ \ \varphi_{l,r}:=\frac{1}{2}(\beta_{+}\Phi_{+}\pm\beta_{-}\Phi_{-})\,,\ \ S_{l,r}:=\frac{1}{2}(S_{+}\pm S_{-})\,,\ \ I_{l,r}:=\frac{1}{2}(I_{+}\pm I_{-})\,.
\end{align}
Explicitly, we find the left-moving sector given by
\begin{align}
\beta_{l} & =2\pi(r_{-}+r_{+})-\frac{8\pi(r_{-}+r_{+})f_{4D}}{5r_{-}r_{+}}\,,\nonumber \\
\varphi_{l} & =4\pi \sigma\sqrt{r_{-}r_{+}}-\frac{8\pi \sigma\sqrt{r_{-}r_{+}}(r_{-}+r_{+}){}^{2}f_{4D}}{5r_{-}^{2}r_{+}^{2}}\,,\nonumber \\
S_{l} & =\frac{\pi(r_{-}^{2}+r_{+}^{2})}{2G_{N}}+\frac{2\pi}{5G_{N}}\big(5(2c_{3}+c_{6})-f_{4D}\frac{r_{-}^{2}+r_{+}^{2}+r_{-}r_{+}}{r_{-}r_{+}}\big)\ ,\nonumber \\
I_{l} & =\frac{\pi(r_{-}^{2}+r_{+}^{2})}{2G_{N}}-\frac{2\pi}{5G_{N}}\big(5(2c_{3}+c_{6})-f_{4D}\frac{r_{-}^{2}+r_{+}^{2}+r_{-}r_{+}}{r_{-}r_{+}}\big)\,.\label{Left4D1}
\end{align}
The right-moving sector is instead
\begin{align}
\beta_{r} & =\frac{2\pi(r_{-}^{2}+r_{+}^{2})}{(r_{+}-r_{-})}-\frac{8\pi\big(f_{4D}(-2r_{+}r_{-}^{3}+r_{-}^{4}-2r_{-}r_{+}^{3}+r_{+}^{4})+10c_{6}r_{-}^{2}r_{+}^{2}\big)}{5r_{-}r_{+}(r_{-}-r_{+}){}^{3}}\,,\nonumber \\
\varphi_{r} & =\frac{4\pi \sigma\sqrt{r_{-}r_{+}}(r_{-}+r_{+})}{(r_{+}-r_{-})}-\frac{8\pi \sigma\sqrt{r_{-}r_{+}}(r_{-}+r_{+})\big((r_{-}^{4}-2r_{+}r_{-}^{3}-2r_{+}^{3}r_{-}+r_{+}^{4})f_{4D}+10c_{6}r_{-}^{2}r_{+}^{2}\big)}{5r_{-}^{2}(r_{-}-r_{+})^{3}r_{+}^{2}}\,,\nonumber \\
S_{r} & =\frac{(r_{+}^{2}-r_{-}^{2})\pi}{2G_{N}}+\frac{2\pi\left(5c_{6}r_{-}r_{+}(r_{-}+r_{+})-(r_{-}^{3}+r_{+}^{3})f_{4D}\right)}{5G_{N}r_{-}(r_{-}-r_{+})r_{+}}\,,\nonumber \\
I_{r} & =\frac{(r_{+}^{2}-r_{-}^{2})\pi}{2G_{N}}-\frac{2\pi\left(5c_{6}r_{-}r_{+}(r_{-}+r_{+})-(r_{-}^{3}+r_{+}^{3})f_{4D}\right)}{5G_{N}r_{-}(r_{-}-r_{+})r_{+}}\,.
\end{align}
These definitions lead to an alternative version of the first law and quantum
statistical relation
\begin{equation}
\beta_{l,r}\delta M=\delta S_{l,r}+\varphi_{l,r}\delta Q\,,\qquad I_{l,r}=\beta_{l,r}\, M - S_{l,r}-\varphi_{l,r}\, Q\ ,\label{LRLaw}
\end{equation}

It should be noted that the establishment of eq.(\ref{LRLaw}) is
based on the premise that the corrected asymptotic charges still satisfy the first law with the chemical potentials defined on each horizon separately. While this is a simple self-consistency check on the corrected solution, which is simply verified in our case, this becomes technically more complicated upon choosing different parametrizations for different horizons. In a sense this is precisely the case we turn to describe now, with the RS method simply leading to a different parametrization of the corrected thermodynamics.

\subsubsection{The Reall-Santos method}

The two-derivative Einstein-Maxwell theory, given by $\mathcal{L}=R-\frac{1}{4}F^{\mu\nu}F_{\mu\nu}$,
admits the static charged black hole solution
\begin{align}
ds_{D}^{2}= & -h_{0}(r)dt^{2}+\frac{dr^{2}}{f_{0}(r)}+r^{2}d\Omega_{D-2}^{2}\,,\ h_{0}=f_{0}=1-\frac{2\bar{m}}{r^{D-3}}+\frac{\bar{q}^{2}}{r^{2(D-3)}}\,,\nonumber \\
A= & \psi_{0}(r)dt\ ,\ \ \ \psi_{0}=-\sqrt{\frac{2(D-2)}{(D-3)}}\frac{\sigma\bar{q}}{r^{D-3}}\ ,\label{2dSolution}
\end{align}
where we were careful to choose different parameters $\bar{m},\bar{q}$
and $\bar{m}$ has been replaced via $\bar{m}=\frac{\bar{q}^{2}}{2\bar{r}_{0}^{D-3}}+\frac{1}{2}\bar{r}_{0}^{D-3}$. We did this to emphasize the crucial fact that these effective ``coordinates'' that parametrize the solution are no longer the same as the $m, q, r_0$ parameters describing the corrected solutions in the previous sections.

In $D=4$, we can obtain the chemical potentials by the Reall-Santos method \cite{Reall:2019sah,Hu:2023gru,Ma:2024ynp}, i.e.\ keeping them at their two-derivative values:
\begin{align}
\beta & =\frac{4\pi\bar{r}_{0}^{3}}{\bar{r}_{0}^{2}-\bar{q}^{2}}\,,\qquad \Phi=\frac{2\sigma \bar{q}}{\bar{r}_{0}^{D-3}}\,,\label{PhiRS}
\end{align}
Up to order $\mathcal{O}(c_{i})$, the on-shell Euclidean action of
black hole in theory (\ref{Action}) could be obtained by evaluating
the total action (\ref{Action}) on the uncorrected solution (\ref{2dSolution}),
\begin{equation}
I=\frac{\pi\bar{r}_{0}^{2}}{G_{N}}+\frac{4\pi[5\bar{r}_{0}^{2}\big((c_{3}+c_{6})\bar{q}^{2}-c_{3}\bar{r}_{0}^{2}\big)-\bar{q}^{4}f_{4D}]}{5G_{N}\bar{r}_{0}^{2}(\bar{r}_{0}^{2}-\bar{q}^{2})}\,.\label{ActionIRS}
\end{equation}
The potential corresponding to the on-shell action (\ref{ActionIRS})
is fixed to the leading order (\ref{PhiRS}). According to the thermodynamic
relations
\begin{align}
Q & =-\frac{1}{\beta}\left(\frac{\partial I}{\partial\Phi_{}}\right)_{\beta},\ \ M=\left(\frac{\partial I}{\partial\beta}\right)_{\Phi}+\Phi Q\ ,\ \ S=\beta M-I-\beta\Phi Q\ ,
\end{align}
we obtain
\begin{align}
S & =\frac{\pi\bar{r}_{0}^{2}}{G_{N}}-\frac{4\pi[5\bar{r}_{0}^{2}\big((c_{3}+c_{6})\bar{q}^{2}-c_{3}\bar{r}_{0}^{2}\big)-\bar{q}^{4}f_{4D}]}{5G_{N}\bar{r}_{0}^{2}(\bar{r}_{0}^{2}-\bar{q}^{2})}\ ,\nonumber \\
Q & =\frac{\bar{q}}{2\sigma G_{N}}+\frac{20\pi c_{6}\bar{q}\bar{r}_{0}^{4}-8\pi\bar{q}^{3}\bar{r}_{0}^{2}f_{4D}+4\pi\bar{q}^{5}f_{4D}}{20\pi \sigma G_{N}\bar{q}^{2}\bar{r}_{0}^{4}-20\pi \sigma G_{N}\bar{r}_{0}^{6}}\,,\nonumber \\
&=\frac{\bar{q}}{2\sigma G_{N}}-\frac{\bar{q}\big(5c_{6}(\sqrt{\bar{m}^{2}-\bar{q}^{2}}+\bar{m})^{4}-2\bar{q}^{2}f_{4D}(\sqrt{\bar{m}^{2}-\bar{q}^{2}}+\bar{m})^{2}+\bar{q}^{4}f_{4D}\big)}{10gG_{N}(\sqrt{\bar{m}^{2}-\bar{q}^{2}}+\bar{m})^{4}(\bar{m}\sqrt{\bar{m}^{2}-\bar{q}^{2}}+\bar{m}^{2}-\bar{q}^{2})}\,,\nonumber\\
M & =\frac{\bar{q}^{2}+\bar{r}_{0}^{2}}{2G_{N}\bar{r}_{0}}+\frac{20\pi c_{6}\bar{q}^{2}\bar{r}_{0}^{4}-8\pi\bar{q}^{4}\bar{r}_{0}^{2}f_{4D}+4\pi\bar{q}^{6}f_{4D}}{10\pi G_{N}\bar{q}^{2}\bar{r}_{0}^{5}-10\pi G_{N}\bar{r}_{0}^{7}}\,,\nonumber\\
&=\frac{\bar{m}}{G_{N}}+\frac{2\bar{q}^{2}\big(5c_{6}(\sqrt{\bar{m}^{2}-\bar{q}^{2}}+\bar{m})^{4}-2\bar{q}^{2}f_{4D}(\sqrt{\bar{m}^{2}-\bar{q}^{2}}+\bar{m})^{2}+\bar{q}^{4}f_{4D}\big)}{5G_{N}(\sqrt{\bar{m}^{2}-\bar{q}^{2}}+\bar{m})^{5}\big(\bar{q}^{2}-(\sqrt{\bar{m}^{2}-\bar{q}^{2}}+\bar{m})^{2}\big)}\,,
\label{SQMIRS}
\end{align}
where $\bar{r}_0=\bar{m}\pm\sqrt{\bar{m}^2-\bar{q}^2}$. Although the thermodynamic results given by the RS method eqs.(\ref{PhiRS})-(\ref{SQMIRS}) do not seem consistent with
those given by the modified solution (\ref{Thermo4D}), they have
equivalent thermodynamic relations. Crucially, we now observe explicitly the subtlety explained in the introduction: naively substituting the values of the two different horizons, $\bar r_\pm$, in the expressions for $M$ and $Q$, leads to the puzzling conclusion that $M_+ \neq M_-$ and $Q_+ \neq Q_-$. This is an artifact of the RS method, and we present two approaches to handle this subtlety. 

First, as a short-cut, we observe that we can already choose the following change of effective coordinates:
\begin{equation}
\bar{r}_{0}=r_{0}\,,\ \ \ \bar{q}=q+\delta q\,,\ \ \delta q=\frac{2\left(5c_{6}qr_{0}^{4}+q^{3}f_{4D}(q^{2}-2r_{0}^{2})\right)}{5r_{0}^{6}-5q^{2}r_{0}^{4}}\,.
\label{dq}
\end{equation}
Substituting eq.(\ref{dq}) into the thermodynamics eqs.(\ref{PhiRS})-(\ref{SQMIRS}), we find the result from Reall-Santos method agrees with the one from the ordinary method (\ref{Thermo4D}). It then follows that upon substitution $r_0=r_\pm$, we find that $M_+=M_-, Q_+=Q_-$. 

Alternatively, we can define a more algorithmic procedure. Since the RS method selects different integration constants at the inner $(m_+, \ q_+)$ and outer $(m_-,\ q_-)$ horizons, this seems to lead to
\begin{equation}
M_{\pm}=M(\bar{m}=m_{\pm},\,\bar{q}=q_{\pm})\ , \qquad Q_{\pm}=Q(\bar{m}=m_{\pm},\,\bar{q}=q_{\pm})\,,
\end{equation} 
Now, we can impose by hand the conditions $M_+ \stackrel{!}{=} M_-, \, Q_+ \stackrel{!}{=} Q_-$, such that the inconsistency in the original RS parametrization is removed at the expense of constraining the parametrization $m_{-} =m_{+}+\delta m_{-},\ q_{-}=q_{+}+\delta q_{-}$,
\begin{align}
\delta m_{-} & =\frac{2\big(5c_{6}q_{+}^{4}(q_{+}^{2}-2m_{+}^{2})+f_{4D}(64m_{+}^{4}q_{+}^{2}-34m_{+}^{2}q_{+}^{4}-32m_{+}^{6}+3q_{+}^{6})\big)}{5q_{+}^{6}\sqrt{m_{+}^{2}-q_{+}^{2}}}\,,\nonumber \\
\delta q_{-} & =-\frac{2m_{+}\big(5c_{6}q_{+}^{4}+f_{4D}(16m_{+}^{4}+11q_{+}^{4}-28m_{+}^{2}q_{+}^{2})\big)}{5q_{+}^{5}\sqrt{m_{+}^{2}-q_{+}^{2}}}\,.
\end{align}
Using these values of $m_-,\,q_-$, we find 
\begin{align}
M_{+} & =M_{-}=\frac{m_{+}}{G_{N}}+\frac{2q_{+}^{2}\big(5c_{6}(\sqrt{m_{+}^{2}-q_{+}^{2}}+m_{+}){}^{4}-2q_{+}^{2}f_{4D}(\sqrt{m_{+}^{2}-q_{+}^{2}}+m_{+})^{2}+q_{+}^{4}f_{4D}\big)}{5G_{N}(\sqrt{m_{+}^{2}-q_{+}^{2}}+m_{+})^{5}\big(q_{+}^{2}-(\sqrt{m_{+}^{2}-q_{+}^{2}}+m_{+})^{2}\big)}\,,\nonumber \\
Q_{+} & =Q_{-}=\frac{q_{+}}{2\sigma G_{N}}-\frac{q_{+}\big(5c_{6}(\sqrt{m_{+}^{2}-q_{+}^{2}}+m_{+}){}^{4}-2q_{+}^{2}f_{4D}(\sqrt{m_{+}^{2}-q_{+}^{2}}+m_{+})^{2}+q_{+}^{4}f_{4D}\big)}{10\sigma G_{N}(\sqrt{m_{+}^{2}-q_{+}^{2}}+m_{+})^{4}\big(m_{+}(\sqrt{m_{+}^{2}-q_{+}^{2}}+m_{+})-q_{+}^{2}\big)}\,,
\end{align}
which restores consistency along both horizons simultaneously, and allows us to address the natural variables definition as below.

In addition, for the sake simplicity, we are again allowed to fix the integration constants $m_+,\ q_+$ such that $M_\pm,\, Q_\pm$ do not receive first-order corrections,
\begin{align}
m_{+} & =m+\delta m_{+}\,,\ \ \ q_{+}=q+\delta q_{+}\,,\nonumber \\
\delta m_{+} & =-\frac{2q^{2}\big(5c_{6}(\sqrt{m^{2}-q^{2}}+m)^{4}-2q^{2}f_{4D}(\sqrt{m^{2}-q^{2}}+m)^{2}+q^{4}f_{4D}\big)}{5(\sqrt{m^{2}-q^{2}}+m)^{5}\big(q^{2}-(\sqrt{m^{2}-q^{2}}+m)^{2}\big)},\nonumber \\
\delta q_{+} & =\frac{q\big(5c_{6}(\sqrt{m^{2}-q^{2}}+m)^{4}-2q^{2}f_{4D}(\sqrt{m^{2}-q^{2}}+m)^{2}+q^{4}f_{4D}\big)}{5(\sqrt{m^{2}-q^{2}}+m)^{4}(m(\sqrt{m^{2}-q^{2}}+m)-q^{2})}\,.
\end{align}
The above choice manifestly brings the results obtained by the RS method to exact agreement with those obtained by the ordinary method (\ref{Thermo4D}).

Once we have obtained this augmented thermodynamics by the Reall-Santos method, we can use the ideas of the previous section to look at the thermodynamics in terms of natural variables.

\subsection{$I_{l,r}(\beta_{l,r},\varphi_{l,r})$}

Next, we will explicitly write the action $I_{l,r}$ in terms of the left- and
right-moving potentials $\beta_{l,r},\,\varphi_{l,r}$. 

\subsubsection*{Left-moving sector}

Since the choice of parametrization, or integration constants, does not affect the thermodynamic
relations $I_{l,r}(\beta_{l,r},\varphi_{l,r})$, we choose new variables $\tilde{r}_{\pm}$ that are chosen to preserve the chemical potentials at leading order $\beta_{l}\rightarrow\beta_{l,0},\,\varphi_{l,}\rightarrow\varphi_{l,0}$~\footnote{Note that this choice can be seen as inspired by the RS trick, but is manifestly different. The parametrization here keeps the left-moving potentials unperturbed, but would instead perturb the right-moving potentials. We make the opposite choice when looking at the right-moving sector, again for the sake of simplifying the calculation.}
\begin{equation}
r_{+}=\tilde{r}_{+}+\delta\tilde{r}_{+}\,,\qquad r_{-}=\tilde{r}_{-}+\delta\tilde{r}_{-}\,,\label{rori4D1}
\end{equation}
where 
\begin{equation}
\delta\tilde{r}_{+}=\frac{4(\tilde{r}_{-}+\tilde{r}_{+})f_{4D}}{5(\tilde{r}_{-}-\tilde{r}_{+})\tilde{r}_{+}}\,,\ \ \delta\tilde{r}_{-}=-\frac{4(\tilde{r}_{-}+\tilde{r}_{+})f_{4D}}{5\tilde{r}_{-}(\tilde{r}_{-}-\tilde{r}_{+})}\,.
\end{equation}
Substituting eq.(\ref{rori4D1}) into eq.(\ref{Left4D1}), we list
all the left-moving sector up to $\mathcal{O}(c_{i})$
\begin{align}
\beta_{l} & =\beta_{l,0}=2\pi(\tilde{r}_{-}+\tilde{r}_{+})\,,\ \ \varphi_{l}=\varphi_{l,0}=4\pi \sigma\sqrt{\tilde{r}_{-}\tilde{r}_{+}}\ ,\nonumber \\
S_{l} & =\frac{(\tilde{r}_{-}^{2}+\tilde{r}_{+}^{2})\pi}{2G_{N}}+\frac{2\pi}{5G_{N}}\big(5(2c_{3}+c_{6})-f_{4D}\frac{\tilde{r}_{-}^{2}+\tilde{r}_{+}\tilde{r}_{-}+\tilde{r}_{+}^{2}}{\tilde{r}_{-}\tilde{r}_{+}}\big)\ ,\nonumber \\
I_{l} & =\frac{(\tilde{r}_{-}^{2}+\tilde{r}_{+}^{2})\pi}{2G_{N}}-\frac{2\pi}{5G_{N}}\big(5(2c_{3}+c_{6})-f_{4D}\frac{\tilde{r}_{-}^{2}+\tilde{r}_{+}\tilde{r}_{-}+\tilde{r}_{+}^{2}}{\tilde{r}_{-}\tilde{r}_{+}}\big)\,,\nonumber \\
M & =\frac{(\tilde{r}_{-}+\tilde{r}_{+})}{2G_{N}}+\frac{2(\tilde{r}_{-}+\tilde{r}_{+})f_{4D}}{5G_{N}\tilde{r}_{-}\tilde{r}_{+}}\,,\ \ Q=\frac{\sqrt{\tilde{r}_{-}\tilde{r}_{+}}}{2\sigma G_{N}}+\frac{(\tilde{r}_{-}+\tilde{r}_{+})^{2}f_{4D}}{5\sigma G_{N}(\tilde{r}_{-}\tilde{r}_{+}){}^{3/2}}\,.\label{LeftThermodynamics}
\end{align}
The parameters characterizing the black hole solution can be explicitly
inverted in terms of the new variables 
\begin{equation}
\tilde{r}_{\pm}=\frac{\sigma\beta_{l}\pm\sqrt{(\sigma^{2}\beta_{l}^{2}-\varphi_{l}^{2})}}{4\pi \sigma}\,.\label{rotrit}
\end{equation}
Substituting eq.(\ref{rotrit}) into the left-moving thermodynamics
eq.(\ref{LeftThermodynamics}), we obtain the following results
\begin{align}
I_{l}(\beta_{l},\varphi_{l}) & =\frac{(2\sigma^{2}\beta_{l}^{2}-\varphi_{l}^{2})}{16\pi G_{N}\sigma^{2}}+\frac{2\pi\big(4\sigma^{2}f_{4D}\beta_{l}^{2}-\varphi_{l}^{2}(10c_{3}+5c_{6}+f_{4D})\big)}{5G_{N}\varphi_{l}^{2}}\,,\nonumber \\
S_{l}(\beta_{l},\varphi_{l}) & =\frac{(2\sigma^{2}\beta_{l}^{2}-\varphi_{l}^{2})}{16\pi G_{N}\sigma^{2}}-\frac{2\pi\big(4\sigma^{2}f_{4D}\beta_{l}^{2}-\varphi_{l}^{2}(10c_{3}+5c_{6}+f_{4D})\big)}{5G_{N}\varphi_{l}^{2}}\,,\nonumber \\
M(\beta_{l},\varphi_{l}) & =\frac{\beta_{l}}{4\pi G_{N}}+\frac{16\pi \sigma^{2}f_{4D}\beta_{l}}{5G_{N}\varphi_{l}^{2}}\,,\ \ Q(\beta_{l},\varphi_{l})=\frac{\varphi_{l}}{8\pi \sigma^{2}G_{N}}+\frac{16\pi \sigma^{2}f_{4D}\beta_{l}^{2}}{5G_{N}\varphi_{l}^{3}}\,.\label{Left4D}
\end{align}
For the left-moving on-shell action $I_l$, we find the relations
\begin{equation}
		\frac{\partial I_{l}}{\partial \beta_{l}} = M \ ,  \qquad \frac{\partial I_{l}}{\partial \varphi_{l}} = - Q\ .
\end{equation}

\subsubsection*{Right-moving sector}

Similarly, we are free to choose new integration constants $\hat{r}_{\pm}$ making
the potential energy fixed at the leading order $\beta_{r}\rightarrow\beta_{r,0},\,\varphi_{r,}\rightarrow\varphi_{r,0}$
\begin{equation}
r_{+}=\hat{r}_{+}+\delta\hat{r}_{+}\,,\ \ r_{-}=\hat{r}_{-}+\delta\hat{r}_{-}\,,
\end{equation}
where 
\begin{align}
\delta\hat{r}_{+} & =-\frac{4\big(10c_{6}\hat{r}_{-}^{2}\hat{r}_{+}^{2}+(\hat{r}_{-}^{4}-2\hat{r}_{+}\hat{r}_{-}^{3}-2\hat{r}_{+}^{3}\hat{r}_{-}+\hat{r}_{+}^{4})f_{4D}\big)}{5(\hat{r}_{-}-\hat{r}_{+})^{2}\hat{r}_{+}(\hat{r}_{-}^{2}+4\hat{r}_{+}\hat{r}_{-}+\hat{r}_{+}^{2})}\,,\nonumber \\
\delta\hat{r}_{-} & =-\frac{4\big(10c_{6}\hat{r}_{-}^{2}\hat{r}_{+}^{2}+(\hat{r}_{-}^{4}-2\hat{r}_{+}\hat{r}_{-}^{3}-2\hat{r}_{+}^{3}\hat{r}_{-}+\hat{r}_{+}^{4})f_{4D}\big)}{5\hat{r}_{-}(\hat{r}_{-}-\hat{r}_{+})^{2}(\hat{r}_{-}^{2}+4\hat{r}_{+}\hat{r}_{-}+\hat{r}_{+}^{2})}\,.
\end{align}
In this parametrization, we list all the right-moving thermodynamics
below
\begin{align}
\beta_{r} & =\beta_{r,0}=\frac{2\pi(\hat{r}_{-}^{2}+\hat{r}_{+}^{2})}{(\hat{r}_{+}-\hat{r}_{-})}\,,\ \ \varphi_{r}=\varphi_{r,0}=\frac{4\pi \sigma\sqrt{\hat{r}_{-}\hat{r}_{+}}(\hat{r}_{-}+\hat{r}_{+})}{(\hat{r}_{+}-\hat{r}_{-})}\,,\nonumber \\
S_{r} & =\frac{(\hat{r}_{+}^{2}-\hat{r}_{-}^{2})\pi}{2G_{N}}+\frac{2\pi\big(5c_{6}\hat{r}_{-}\hat{r}_{+}(\hat{r}_{-}+\hat{r}_{+})-(\hat{r}_{-}^{3}+\hat{r}_{+}^{3})f_{4D}\big)}{5G_{N}\hat{r}_{-}(\hat{r}_{-}-\hat{r}_{+})\hat{r}_{+}}\,,\nonumber \\
I_{r} & =\frac{(\hat{r}_{+}^{2}-\hat{r}_{-}^{2})\pi}{2G_{N}}-\frac{2\pi\big(5c_{6}\hat{r}_{-}\hat{r}_{+}(\hat{r}_{-}+\hat{r}_{+})-(\hat{r}_{-}^{3}+\hat{r}_{+}^{3})f_{4D}\big)}{5G_{N}\hat{r}_{-}(\hat{r}_{-}-\hat{r}_{+})\hat{r}_{+}}\,.\nonumber \\
M & =\frac{(\hat{r}_{-}+\hat{r}_{+})}{2G_{N}}-\frac{2(\hat{r}_{-}+\hat{r}_{+})\big(10c_{6}\hat{r}_{-}^{2}\hat{r}_{+}^{2}+(\hat{r}_{-}^{4}-2\hat{r}_{+}\hat{r}_{-}^{3}-2\hat{r}_{+}^{3}\hat{r}_{-}+\hat{r}_{+}^{4})f_{4D}\big)}{5G_{N}\hat{r}_{-}(\hat{r}_{-}-\hat{r}_{+})^{2}\hat{r}_{+}(\hat{r}_{-}^{2}+4\hat{r}_{+}\hat{r}_{-}+\hat{r}_{+}^{2})}\,,\nonumber \\
Q & =\frac{\sqrt{\hat{r}_{-}\hat{r}_{+}}}{2\sigma G_{N}}-\frac{(\hat{r}_{-}+\hat{r}_{+})\big(10c_{6}\hat{r}_{-}^{2}\hat{r}_{+}^{2}+(\hat{r}_{-}^{4}-2\hat{r}_{+}\hat{r}_{-}^{3}-2\hat{r}_{+}^{3}\hat{r}_{-}+\hat{r}_{+}^{4})f_{4D}\big)}{5\sigma G_{N}(\hat{r}_{-}-\hat{r}_{+})^{2}(\hat{r}_{-}\hat{r}_{+}){}^{3/2}(\hat{r}_{-}^{2}+4\hat{r}_{+}\hat{r}_{-}+\hat{r}_{+}^{2})}\,.
\end{align}
The parameters characterizing the black hole solution can be explicitly
inverted in terms of the new variables
\begin{align}
\hat{r}_{\pm} & =\frac{1}{8\pi \sigma}\big(\sqrt{(f_{\beta,\varphi}+\sigma\beta_{r})^{2}-4\varphi_{r}^{2}}\pm(3\sigma \beta_{r}-f_{\beta,\varphi})\big)\,,
\end{align}
where $f_{\beta,\varphi}:=\sqrt{\sigma^{2}\beta_{r}^{2}+2\varphi_{r}^{2}}$.
The right-moving sector is
\begin{align}
I_{r} & =\frac{(3\sigma \beta_{r}-f_{\beta,\varphi})^{\frac{3}{2}}\sqrt{f_{\beta,\varphi}+\sigma\beta_{r}}}{32\pi \sigma^{2}G_{N}}+\frac{2\pi\sqrt{(f_{\beta,\varphi}+\sigma\beta_{r})^{2}-4\varphi_{r}^{2}}}{5G_{N}(8\sigma ^{2}\beta_{r}^{2}f_{\beta,\varphi}-8\sigma ^{3}\beta_{r}^{3}+\varphi_{r}^{2}f_{\beta,\varphi}-7\sigma \beta_{r}\varphi_{r}^{2})}\big(\nonumber \\
 & -2\sigma ^{2}\beta_{r}^{2}(5c_{6}+4f_{4D})-\varphi_{r}^{2}(5c_{6}+f_{4D})+2(5c_{6}+2f_{4D})\sigma\beta_{r}f_{\beta,\varphi}\big)\,,\nonumber \\
S_{r} & =\frac{(3\sigma \beta_{r}-f_{\beta,\varphi})^{\frac{3}{2}}\sqrt{f_{\beta,\varphi}+\sigma\beta_{r}}}{32\pi \sigma^{2}G_{N}}-\frac{2\pi\sqrt{(f_{\beta,\varphi}+\sigma\beta_{r})^{2}-4\varphi_{r}^{2}}}{5G_{N}(8\sigma ^{2}\beta_{r}^{2}f_{\beta,\varphi}-8\sigma ^{3}\beta_{r}^{3}+\varphi_{r}^{2}f_{\beta,\varphi}-7\sigma \beta_{r}\varphi_{r}^{2})}\big(\nonumber \\
 & -2\sigma ^{2}\beta_{r}^{2}\left(5c_{6}+4f_{4D}\right)-\varphi_{r}^{2}\left(5c_{6}+f_{4D}\right)+2(5c_{6}+2f_{4D})\sigma\beta_{r}f_{\beta,\varphi}\big)\,,\nonumber \\
M & =\frac{\sqrt{3\sigma \beta_{r}-f_{\beta,\varphi}}\sqrt{f_{\beta,\varphi}+\sigma\beta_{r}}}{8\pi \sigma G_{N}}+\frac{2\pi \sigma\sqrt{(f_{\beta,\varphi}+\sigma\beta_{r})^{2}-4\varphi_{r}^{2}}}{5G_{N}f_{\beta,\varphi}\left(\varphi_{r}^{3}-4\sigma ^{2}\beta_{r}^{2}\varphi_{r}\right){}^{2}}\Big(10\sigma ^{2}\varphi_{r}^{2}\beta_{r}^{2}(c_{6}-f_{4D})(\sigma\beta_{r}\nonumber \\
 & -f_{\beta,\varphi})+\varphi_{r}^{4}\left(f_{4D}-5c_{6}\right)f_{\beta,\varphi}+\sigma\beta_{r}\varphi_{r}^{4}\left(9f_{4D}-25c_{6}\right)-32\sigma ^{4}f_{4D}\beta_{r}^{4}(\sigma\beta_{r}+f_{\beta,\varphi})\Big)\,,\nonumber \\
Q & =\frac{\varphi _r \sqrt{3 \sigma  \beta _r-f_{\beta ,\varphi }}}{8 \pi  \sigma ^2 G_N \sqrt{f_{\beta ,\varphi }+\sigma  \beta _r}}+[5G_{N}\big(2\sigma \beta_{r}(f_{\beta,\varphi}-\sigma\beta_{r})-\varphi_{r}^{2}\big)^{\frac{3}{2}}\big(\sigma\beta_{r}(3f_{\beta,\varphi}-\sigma\beta_{r})\nonumber \\
 & -2\varphi_{r}^{2}\big)(f_{\beta,\varphi}-3\sigma \beta_{r})]^{-1}\times8\pi \sigma\beta_{r}\big(4\sigma ^{4}\beta_{r}^{4}(10c_{6}+7f_{4D})+6\sigma ^{2}\beta_{r}^{2}\varphi_{r}^{2}(10c_{6}+f_{4D})\nonumber \\
 & +\varphi_{r}^{4}(5c_{6}-f_{4D})+2\sigma \beta_{r}\varphi_{r}^{2}f_{\beta,\varphi}(f_{4D}-10c_{6})-20\sigma ^{3}\beta_{r}^{3}f_{\beta,\varphi}(2c_{6}+f_{4D})\big)\label{Ir4D}\ .
\end{align}
For the right-moving on-shell actions $I_r$, we find the relations
\begin{equation}
		\frac{\partial I_{r}}{\partial \beta_{r}} = M \ ,  \qquad \frac{\partial I_{r}}{\partial \varphi_{r}} = - Q\ .
\end{equation}

\subsection{$D=4$ supergravity}

For a given theory of gravity, the above thermodynamics can be simplified.
For example, we consider the bosonic four-derivative action of the
supergravity with zero cosmological constant \cite{Bergshoeff:1980is,Butter:2013lta,Bobev:2020egg},~\footnote{Note that we can also consider a third susy-invariant, which corresponds to the fermionic completion of the $R^2$ term, \cite{deWit:2006gn,Kuzenko:2015jxa}. It is easy to see that this contribution simply vanishes, since the coefficient $c_1$ does not enter the corrected thermodynamics.}
\begin{align}
\Delta\mathcal{L} & =(d_{1}-d_{2})\mathcal{L}_{{\rm W}^{2}}+d_{2}\ensuremath{\mathcal{L}_{{\rm GB}^{2}}}\ ,\nonumber \\
\mathcal{L}_{{\rm W}^{2}} & =C^{\mu\nu\rho\sigma}C_{\mu\nu\rho\sigma}-\frac{1}{8\sigma ^{4}}(F^{\mu\nu}F_{\mu\nu})^{2}+\frac{1}{2\sigma ^{4}}F_{\mu\nu}F^{\nu\lambda}F_{\lambda\delta}F^{\delta\mu}\nonumber \\
 & -\frac{2}{\sigma^{2}}R_{\mu\nu}F^{\mu\lambda}F_{\ \lambda}^{\nu}+\frac{1}{2\sigma ^{2}}RF^{\mu\nu}F_{\mu\nu}+\frac{2}{\sigma^{2}}\nabla^{\mu}F_{\mu\rho}\nabla_{\nu}F^{\nu\rho}\,,\nonumber \\
\ensuremath{\mathcal{L}_{{\rm GB}^{2}}} & =R^{2}-4R_{\mu\nu}R^{\mu\nu}+R^{\mu\nu\rho\sigma}R_{\mu\nu\rho\sigma}\,.\label{Sugra4D}
\end{align}
In this parametrization, $d_1$ is the coefficient in front of the Weyl$^2$ invariant, \cite{Bergshoeff:1980is}, while $d_2$ is the coefficient in front of the $\log$-invariant, \cite{Butter:2013lta}.

If we impose the following constraints on eq.(\ref{dL}), we find
\begin{align}
c_{4}&=\frac{1}{4}(c_{2}+4c_{3}),\ c_{5}=-(c_{2}+4c_{3}),\ c_{6}=0,\ c_{7}=-\frac{1}{16}(c_{2}+4c_{3}),\ c_{8}=\frac{1}{4}(c_{2}+4c_{3})\,,\nonumber\\
c_{1}&=-\frac{1}{3}(c_{2}+c_{3}),\ c_{9}=c_{2}+4c_{3},\ c_{2}=-2(d_{1}+d_{2}),\ c_{3}=d_{1}\,,
\label{4DSUGRAConstraints}
\end{align}
the 4-derivative action (\ref{dL}) can be written explicitly in the
form of eq.(\ref{Sugra4D}). These constraints lead to
\begin{align*}
f_{4D} & =c_{2}+4c_{3}+2c_{5}+7c_{6}+8(2c_{7}+c_{8})=0\,.
\end{align*}
It can be checked that in this case the two-derivative solutions do not receive any modifications, such that the chemical potentials remain unchanged. We can use the field redefinitions to prove it. First, we consider the following coefficient constraints \cite{Hu:2023qhs}
\begin{equation}
c_{4}=\frac{1}{4}(c_{2}+4c_{3}),\ c_{5}=-(c_{2}+4c_{3}),\ c_{6}=0,\ c_{7}=-\frac{1}{16}(c_{2}+4c_{3}),\ c_{8}=\frac{1}{4}(c_{2}+4c_{3})\,,\label{5c}
\end{equation}
If we impose the above constraints on \eqref{dL}, we find 
\begin{align}
\Delta\mathcal{L}_{1} & =c_{1}'R^{2}+c_{2}'\mathcal{L}_{W^{2}}+c_{3}'\mathcal{L}_{{\rm GB}}+\frac{c_{4}'}{\sigma^2}\nabla^{\mu}F_{\mu\rho}\nabla_{\nu}F^{\nu\rho}\,,\label{dL4}
\end{align}
where the redefined coefficients $c_{i}'$ are $c_{1}'=c_{1}+\frac{1}{3}(c_{2}+c_{3}),\ c_{2}'=\frac{1}{2}(c_{2}+4c_{3}),\ c_{3}'=-\frac{1}{2}(c_{2}+2c_{3}),\ c_{4}'=c_{9}-c_{2}-4c_{3}.$
Note that the constraints \eqref{4DSUGRAConstraints} of supergravity  are stronger than constraints \eqref{5c}. Next, after performing the field redefinition for \eqref{dL4}
\begin{align}
g_{\mu\nu} & \rightarrow g_{\mu\nu}+\delta g_{\mu\nu},\ A_{\mu}\rightarrow A_{\mu}+\delta A_{\mu}\,,\ \delta A_{\mu}=-c_9\nabla^{\mu}F_{\mu\nu\,,}\nonumber \\
\delta g_{\mu\nu} & =(c_{2}+4c_{3})(R_{\mu\nu}-\frac{1}{2}Rg_{\mu\nu}-\frac{1}{2}T_{\mu\nu})+(c_{3}-c_{1})Rg_{\mu\nu}\,,
\end{align}
the 4-derivative action (\ref{dL4}) becomes
\begin{equation}
\Delta\mathcal{L}_{2}=c_{3}\mathcal{L}_{{\rm GB}}\,,
\end{equation}
Moreover, we can  check that $\delta g_{\mu\nu}|_{{\rm on-shell}}=\delta A_{\mu}|_{{\rm on-shell}}=0$, which means the solutions are not modified by the field redefinitions above, the 4-derivative actions $\Delta\mathcal{L}_{1}$ and $\Delta\mathcal{L}_{2}$ will give the same solutions $g_{\mu\nu}, A_\mu$. Since the GB term does not affect the field equations, it is clear that the general solutions remain uncorrected up to first order  $\mathcal{O}(c_{i})$ in the the 4-derivative action $\Delta\mathcal{L}_{1}$. For  supergravity \eqref{Sugra4D} in $D=4$, since constraints \eqref{4DSUGRAConstraints} are stronger than constraints (\ref{5c}), we find the 4-derivative solutions should be the same as the 2-derivative ones.

Imposing the constraints \eqref{4DSUGRAConstraints} on \eqref{eq:onshellinPhibeta} , the on-shell action and  entropy  in terms of the chemical potential at each horizon is
\begin{equation}
   I_\pm=\frac{\beta _{\pm }^2 \left(\Phi _{\pm }^2-4 \sigma^2\right){}^2}{256 \pi  \sigma^4 G_N}-\frac{4 \pi  d_1}{G_N}\ ,\ \ \  S_\pm=\frac{\beta _{\pm }^2 \left(\Phi _{\pm }^2-4 \sigma^2\right){}^2}{256 \pi  \sigma^4 G_N}+\frac{4 \pi  d_1}{G_N}\ .
\end{equation}
The above results show that the thermodynamic results (\ref{Left4D})
and (\ref{Ir4D}) are simplified for the case of supergravity (\ref{Sugra4D}),
which is only related to the coefficient $d_1$
\begin{align*}
I_{l}(\beta_{l},\varphi_{l}) & =\frac{(2\sigma ^{2}\beta_{l}^{2}-\varphi_{l}^{2})}{16\pi G_{N}\sigma^{2}}-\frac{4\pi d_{1}}{G_{N}}\,,\ \ S_{l}(\beta_{l},\varphi_{l})=\frac{(2\sigma ^{2}\beta_{l}^{2}-\varphi_{l}^{2})}{16\pi G_{N}\sigma^{2}}+\frac{4\pi d_{1}}{G_{N}}\,,\\
M(\beta_{l}) & =\frac{\beta_{l}}{4\pi G_{N}}\,,\ \ Q(\varphi_{l})=\frac{\varphi_{l}}{8\pi \sigma^{2}G_{N}}\,.\\
I_{r}(\beta_{r},\varphi_{r}) & =S_{r}(\beta_{r},\varphi_{r})=\frac{(3\sigma \beta_{r}-f_{\beta,\varphi})^{\frac{3}{2}}\sqrt{f_{\beta,\varphi}+\sigma\beta_{r}}}{32\pi \sigma^{2}G_{N}}\ .\\
M(\beta_{r},\varphi_{r}) & =\frac{\sqrt{3\sigma \beta_{r}-f_{\beta,\varphi}}\sqrt{f_{\beta,\varphi}+\sigma\beta_{r}}}{8\pi \sigma G_{N}}\,,\ \ Q(\beta_{r},\varphi_{r})=\frac{\varphi _r \sqrt{3 \sigma  \beta _r-f_{\beta ,\varphi }}}{8 \pi  \sigma ^2 G_N \sqrt{f_{\beta ,\varphi }+\sigma  \beta _r}}\,.
\end{align*}
In particular, the right-moving sector does not receive any correction, while the left-moving sector only get shifted by a constant, related to the Weyl$^2$ invariant. This is in exact agreement with \cite{Hristov:2023cuo}, where the reader can find further comments on their significance for the natural variables approach.

\section{Thermodynamics in $D=5$}
\label{sec:5dstatic}

\subsection{Natural variables}

Specifying to $D=5$, we obtain the corrected black hole correction, c.f.\ \eqref{RNsolution}
\begin{align}
\Delta h & =\frac{4}{3r^{6}}\left((2c_{1}-5c_{2}-22c_{3}+12c_{4}-3c_{5}-12c_{6})q^{2}+6c_{3}m^{2}\right)\nonumber \\
 & +\frac{1}{6r^{10}}\left(47c_{1}+13c_{2}+17c_{3}+276c_{4}+48c_{5}-24(c_{6}+6c_{7}+3c_{8})\right)q^{4}\nonumber \\
 & -\frac{8}{3r^{8}}(4c_{1}-c_{2}-5c_{3}+24c_{4}+3c_{5}-6c_{6})mq^{2}\ ,\nonumber \\
\Delta f & =\Delta h-\frac{8q^{2}}{3r^{10}}(7c_{1}+5c_{2}+13c_{3}+42c_{4}+12c_{5}+12c_{6})(r^{4}-2mr^{2}+q^{2})\ ,\nonumber \\
\Delta\psi & =-\frac{\sigma q}{\sqrt{3}r^{8}}\left(48c_{6}mr^{2}+(7c_{1}+5c_{2}+13c_{3}+36c_{4}-48c_{6}-144c_{7}-72c_{8})q^{2}\right)\ .
\end{align}
 The corrected thermodynamic quantities are given by
\begin{align}
\beta & =\frac{2\pi r_{0}}{1-\frac{q^{2}}{r_{0}^{4}}}+\frac{12\pi r_{0}^{4}\left(4\left(c_{3}+c_{6}\right)q^{4}-3\left(c_{3}-4c_{6}\right)q^{2}r_{0}^{4}+3c_{3}r_{0}^{8}\right)+\pi q^{4}f_{5D}\left(3q^{2}-7r_{0}^{4}\right)}{6r_{0}\left(r_{0}^{4}-q^{2}\right){}^{3}}\,,\nonumber \\
\Phi & =\frac{\sqrt{3}\sigma q}{r_{0}^{2}}+\frac{12\left(c_{3}+4c_{6}\right)\sigma qr_{0}^{8}+\sigma q^{3}f_{5D}\left(q^{2}-2r_{0}^{4}\right)}{2\sqrt{3}r_{0}^{8}\left(r_{0}^{4}-q^{2}\right)}\,,\nonumber \\
S & =\frac{\pi^{2}r_{0}^{3}}{2G_{N}}+\frac{\pi^{2}\left(-48\left(c_{3}+c_{6}\right)q^{2}r_{0}^{4}+36c_{3}r_{0}^{8}+q^{4}f_{5D}\right)}{8r_{0}^{3}G_{N}\left(r_{0}^{4}-q^{2}\right)}\,,\nonumber \\
I & =\frac{\pi^{2}r_{0}^{3}}{4G_{N}}+\frac{\pi^{2}\left(48\left(c_{3}+c_{6}\right)q^{2}r_{0}^{4}-36c_{3}r_{0}^{8}-q^{4}f_{5D}\right)}{16r_{0}^{3}G_{N}\left(r_{0}^{4}-q^{2}\right)}\,,\nonumber \\
M & =\frac{3\pi\left(q^{2}+r_{0}^{4}\right)}{8r_{0}^{2}G_{N}}\,,\ \ \ Q=\frac{\sqrt{3}\pi q}{4\sigma G_{N}}\,.\label{Thermo5D}
\end{align}
where $\omega_{D-2}=\omega_3=2\pi^{2}$, $f_{5D}:=c_{1}+11c_{2}+43c_{3}+12\left(c_{4}+2c_{5}+6(c_{6}+2c_{7}+c_{8})\right)$. From $f_{0}(r_{\pm})=0$, we find
\begin{equation}
m=\frac{1}{2}(r_{-}^{2}+r_{+}^{2})\ ,\ \ \ q=r_{-}r_{+}\ .\label{mq5D}
\end{equation}
Substituting eq.(\ref{mq5D}) into the corrected thermodynamics eq.(\ref{Thermo5D}),
we obtain the expected first law,  \eqref{eq:firstlaw}. Then, we can compute the left- and right-moving variables. The left-moving
thermodynamics is
\begin{align}
\beta_{l} & =\pi(\frac{r_{-}^{2}}{r_{-}+r_{+}}+r_{+})+\frac{1}{12r_{-}r_{+}\left(r_{-}+r_{+}\right){}^{3}}\Big(12\pi r_{-}r_{+}\left(c_{3}(3r_{-}^{2}+5r_{+}r_{-}+3r_{+}^{2})-4c_{6}r_{-}r_{+}\right)\nonumber \\
 & -\pi(3r_{-}^{4}+9r_{+}r_{-}^{3}+11r_{+}^{2}r_{-}^{2}+9r_{+}^{3}r_{-}+3r_{+}^{4})f_{5D}\Big)\,,\nonumber \\
\varphi_{l} & =\frac{\pi\sqrt{3}\sigma r_{-}r_{+}}{r_{-}+r_{+}}+\frac{\pi \sigma}{4\sqrt{3}r_{-}^{2}r_{+}^{2}\left(r_{-}+r_{+}\right){}^{3}}\Big(12r_{-}^{2}r_{+}^{2}\big(-4(c_{3}+c_{6})r_{-}^{2}-(7c_{3}+4c_{6})r_{+}r_{-}\nonumber \\
 & -4(c_{3}+c_{6})r_{+}^{2}\big)-(r_{-}^{3}+2r_{+}r_{-}^{2}+r_{+}^{2}r_{-}+r_{+}^{3})(r_{-}^{3}+r_{+}r_{-}^{2}+2r_{+}^{2}r_{-}+r_{+}^{3})f_{5D}\Big)\,,\nonumber \\
S_{l} & =\frac{\pi^{2}(r_{-}^{3}+r_{+}^{3})}{4G_{N}}+\frac{\pi^{2}}{16r_{-}r_{+}\left(r_{-}+r_{+}\right)G_{N}}\Big(-(r_{-}^{4}+r_{+}r_{-}^{3}+r_{+}^{2}r_{-}^{2}+r_{+}^{3}r_{-}+r_{+}^{4})f_{5D}\nonumber \\
 & +12r_{-}r_{+}\left(c_{3}(3r_{-}^{2}+7r_{+}r_{-}+3r_{+}^{2})+4c_{6}r_{-}r_{+}\right)\Big)\ ,\nonumber \\
I_{l} & =\frac{\pi^{2}(r_{-}^{3}+r_{+}^{3})}{8G_{N}}+\frac{\pi^{2}}{32r_{-}r_{+}\left(r_{-}+r_{+}\right)G_{N}}\Big((r_{-}^{4}+r_{+}r_{-}^{3}+r_{+}^{2}r_{-}^{2}+r_{+}^{3}r_{-}+r_{+}^{4})f_{5D}\nonumber \\
 & -12r_{-}r_{+}\left(c_{3}(3r_{-}^{2}+7r_{+}r_{-}+3r_{+}^{2})+4c_{6}r_{-}r_{+}\right)\Big)\,,\nonumber \\
M & =\frac{3\pi\left(r_{-}^{2}+r_{+}^{2}\right)}{8G_{N}}\,,\ \ Q=\frac{\sqrt{3}\pi r_{-}r_{+}}{4\sigma G_{N}}\,.
\end{align}

The right-moving thermodynamics is
\begin{align}
\beta_{r} & =\pi(\frac{r_{-}^{2}}{r_{+}-r_{-}}+r_{+})+\frac{1}{12r_{-}\left(r_{-}-r_{+}\right){}^{3}r_{+}}\big(\pi(-3r_{-}^{4}+9r_{+}r_{-}^{3}-11r_{+}^{2}r_{-}^{2}+9r_{+}^{3}r_{-}-3r_{+}^{4})f_{5D}\nonumber \\
 & -12\pi r_{-}r_{+}\left(c_{3}(3r_{-}^{2}-5r_{+}r_{-}+3r_{+}^{2})+4c_{6}r_{-}r_{+}\right)\big)\,,\nonumber \\
\varphi_{r} & =\frac{\sqrt{3}\pi \sigma r_{-}r_{+}}{r_{+}-r_{-}}+\frac{\pi \sigma}{4\sqrt{3}r_{-}^{2}\left(r_{-}-r_{+}\right){}^{3}r_{+}^{2}}\big(12r_{-}^{2}r_{+}^{2}\big(-4(c_{3}+c_{6})r_{-}^{2}+(7c_{3}+4c_{6})r_{+}r_{-}\nonumber \\
 & -4(c_{3}+c_{6})r_{+}^{2}\big)-(r_{-}^{3}-2r_{+}r_{-}^{2}+r_{+}^{2}r_{-}-r_{+}^{3})(r_{-}^{3}-r_{+}r_{-}^{2}+2r_{+}^{2}r_{-}-r_{+}^{3})f_{5D}\big)\,,\nonumber \\
S_{r} & =\frac{\pi^{2}\left(r_{+}^{3}-r_{-}^{3}\right)}{4G_{N}}-\frac{\pi^{2}}{16G_{N}r_{-}\left(r_{-}-r_{+}\right)r_{+}}\big(12r_{-}r_{+}\left(c_{3}(3r_{-}^{2}-7r_{+}r_{-}+3r_{+}^{2})-4c_{6}r_{-}r_{+}\right)\nonumber \\
 & +(r_{-}^{4}-r_{+}r_{-}^{3}+r_{+}^{2}r_{-}^{2}-r_{+}^{3}r_{-}+r_{+}^{4})f_{5D}\big)\,,\nonumber \\
I_{r} & =\frac{\pi^{2}\left(r_{+}^{3}-r_{-}^{3}\right)}{8G_{N}}+\frac{\pi^{2}}{32G_{N}r_{-}\left(r_{-}-r_{+}\right)r_{+}}\big(12r_{-}r_{+}\left(c_{3}(3r_{-}^{2}-7r_{+}r_{-}+3r_{+}^{2})-4c_{6}r_{-}r_{+}\right)\nonumber \\
 & +(r_{-}^{4}-r_{+}r_{-}^{3}+r_{+}^{2}r_{-}^{2}-r_{+}^{3}r_{-}+r_{+}^{4})f_{5D}\big)\,.
\end{align}
They give rise to the first law and quantum statistical relation
\begin{equation}
\beta_{l,r}\delta M=\delta S_{l,r}+\varphi_{l,r}\delta Q\,,\qquad  I_{l,r}=\beta_{l,r}\, M - S_{l,r}-\varphi_{l,r}\, Q\ .
\end{equation}

\subsection{$I_{l,r}(\beta_{l,r},\varphi_{l,r})$}
\subsubsection*{Left-moving sector}
The calculation process for $D=5$ is similar to that for $D=4$.
For the left-moving sector, we choose new integration constants $\tilde{r}_{\pm}$
making the potential energy fixed at the leading order $\beta_{l}\rightarrow\beta_{l,0},\,\varphi_{l,}\rightarrow\varphi_{l,0}$
\begin{equation}
r_{+}=\tilde{r}_{+}+\delta\tilde{r}_{+}\,,\ \ r_{-}=\tilde{r}_{-}+\delta\tilde{r}_{-}\,,
\end{equation}
where 
\begin{align}
\delta\tilde{r}_{+} & =\frac{\left(12c_{3}\tilde{r}_{-}\left(4\tilde{r}_{-}+3\tilde{r}_{+}\right)\tilde{r}_{+}^{2}+48c_{6}\tilde{r}_{-}^{2}\tilde{r}_{+}^{2}+\left(\tilde{r}_{-}^{4}+2\tilde{r}_{+}\tilde{r}_{-}^{3}+2\tilde{r}_{+}^{2}\tilde{r}_{-}^{2}+\tilde{r}_{+}^{3}\tilde{r}_{-}-\tilde{r}_{+}^{4}\right)f_{5D}\right)}{12\tilde{r}_{-}\left(\tilde{r}_{-}-\tilde{r}_{+}\right)\tilde{r}_{+}^{2}\left(\tilde{r}_{-}+\tilde{r}_{+}\right)}\,,\nonumber \\
\delta\tilde{r}_{-} & =-\frac{12c_{3}\tilde{r}_{+}\left(3\tilde{r}_{-}+4\tilde{r}_{+}\right)\tilde{r}_{-}^{2}+48c_{6}\tilde{r}_{+}^{2}\tilde{r}_{-}^{2}+\left(-\tilde{r}_{-}^{4}+\tilde{r}_{+}\tilde{r}_{-}^{3}+2\tilde{r}_{+}^{2}\tilde{r}_{-}^{2}+2\tilde{r}_{+}^{3}\tilde{r}_{-}+\tilde{r}_{+}^{4}\right)f_{5D}}{12\tilde{r}_{-}^{2}\left(\tilde{r}_{-}-\tilde{r}_{+}\right)\tilde{r}_{+}\left(\tilde{r}_{-}+\tilde{r}_{+}\right)}\,.
\end{align}
In this parametrization, we list all the left-moving sector below
\begin{align}
\beta_{l} & =\beta_{l,0}=\pi(\frac{\tilde{r}_{-}^{2}}{\tilde{r}_{+}-\tilde{r}_{-}}+\tilde{r}_{+})\,,\ \ \varphi_{l}=\varphi_{l,0}=\frac{\pi\sqrt{3}\sigma\tilde{r}_{-}\tilde{r}_{+}}{\tilde{r}_{-}+\tilde{r}_{+}}\,,\ \ S_{l}=\frac{\pi^{2}(\tilde{r}_{-}^{3}+\tilde{r}_{+}^{3})}{4G_{N}}\,,\nonumber \\
I_{l} & =\frac{\pi^{2}\left(\tilde{r}_{-}^{3}+\tilde{r}_{+}^{3}\right)}{8G_{N}}+\frac{\pi^{2}}{16G_{N}\tilde{r}_{-}\tilde{r}_{+}\left(\tilde{r}_{-}+\tilde{r}_{+}\right)}\Big((\tilde{r}_{-}^{4}+\tilde{r}_{+}\tilde{r}_{-}^{3}+\tilde{r}_{+}^{2}\tilde{r}_{-}^{2}+\tilde{r}_{+}^{3}\tilde{r}_{-}+\tilde{r}_{+}^{4})f_{5D}\nonumber \\
 & -12\tilde{r}_{-}\tilde{r}_{+}\left(c_{3}(3\tilde{r}_{-}^{2}+7\tilde{r}_{+}\tilde{r}_{-}+3\tilde{r}_{+}^{2})+4c_{6}\tilde{r}_{-}\tilde{r}_{+}\right)\Big)\,,\nonumber \\
M & =\frac{3\pi(\tilde{r}_{-}^{2}+\tilde{r}_{+}^{2})}{8G_{N}}+\frac{\pi\left((2\tilde{r}_{-}^{2}+\tilde{r}_{+}\tilde{r}_{-}+2\tilde{r}_{+}^{2})f_{5D}-36c_{3}\tilde{r}_{-}\tilde{r}_{+}\right)}{16G_{N}\tilde{r}_{-}\tilde{r}_{+}}\,,\nonumber \\
Q & =\frac{\pi\sqrt{3}\tilde{r}_{-}\tilde{r}_{+}}{4\sigma G_{N}}+\frac{48\pi(c_{3}+c_{6})\tilde{r}_{-}^{2}\tilde{r}_{+}^{2}+\pi(\tilde{r}_{-}^{4}+2\tilde{r}_{+}\tilde{r}_{-}^{3}+4\tilde{r}_{+}^{2}\tilde{r}_{-}^{2}+2\tilde{r}_{+}^{3}\tilde{r}_{-}+\tilde{r}_{+}^{4})f_{5D}}{16\sqrt{3}\sigma G_{N}\tilde{r}_{-}^{2}\tilde{r}_{+}^{2}}\,.\label{LeftThermodynamics5D}
\end{align}
The parameters $\tilde{r}_{\pm}$ can be explicitly inverted in terms
of the new variables 
\begin{align}
\tilde{r}_{\pm} & =\frac{\sqrt{3}\sigma\beta_{l}+\varphi_{l}\pm\sqrt{(\sqrt{3}\sigma\beta_{l}-\varphi_{l})^{2}-4\varphi_{l}^{2}}}{2\sqrt{3}\pi \sigma}\,,\label{rotrit5D}
\end{align}
Substituting eq.(\ref{rotrit5D}) into eq.(\ref{LeftThermodynamics5D}),
we obtain the following results
\begin{align}
I_{l}(\beta_{l},\varphi_{l}) & =\frac{\left(\sqrt{3}\sigma\beta_{l}-2\varphi_{l}\right)\left(\sqrt{3}\sigma\beta_{l}+\varphi_{l}\right){}^{2}}{24\sqrt{3}\pi \sigma^{3}G_{N}}\nonumber \\
 & +\frac{\pi\left(-3\sigma \beta_{l}\varphi_{l}(36c_{3}+f_{5D})-\sqrt{3}\varphi_{l}^{2}\left(48(c_{3}+c_{6})+f_{5D}\right)+3\sqrt{3}\sigma^{2}f_{5D}\beta_{l}^{2}\right)}{48\sigma G_{N}\varphi_{l}}\,,\nonumber \\
S_{l}(\beta_{l},\varphi_{l}) & =\frac{\left(\sqrt{3}\sigma\beta_{l}-2\varphi_{l}\right)\left(\sqrt{3}\sigma\beta_{l}+\varphi_{l}\right){}^{2}}{12\sqrt{3}\pi \sigma^{3}G_{N}}\,,\nonumber \\
M(\beta_{l},\varphi_{l}) & =\frac{3\sigma ^{2}\beta_{l}^{2}-\varphi_{l}^{2}}{8\pi \sigma^{2}G_{N}}-\frac{\pi\left(\varphi_{l}\left(36c_{3}+f_{5D}\right)-2\sqrt{3}\sigma f_{5D}\beta_{l}\right)}{16G_{N}\varphi_{l}}\,,\nonumber \\
Q(\beta_{l},\varphi_{l}) & =\frac{\varphi_{l}\left(3\sigma \beta_{l}+\sqrt{3}\varphi_{l}\right)}{12\pi \sigma^{3}G_{N}}+\frac{\pi\varphi_{l}^{2}\left(48\left(c_{3}+c_{6}\right)+f_{5D}\right)+3\pi \sigma^{2}f_{5D}\beta_{l}^{2}}{16\sqrt{3}\sigma G_{N}\varphi_{l}^{2}}\,.\label{Left5D}
\end{align}
For the left-moving on-shell actions $I_l$, we find the relations
\begin{equation}
		\frac{\partial I_{l}}{\partial \beta_{l}} = M \ ,  \qquad \frac{\partial I_{l}}{\partial \varphi_{l}} = - Q\ .
\end{equation}
\subsubsection*{Right-moving sector}
For the right-moving sector, we choose new integration constants $\hat{r}_{\pm}$
making the potential energy fixed at the leading order $\beta_{r}\rightarrow\beta_{r,0},\,\varphi_{r,}\rightarrow\varphi_{r,0}$
\begin{equation}
r_{+}=\hat{r}_{+}+\delta\hat{r}_{+}\,,\ \ r_{-}=\hat{r}_{-}+\delta\hat{r}_{-}\,,
\end{equation}
where 
\begin{align}
\delta\hat{r}_{+} & =-\frac{12c_{3}\hat{r}_{-}\left(4\hat{r}_{-}-3\hat{r}_{+}\right)\hat{r}_{+}^{2}+48c_{6}\hat{r}_{-}^{2}\hat{r}_{+}^{2}+\left(\hat{r}_{-}^{4}-2\hat{r}_{+}\hat{r}_{-}^{3}+2\hat{r}_{+}^{2}\hat{r}_{-}^{2}-\hat{r}_{+}^{3}\hat{r}_{-}-\hat{r}_{+}^{4}\right)f_{5D}}{12\hat{r}_{-}\left(\hat{r}_{-}-\hat{r}_{+}\right)\hat{r}_{+}^{2}\left(\hat{r}_{-}+\hat{r}_{+}\right)}\,,\nonumber \\
\delta\hat{r}_{-} & =-\frac{12c_{3}\hat{r}_{+}\left(3\hat{r}_{-}-4\hat{r}_{+}\right)\hat{r}_{-}^{2}-48c_{6}\hat{r}_{+}^{2}\hat{r}_{-}^{2}+\left(\hat{r}_{-}^{4}+\hat{r}_{+}\hat{r}_{-}^{3}-2\hat{r}_{+}^{2}\hat{r}_{-}^{2}+2\hat{r}_{+}^{3}\hat{r}_{-}-\hat{r}_{+}^{4}\right)f_{5D}}{12\hat{r}_{-}^{2}\left(\hat{r}_{-}-\hat{r}_{+}\right)\hat{r}_{+}\left(\hat{r}_{-}+\hat{r}_{+}\right)}\,.
\end{align}
In this parametrization, we list all the right-moving thermodynamics
below
\begin{align}
\beta_{r} & =\beta_{r,0}=\pi(\frac{r_{-}^{2}}{r_{+}-r_{-}}+r_{+})\,,\ \ \varphi_{r}=\varphi_{r,0}=\frac{\sqrt{3}\pi \sigma r_{-}r_{+}}{r_{+}-r_{-}}\,,\ \ S_{r}=\frac{\pi^{2}\left(\hat{r}_{+}^{3}-\hat{r}_{-}^{3}\right)}{4G_{N}}\,,\nonumber \\
I_{r} & =\frac{\pi^{2}\left(\hat{r}_{+}^{3}-\hat{r}_{-}^{3}\right)}{8G_{N}}+\frac{\pi^{2}}{16\hat{r}_{-}\left(\hat{r}_{-}-\hat{r}_{+}\right)\hat{r}_{+}G_{N}}\Big(12\hat{r}_{-}\hat{r}_{+}\left(c_{3}(3\hat{r}_{-}^{2}-7\hat{r}_{+}\hat{r}_{-}+3\hat{r}_{+}^{2})-4c_{6}\hat{r}_{-}\hat{r}_{+}\right)\nonumber \\
 & +(\hat{r}_{-}^{4}-\hat{r}_{+}\hat{r}_{-}^{3}+\hat{r}_{+}^{2}\hat{r}_{-}^{2}-\hat{r}_{+}^{3}\hat{r}_{-}+\hat{r}_{+}^{4})f_{5D}\Big)\ ,\nonumber \\
M & =\frac{3\pi\left(\hat{r}_{-}^{2}+\hat{r}_{+}^{2}\right)}{8G_{N}}-\frac{\pi\left(36c_{3}\hat{r}_{-}\hat{r}_{+}+(2\hat{r}_{-}^{2}-\hat{r}_{+}\hat{r}_{-}+2\hat{r}_{+}^{2})f_{5D}\right)}{16\hat{r}_{-}\hat{r}_{+}G_{N}}\,,\nonumber \\
Q & =\frac{\pi\sqrt{3}\hat{r}_{-}\hat{r}_{+}}{4\sigma G_{N}}-\frac{48\pi\left(c_{3}+c_{6}\right)\hat{r}_{-}^{2}\hat{r}_{+}^{2}+\pi\left(\hat{r}_{-}^{4}-2\hat{r}_{+}\hat{r}_{-}^{3}+4\hat{r}_{+}^{2}\hat{r}_{-}^{2}-2\hat{r}_{+}^{3}\hat{r}_{-}+\hat{r}_{+}^{4}\right)f_{5D}}{16\sqrt{3}\sigma\hat{r}_{-}^{2}\hat{r}_{+}^{2}G_{N}}.\label{RightThermodynamics5D}
\end{align}
The parameters characterizing the black hole solution can be explicitly
inverted in terms of the new variables
\begin{align}
\hat{r}_{\pm} & =\frac{1}{8\pi \sigma}\big(\sqrt{2\sigma \beta_{r}(\sqrt{\sigma^{2}\beta_{r}^{2}+2\varphi_{r}^{2}}+\sigma\beta_{r})-2\varphi_{r}^{2}}\pm(3\sigma \beta_{r}-\sqrt{\sigma^{2}\beta_{r}^{2}+2\varphi_{r}^{2}})\big)\,,\label{r0ri5D}
\end{align}
Substituting eq.(\ref{r0ri5D}) into eq.(\ref{RightThermodynamics5D}),
we obtain the following results
\begin{align}
I_{r}(\beta_{r},\varphi_{r}) & =\frac{\left(\sqrt{3}\sigma\beta_{r}-\varphi_{r}\right){}^{2}\left(\sqrt{3}\sigma\beta_{r}+2\varphi_{r}\right)}{24\sqrt{3}\pi \sigma^{3}G_{N}}\nonumber \\
 & +\frac{\pi\left(-3\sigma \beta_{r}\varphi_{r}(36c_{3}+f_{5D})+\sqrt{3}\varphi_{r}^{2}\left(48(c_{3}+c_{6})+f_{5D}\right)-3\sqrt{3}\sigma^{2}f_{5D}\beta_{r}^{2}\right)}{48\sigma G_{N}\varphi_{r}}\nonumber \\
S_{r}(\beta_{r},\varphi_{r}) & =\frac{\left(\sqrt{3}\sigma\beta_{r}-\varphi_{r}\right){}^{2}\left(\sqrt{3}\sigma\beta_{r}+2\varphi_{r}\right)}{12\sqrt{3}\pi \sigma^{3}G_{N}}\,,\nonumber \\
M(\beta_{r},\varphi_{r}) & =\frac{\varphi_{r}^{2}-3\sigma ^{2}\beta_{r}^{2}}{8\pi \sigma^{2}G_{N}}+\frac{\pi\left(\sqrt{3}\sigma\beta_{r}\varphi_{r}(f_{5D}-36c_{3})+\varphi_{r}^{2}(36c_{3}+f_{5D})-6\sigma ^{2}f_{5D}\beta_{r}^{2}\right)}{16G_{N}\varphi_{r}\left(\sqrt{3}\sigma\beta_{r}-\varphi_{r}\right)}\,,\nonumber \\
Q(\beta_{r},\varphi_{r}) & =\frac{\varphi_{r}\left(3\sigma \beta_{r}-\sqrt{3}\varphi_{r}\right)}{12\pi \sigma^{3}G_{N}}-\frac{\pi\left(\varphi_{r}^{2}\left(48(c_{3}+c_{6})+f_{5D}\right)+3\sigma ^{2}f_{5D}\beta_{r}^{2}\right)}{16\sqrt{3}\sigma G_{N}\varphi_{r}^{2}}\,.\label{Left5DLast}
\end{align}
For the right-moving on-shell actions $I_r$, we once again find
\begin{equation}
		\frac{\partial I_{r}}{\partial \beta_{r}} = M \ ,  \qquad \frac{\partial I_{r}}{\partial \varphi_{r}} = - Q\ .
\end{equation}
We remark that, unlike the four-dimensional case, the left- and right-moving sectors in $D=5$ exhibit a symmetry, as already noticed in \cite{Hristov:2023cuo}. The present results confirm this observation in the higher-derivative setting as well.

\subsection{$D=5$ supergravity}

We consider the parity even terms in 5-dimensional supergravity actions
with zero cosmological constant upon eliminating the auxiliary fields
\cite{Ozkan:2013nwa,Ozkan:2013uk,Butter:2014xxa,Gold:2023ymc,Ma:2024ynp}
\begin{align}
\Delta\mathcal{L} & =d_{1}\mathcal{L}_{{\rm Weyl}^{2}}+d_{2}\ensuremath{\mathcal{L}_{{\rm Ricci}^{2}}}+d_{3}\ensuremath{\mathcal{L}_{R^{2}}}\ ,\nonumber \\
\mathcal{L}_{{\rm Weyl}^{2}} & =C^{\mu\nu\rho\sigma}C_{\mu\nu\rho\sigma}+\frac{R^{2}}{6}-\frac{1}{2\sigma ^{2}}R_{\mu\nu\rho\sigma}F^{\mu\nu}F^{\rho\sigma}-\frac{4}{3\sigma ^{2}}R_{\mu\nu}F^{\mu\lambda}F_{\ \lambda}^{\nu}\nonumber \\
 & +\frac{2}{9\sigma ^{2}}RF^{\mu\nu}F_{\mu\nu}-\frac{2}{9\sigma ^{2}}\nabla^{\mu}F_{\mu\rho}\nabla_{\nu}F^{\nu\rho}-\frac{61}{432\sigma ^{4}}(F^{\mu\nu}F_{\mu\nu})^{2}\nonumber \\
 & +\frac{5}{8\sigma ^{4}}F_{\mu\nu}F^{\nu\lambda}F_{\lambda\delta}F^{\delta\mu}+\frac{5}{72\sqrt{3}\sigma^{3}}\epsilon_{\mu\nu\rho\sigma\alpha}F^{\mu\nu}F^{\rho\sigma}\nabla_{\beta}F^{\beta\alpha}\nonumber \\
 & +\frac{1}{2\sqrt{3}\sigma}\epsilon^{\mu\nu\rho\sigma\alpha}A_{\mu}R_{\nu\rho}^{\ \ \ \beta\gamma}R_{\sigma\alpha\beta\gamma}\,,\nonumber \\
\ensuremath{\mathcal{L}_{{\rm Ricci}^{2}}} & =R^{2}-4R_{\mu\nu}R^{\mu\nu}-\frac{5}{6\sigma ^{2}}RF^{\mu\nu}F_{\mu\nu}-\frac{2}{\sigma^{2}}\nabla^{\mu}F_{\mu\rho}\nabla_{\nu}F^{\nu\rho}+\frac{4}{\sigma^{2}}R_{\mu\nu}F^{\mu\lambda}F_{\ \lambda}^{\nu}\nonumber \\
 & +\frac{5}{16\sigma ^{4}}(F^{\mu\nu}F_{\mu\nu})^{2}-\frac{11}{9\sigma ^{4}}F_{\mu\nu}F^{\nu\lambda}F_{\lambda\delta}F^{\delta\mu}-\frac{2}{3\sqrt{3}\sigma^{3}}\epsilon_{\mu\nu\rho\sigma\alpha}F^{\mu\nu}F^{\rho\sigma}\nabla_{\beta}F^{\beta\alpha}\,,\nonumber \\
\mathcal{L}_{R^{2}} & =R^{2}-\frac{1}{6\sigma ^{2}}RF^{\mu\nu}F_{\mu\nu}+\frac{1}{144\sigma ^{4}}(F^{\mu\nu}F_{\mu\nu})^{2}\ .\label{5DSugra}
\end{align}
For charged black holes (\ref{RNsolution}), the term involving $\epsilon_{\mu\nu\rho\sigma\tau}$
makes no contribution and can therefore be disregarded. We impose
the following constraints on eqs.(\ref{dL})
\begin{align}
c_{4}&=\frac{1}{6}(c_{2}-c_{1}+3c_{3}),\ c_{5}=-c_{2}-\frac{8c_{3}}{3},\ c_{6}=-\frac{c_{3}}{2},\nonumber \\
c_{7}&=\frac{1}{432}(3c_{1}-33c_{2}-106c_{3}),\ c_{8}=\frac{1}{216}(66c_{2}+223c_{3}),\ c_{9}=\frac{c_{2}}{2}+\frac{4c_{3}}{9},\nonumber \\
c_{1}&=\frac{d_{1}}{3}+d_{2}+d_{3},\ c_{2}=-\frac{4d_{1}}{3}-4d_{2},\ c_{3}=d_{1}\,,
\label{5Dc}
\end{align}
we find the 4-derivative interactions (\ref{dL}) can be written explicitly
in the form of eq.(\ref{5DSugra}). These constraints lead to
\begin{align}
f_{5D} & =c_{1}+11c_{2}+43c_{3}+12\left(c_{4}+2c_{5}+6(c_{6}+2c_{7}+c_{8})\right)=-12d_{1}\ .
\end{align}
Note that, unlike the four-dimensional case, now the black hole solutions do get corrected at a four-derivative order, and the present work lists these corrections for the first time in literature.

Imposing the constraints \eqref{5Dc} on \eqref{eq:onshellinPhibeta} , the on-shell action and  entropy  in terms of the chemical potential at each horizon is
\begin{align}
    I_{\pm}&=	\frac{\beta_{\pm}^{3}(3\sigma^{2}-\Phi_{\pm}^{2})^{3}}{864\pi \sigma^{6}G_{N}}+\frac{\pi d_{1}\beta_{\pm}\left(-27\sigma^{4}+6\sigma^{2}\Phi_{\pm}^{2}+\Phi_{\pm}^{4}\right)}{12\sigma^{4}G_{N}}\,, \nonumber\\
S_{\pm}&=	\frac{\beta_{\pm}^{3}\left(3\sigma^{2}-\Phi_{\pm}^{2}\right){}^{3}}{432\pi \sigma^{6}G_{N}}\,.
\end{align}
Combining \eqref{Left5D}, \eqref{Left5DLast}, and (\ref{5Dc}), the thermodynamic
results are simplified:
\begin{align}
I_{l}(\beta_{l},\varphi_{l}) & =\frac{\sqrt{3}\pi(\sqrt{3}\sigma\beta_{l}+\varphi_{l})^{2}}{4\sigma G_{N}\varphi_{l}}\big(\frac{\varphi_{l}(\sqrt{3}\sigma\beta_{l}-2\varphi_{l})}{18\pi^{2}\sigma^{2}}-d_{1}\big)\,,\nonumber \\
S_{l}(\beta_{l},\varphi_{l}) & =\frac{(\sqrt{3}\sigma\beta_{l}-2\varphi_{l})(\sqrt{3}\sigma\beta_{l}+\varphi_{l})^{2}}{12\sqrt{3}\pi \sigma^{3}G_{N}}\,,\nonumber \\
M(\beta_{l},\varphi_{l}) & =\frac{3\pi(\sqrt{3}\sigma\beta_{l}+\varphi_{l})}{2G_{N}\varphi_{l}}\big(\frac{\varphi_{l}(\sqrt{3}\sigma\beta_{l}-\varphi_{l})}{12\pi^{2}\sigma^{2}}-d_{1}\big)\,,\nonumber \\
Q(\beta_{l},\varphi_{l}) & =\frac{\pi\sqrt{3}(\sqrt{3}\sigma\beta_{l}+\varphi_{l})}{4\sigma \varphi_{l}^{2}G_{N}}\big(\frac{\varphi_{l}^{3}}{3\pi^{2}\sigma^{2}}+d_{1}(\varphi_{l}-\sqrt{3}\sigma\beta_{l})\big)\,,
\end{align}
and 
\begin{align}
I_{r}(\beta_{r},\varphi_{r}) & =\frac{\sqrt{3}\pi(\sqrt{3}\sigma\beta_{r}-\varphi_{r})^{2}}{4\sigma G_{N}\varphi_{r}}\big(\frac{\varphi_{r}(\sqrt{3}\sigma\beta_{r}+2\varphi_{r})}{18\pi^{2}\sigma^{2}}+d_{1}\big)\,,\nonumber \\
S_{r}(\beta_{r},\varphi_{r}) & =\frac{(\sqrt{3}\sigma\beta_{r}-\varphi_{r})^{2}(\sqrt{3}\sigma\beta_{r}+2\varphi_{r})}{12\sqrt{3}\pi \sigma^{3}G_{N}}\,,\nonumber \\
M(\beta_{r},\varphi_{r}) & =\frac{3\pi(\sqrt{3}\sigma\beta_{r}-\varphi_{r})}{2G_{N}\varphi_{r}}\big(\frac{\varphi_{r}(\sqrt{3}\sigma\beta_{r}+\varphi_{r})}{12\pi^{2}\sigma^{2}}+d_{1}\big)\,,\nonumber \\
Q(\beta_{r},\varphi_{r}) & =\frac{\sqrt{3}\pi(\sqrt{3}\sigma\beta_{r}-\varphi_{r})}{4\sigma \varphi_{r}^{2}G_{N}}\big(\frac{\varphi_{r}^{3}}{3\pi^{2}\sigma^{2}}+d_{1}(\varphi_{r}+\sqrt{3}\sigma\beta_{r})\big)\,.
\end{align}
Once again, it is only the Weyl$^2$ invariant that gives a correction to the black hole thermodynamics. Unlike the four-derivative case, here it corrects both the left- and the right-moving sector, which has a bearing on the BPS and almost BPS limits that we discuss in due course. 

\part{Rotating black holes and BPS limits}

\section{Thermodynamics in $D=4$}
\label{sec:4drotating}

\subsection{Einstein-Maxwell theory}

The 2-derivative Einstein-Maxwell theory
\begin{align}
I_{0} & =\frac{1}{16\pi G_{N}}\int\sqrt{-g}(R-\frac{1}{4\sigma ^{2}}F^{\mu\nu}F_{\mu\nu})\,,
\end{align}
contains a charged rotating black hole
\begin{align}
ds^{2} & =-\frac{\Delta_{r}}{\rho^{2}}\left(dt-a\sin^{2}\theta d\phi\right)^{2}+\frac{\rho^{2}}{\Delta_{r}}dr^{2}+\rho^{2}d\theta^{2}+\frac{\sin^{2}\theta}{\rho^{2}}\left(adt-(r^{2}+a^{2})d\phi\right)^{2}\nonumber \\
A_{(1)} & =-\frac{2\sigma qr}{\rho^{2}}\left(dt-a\sin^{2}\theta d\phi\right)\,,
\end{align}
where
\begin{align}
\Delta_{r} & =r^{2}+a^{2}-2mr+q^{2}\,,\ \ \ \rho^{2}=r^{2}+a^{2}\cos^{2}\theta\,.
\end{align}
The black hole horizon $r=r_{0}$ is located at $\Delta_{r}(r_{0})=0$
\begin{equation}
r_{0}^{2}+a^{2}-2mr_{0}+q^{2}=0\,,
\end{equation}
with inner and outer horizon given by
\be
	r_\pm = m \pm \sqrt{m^2 - q^2 - a^2}\ .
\ee

The 2-derivative thermodynamics
are
\begin{align}
\beta_{0} & =\frac{1}{T_{0}}=\frac{2\pi}{\kappa_{{\rm surf}}}=\frac{4\pi r_{0}\left(r_{0}^{2}+a^{2}\right)}{r_{0}^{2}-a^{2}-q^{2}}\ ,\quad S_{0}=\frac{A_{{\rm horizon}}}{4G_{N}}=\pi\frac{r_{0}^{2}+a^{2}}{G_{N}}\ ,\nonumber \\
J_{0} & =\frac{1}{16\pi G_{N}}\int\star d\xi_{\phi}=\frac{ma}{G_{N}}\ ,\ \ \Omega_{0}=-\frac{g_{t\phi}}{g_{\phi\phi}}|_{r=r_{0}}=\frac{a}{r_{0}^{2}+a^{2}}\ ,\nonumber \\
Q_{0} & =\frac{1}{16\pi G_{N}}\int\frac{\star F_{(2)}}{\sigma^{2}}=\frac{q}{2\sigma G_{N}}\ ,\ \ \Phi_{0}=\xi^{\mu}A_{\mu}|_{r_{0}}^{\infty}=\frac{2\sigma qr_{0}}{r_{0}^{2}+a^{2}}\ ,\nonumber \\
M_{0} & =-\frac{1}{16\pi G_{N}}\int\star d\xi_{t}=\frac{m}{G_{N}}\,,
\end{align}
where $\xi=\xi_{t}+\Omega_{0}\xi_{\phi},\,\xi_{t}=\partial_{t}\,,\ \xi_{\phi}=\partial_{\phi}$.
Using above results, we can compute the free energy
\begin{equation}
G_{0}=M_{0}-T_{0}S_{0}-\Phi_{0}Q_{0}-\Omega_{0}J_{0}=\frac{a^{2}(\frac{2q^{2}}{a^{2}+r_{0}^{2}}+1)-q^{2}+r_{0}^{2}}{4r_{0}G_{N}}\,,
\end{equation}
The on-shell action at fixed potential $\beta_{0},\Phi_{0},\Omega_{0}$
is given by the background subtraction method
\begin{align}
I_{0}(\beta_{0},\Phi_{0},\Omega_{0}) & =\beta_{0}G_{0}=\pi\frac{a^{4}+a^{2}(q^{2}+2r_{0}^{2})-q^{2}r_{0}^{2}+r_{0}^{4}}{G_{N}\left(r_{0}^{2}-(a^{2}+q^{2})\right)}\,.
\end{align}

\subsection{General 4-derivative corrections}

Next, let us apply the RS method \cite{Reall:2019sah} to obtain the
action of the Kerr-Newman black hole in higher derivative gravity
(\ref{Action}, \ref{dL}). After some calculations, we find the corrected
on-shell action $I(\beta_{0},\Phi_{0},\Omega_{0})$ at fixed potential
$\beta_{0},\Phi_{0},\Omega_{0}$
\begin{align}
I & =\beta_{0}G(\beta_{0},\Phi_{0},\Omega_{0})\,,\ \ G=G_{0}+\Delta G\,,\nonumber \\
\Delta G & =\frac{c_{3}(a^{2}+q^{2}-r_{0}^{2})}{r_{0}(a^{2}+r_{0}^{2})G_{N}}+\frac{c_{6}q^{2}\left(a^{4}+a^{2}(q^{2}-10r_{0}^{2})-3q^{2}r_{0}^{2}+5r_{0}^{4}\right)}{5r_{0}\left(a^{2}+r_{0}^{2}\right){}^{3}G_{N}}\nonumber \\
 & -\frac{2(4c_{7}+c_{8})q^{4}r_{0}(3a^{4}-10a^{2}r_{0}^{2}+3r_{0}^{4})}{15(a^{2}+r_{0}^{2})^{5}G_{N}}-\frac{q^{4}}{64a^{5}r_{0}^{4}\left(a^{2}+r_{0}^{2}\right)G_{N}}\Big[\nonumber \\
 & (c_{2}+4c_{3}+2c_{5}+4c_{6}+8c_{7}+6c_{8})\big(3\pi a^{6}+6a^{5}r_{0}+3\pi a^{4}r_{0}^{2}+4a^{3}r_{0}^{3}\nonumber \\
 & -3\pi a^{2}r_{0}^{4}+6ar_{0}^{5}-3\pi r_{0}^{6}-6(a^{2}-r_{0}^{2})(a^{2}+r_{0}^{2})^{2}{\rm arctan}(\frac{r_{0}}{a})\big)\Big]\,.
\end{align}
According to the thermodynamic relations
\begin{align}
Q & =-\frac{1}{\beta_{0}}\left(\frac{\partial I}{\partial\Phi_{0}}\right)_{\beta_{0},\Omega_{0}},\ \ \ M=\left(\frac{\partial I}{\partial\beta}\right)_{\Phi_{0},\Omega_{0}}+\Phi_{0}Q+\Omega_{0}J\,,\nonumber \\
J & =-\frac{1}{\beta_{0}}\left(\frac{\partial I}{\partial\Omega_{0}}\right)_{\beta_{0},\Phi_{0}},\ \ \ S=\beta_{0}M-I_{}-\beta_{0}\Phi_{0}Q-\beta_{0}\Omega_{0}J\ ,\label{5DKNbfo}
\end{align}
We can obtain the corrected thermodynamics at fixed potential $\beta_{0},\Phi_{0},\Omega_{0}$.
Next, we perform the following redefinitions of the integral parameters
\begin{equation}
r_{0}\rightarrow r_{0}+\delta r_{0}\,,\ \ a\rightarrow a+\delta a\,,\ \ q\rightarrow q+\delta q\,,\label{drdadq}
\end{equation}
where $\delta r_{0},\,\delta a,\,\delta q$ are listed in appendix
\ref{AppendixB}. Combine eqs.(\ref{5DKNbfo}, \ref{drdadq}), we
obtain the thermodynamics of Kerr-Newman black hole at fixed $M_{0},Q_{0},J_{0}$
\begin{align}
M & =M_{0}=\frac{m}{G_{N}}\,,\ \ Q=Q_{0}=\frac{q}{2\sigma G_{N}}\,,\ \ J=J_{0}=\frac{ma}{G_{N}}\,,\nonumber \\
\beta & =\frac{4\pi r_{0}\left(r_{0}^{2}+a^{2}\right)}{r_{0}^{2}-a^{2}-q^{2}}+\Delta\beta\,,\ \ S=\pi\frac{r_{0}^{2}+a^{2}}{G_{N}}+\Delta S\,,\nonumber \\
\Omega & =\frac{a}{r_{0}^{2}+a^{2}}+\Delta\Omega\,,\ \ \Phi=\frac{2\sigma qr_{0}}{r_{0}^{2}+a^{2}}+\Delta\Phi\,,\nonumber \\
I & =\pi\frac{a^{4}+a^{2}(q^{2}+2r_{0}^{2})-q^{2}r_{0}^{2}+r_{0}^{4}}{G_{N}\left(r_{0}^{2}-(a^{2}+q^{2})\right)}+\Delta I\,,\label{5DKNmqj}
\end{align}
where the corrected thermodynamics $\Delta\beta,\,\Delta S,\,\Delta\Omega,\ \Delta\Phi,\,\Delta I$
with parameters $(r_{0},q,a)$ are listed in appendix \ref{AppendixB}.
For computational convenience, analogous to the method employed in
preceding sections, thermodynamics can be expressed in terms of the
parameter $(r_{+},r_{-},a)$ via
\begin{equation}
m=\frac{r_{+}+r_{-}}{2}\,,\ \ q=\sqrt{r_{+}r_{-}-a^{2}}\,.
\end{equation}

The complete thermodynamics of black holes is complex. We can discuss
the simple cases. If we impose the following constraints on the higher derivative gravity
(\ref{Action}, \ref{dL}) and thermodynamics (\ref{5DKNmqj}),  there is still a correction to the black hole solution, but the thermodynamics  will be greatly simplified
\begin{equation}
c_{6}=0\,,\ \ c_{7}=-\frac{1}{4}c_{8}\,,\ \ f_{4D}=c_{2}+4c_{3}+2c_{5}+7c_{6}+8(2c_{7}+c_{8})=0\,, \label{4DConstraint2}
\end{equation}

Then, the simplified left- and right-moving thermodynamics is
\begin{align}
I_{l} & =\frac{2\sigma ^{2}\beta_{l}^{2}-\varphi_{l}^{2}}{16\pi \sigma^{2}G_{N}}-\frac{4\pi c_{3}}{G_{N}},\ S_{l}=\frac{2\sigma ^{2}\beta_{l}^{2}-\varphi_{l}^{2}}{16\pi \sigma^{2}G_{N}}+\frac{4\pi c_{3}}{G_{N}},\ M=\frac{\beta_{l}}{4\pi G_{N}},\ Q=\frac{\varphi_{l}}{8\pi \sigma^{2}G_{N}}\,,\nonumber \\
I_{r} & =\frac{\sqrt{f_{\beta,\varphi}+\sigma\beta_{r}}(3\sigma \beta_{r}-f_{\beta,\varphi})^{\frac{3}{2}}}{16\sigma ^{2}G_{N}\sqrt{\omega_{r}^{2}+4\pi^{2}}},\ S_{r}=\frac{\pi^{2}\sqrt{f_{\beta,\varphi}+\sigma\beta_{r}}(3\sigma \beta_{r}-f_{\beta,\varphi})^{\frac{3}{2}}}{4\sigma ^{2}G_{N}(\omega_{r}^{2}+4\pi^{2})^{\frac{3}{2}}}\,,\nonumber \\
M & =\frac{\sqrt{(3\sigma\beta_{r}-f_{\beta,\varphi})(f_{\beta,\varphi}+\sigma\beta_{r})}}{4\sigma G_{N}\sqrt{\omega_{r}^{2}+4\pi^{2}}},\ Q=\frac{\varphi_{r}\sqrt{3\sigma\beta_{r}-f_{\beta,\varphi}}}{4\sigma^{2}G_{N}\sqrt{\omega_{r}^{2}+4\pi^{2}}\sqrt{f_{\beta,\varphi}+\sigma\beta_{r}}},\ \nonumber\\
J&=\frac{\omega_{r}\sqrt{f_{\beta,\varphi}+\sigma\beta_{r}}(3\sigma \beta_{r}-f_{\beta,\varphi})^{\frac{3}{2}}}{16\sigma ^{2}G_{N}(\omega_{r}^{2}+4\pi^{2})^{\frac{3}{2}}}, \label{4D3c}
\end{align}
where $f_{\beta,\varphi}:=\sqrt{\sigma^{2}\beta_{r}^{2}+2\varphi_{r}^{2}}$, and the left- and right-moving variables are $\omega_{l,r}:=\frac{1}{2}(\beta_{+}\Omega_{+}\pm\beta_{-}\Omega_{-})$.
Since the supergravity constraints \eqref{4DSUGRAConstraints} are stronger than constraints \eqref{4DConstraint2}, its thermodynamics remains \eqref{4D3c}. The above expressions contain the two-derivative results of \cite{Hristov:2023sxg} for $\sigma = 1/2$.
We also note that the on-shell actions satisfy
 \begin{equation}
		\frac{\partial I_{l,r}}{\partial \beta_{l,r}}= M \ ,  \qquad \frac{\partial I_{l,r}}{\partial \varphi_{l,r}}= - Q\ , \qquad \frac{\partial I_{r}}{\partial \omega_{r}} = - J\ ,
\end{equation}
as expected.

\subsection{BPS-like limit}
In the supergravity case, the black holes exhibit a BPS limit, \cite{Hristov:2022pmo,Hristov:2023cuo}
\begin{equation}
M-2\sigma Q=0\,,
\end{equation}
which leads to $a=\pm\frac{i}{2}(r_{+}-r_{-})$. We can still consider the same limit, but now additionally check if it leads to a simplification of a larger set of four-derivative theories, potentially not preserving supersymmetry. This is why we call this limit BPS-like in this general case.

At the leading order $\mathcal{O}(c_{i}^{0})$ , the thermodynamic potentials in turn can
be shown to satisfy
\begin{equation}
\beta_{l}=\frac{1}{\sigma}\varphi_{l}\,,\ \ \ \omega_{l}=0\,,\ \ \ \beta_{r}=\frac{1}{2\sigma }\varphi_{r}\,,\ \ \ \omega_{r}=\pm2\pi i\,.\label{betafai}
\end{equation}
At the $\mathcal{O}(c_{i})$ order, the corrected thermodynamic potentials
(\ref{Potential}) may lead to an alteration in the previously established
relation (\ref{betafai}). To ensure the invariance of eq.(\ref{betafai}),
we impose the  constraints \eqref{4DConstraint2} on the thermodynamics (\ref{5DKNmqj}). Then, the simplified left-moving thermodynamics is
\begin{align}
\beta_{l} & =\frac{1}{\sigma}\varphi_{l}=2\pi(r_{-}+r_{+})\,,\ \ \omega_{l}=0\,,\ \ J=\pm\frac{i(r_{+}^{2}-r_{-}^{2})}{4G_{N}}\,,\ \ M=2\sigma Q=\frac{r_{-}+r_{+}}{2G_{N}}\,,\nonumber \\
I_{l}^\text{BPS} & =\frac{\pi(r_{-}+r_{+})^{2}}{4G_{N}}-\frac{4\pi c_{3}}{G_{N}}=\frac{\varphi_{l}^{2}}{16\pi \sigma^{2}G_{N}}-\frac{4\pi c_{3}}{G_{N}}\,,\nonumber \\
S_{l}^\text{BPS} & =\frac{\pi(r_{-}+r_{+})^{2}}{4G_{N}}+\frac{4\pi c_{3}}{G_{N}}=\frac{\varphi_{l}^{2}}{16\pi \sigma^{2}G_{N}}+\frac{4\pi c_{3}}{G_{N}}\,.\label{4DKNThermodynamicsLeft}
\end{align}
and the right-moving thermodynamics is
\begin{align}
\beta_{r} & =\frac{1}{2\sigma }\varphi_{r}=\frac{\pi(r_{-}+r_{+})^{2}}{r_{+}-r_{-}}\,,\ \ \omega_{r}=\pm2\pi i\,,\ \ J=\pm\frac{i(r_{+}^{2}-r_{-}^{2})}{4G_{N}}\,,\nonumber \\
I_{r}^\text{BPS} & =0\,,\ \ S_{r}^\text{BPS}=\frac{\pi(r_{+}^{2}-r_{-}^{2})}{2G_{N}}\,,\ \ M=2\sigma Q=\frac{r_{-}+r_{+}}{2G_{N}}\,.\label{4DKNThermodynamicsRight}
\end{align}
The quantum statistical relation and first law of BPS thermodynamics become
\be
\begin{split}
 I_{l}^\text{BPS}& = \beta_l\, M -S_l^\text{BPS} - \varphi_{l}\, Q = - S_l^\text{BPS} +  \varphi_{l}\,Q\,, \qquad\qquad\quad \delta I_l^{\rm BPS} = \delta\varphi_l Q\ ,\\ 
I_{r}^\text{BPS}& =\beta_r\, M - S_r^\text{BPS}-\varphi_{r}\, Q - \omega_{r}\, J = -S_r^\text{BPS} \mp 2 \pi i\, J  \ , \quad\, \delta I_r^{\rm BPS} = 0\,.
\end{split}
\ee
For the supergravity (\ref{Sugra4D}) in $D=4$, since constraints
(\ref{4DSUGRAConstraints}) are stronger than constraints (\ref{4DConstraint2}),
the thermodynamics remains as in \eqref{4DKNThermodynamicsLeft} - \eqref{4DKNThermodynamicsRight}. This result was discussed more carefully in \cite{Hristov:2023cuo}[Sec. 3], where the reader can find the relation with previously established results relying directly on supersymmetry, \cite{LopesCardoso:1998tkj,LopesCardoso:1999fsj,Ooguri:2004zv}. We note that the direct comparison with previous references requires a redefinition of the chemical potentials due to the non-conventional sign in the Legendre transform between the on-shell action and the entropy above.

\section{Thermodynamics in $D=5$}
\label{sec:5drotating}

In the five-dimensional case the expressions are technically rather involved, which prompts us to directly focus on the simpler case of supergravity that constrains the four-derivative corrections.

\subsection{$2$-derivative supergravity}

The 2-derivative Einstein-Maxwell-Chern-Simons theory
\begin{align}
I_{0} & =\frac{1}{16\pi G_{N}}\int\sqrt{-g}(R-\frac{1}{4\sigma ^{2}}F^{\mu\nu}F_{\mu\nu}+\frac{\epsilon^{\mu\nu\rho\sigma\tau}}{12\sqrt{3}\sigma^{3}}A_{\mu}F_{v\rho}F_{\sigma\tau})\,,\label{5DAction}
\end{align}
contains a charged rotating black hole \cite{Chong:2005hr}
\begin{align}
ds^{2}= & -\frac{(\rho^{2}dt+2q\nu)dt}{\rho^{2}}+\frac{2q\nu\omega}{\rho^{2}}+\frac{2m\rho^{2}-q^{2}}{\rho^{4}}\big(dt-\omega\big)^{2}\nonumber \\
 & +\frac{\rho^{2}dr^{2}}{\Delta_{r}}+\rho^{2}d\theta^{2}+(r^{2}+a^{2})\sin^{2}\theta d\phi^{2}+(r^{2}+b^{2})\cos^{2}\theta d\psi^{2}\ ,\nonumber \\
A_{(1)}= & -\frac{\sqrt{3}\sigma q}{\rho^{2}}\big(dt-\omega\big)+\bar{X}dt\ ,\ \ \ \bar{X}=\Phi_{0}\,,\label{CCLP}
\end{align}
where 
\begin{align}
\nu & =b\sin^{2}\theta d\phi+a\cos^{2}\theta d\psi\ ,\quad\omega=a\sin^{2}\theta d\phi+b\cos^{2}\theta d\psi\ ,\nonumber \\
\Delta_{r} & =\frac{(r^{2}+a^{2})(r^{2}+b^{2})+q^{2}+2abq}{r^{2}}-2m\ ,\ \ \ \rho^{2}=r^{2}+a^{2}\cos^{2}\theta+b^{2}\sin^{2}\theta\ ,
\end{align}
The black hole horizon $r=r_{0}$ is located at $\Delta_{r}(r_{0})=0$
\begin{equation}
m=\frac{1}{2r_{0}^{2}}\big((r_{0}^{2}+a^{2})(r_{0}^{2}+b^{2})+q^{2}+2abq\big)\,,
\end{equation}
The 2-derivative thermodynamics is
\begin{align}
\beta_{0} & =\frac{2\pi}{\kappa_{{\rm surf}}}=\frac{2\pi r_{0}\left(abq+(r_{0}^{2}+a^{2})(r_{0}^{2}+b^{2})\right)}{r_{0}^{4}-(ab+q)^{2}}\ ,\nonumber \\
J_{a,0} & =\frac{1}{16\pi G_{N}}\int\star d\xi_{\phi}=\frac{(2ma+bq)\pi}{4G_{N}}\,,\ \ \ \Omega_{a,0}=\frac{a(r_{0}^{2}+b^{2})+bq}{abq+(r_{0}^{2}+a^{2})(r_{0}^{2}+b^{2})}\,,\nonumber \\
J_{b,0} & =\frac{1}{16\pi G_{N}}\int\star d\xi_{\psi}=\frac{(2mb+aq)\pi}{4G_{N}}\,,\ \ \ \Omega_{b,0}=\Omega_{a,0}|_{a\leftrightarrow b}\,,\nonumber \\
Q_{0} & =\frac{1}{16\pi G_{N}}\int\frac{\star F_{(2)}}{\sigma^{2}}=\frac{\sqrt{3}\pi q}{4\sigma G_{N}}\ ,\ \ \Phi_{0}=\xi^{\mu}A_{\mu}|_{r_{0}}^{\infty}=\frac{\sqrt{3}\sigma qr_{0}^{2}}{abq+(r_{0}^{2}+a^{2})(r_{0}^{2}+b^{2})}\,,\nonumber \\
M_{0} & =-\frac{1}{16\pi G_{N}}\int\star d\xi_{t}=\frac{3\pi m}{4G_{N}}\,,\ \ S_{0}=\frac{A_{{\rm horizon}}}{4G_{N}}=\frac{\pi^{2}}{2G_{N}r_{0}}\big((r_{0}^{2}+a^{2})(r_{0}^{2}+b^{2})+abq\big)\ ,
\end{align}
where $\xi=\xi_{t}+\Omega_{a,0}\xi_{\phi}+\Omega_{b,0}\xi_{\psi},\,\xi_{t}=\partial_{t}\,,\ \xi_{\phi}=\partial_{\phi}\,,\ \xi_{\psi}=\partial_{\psi}$.
Using above results, we can compute the free energy
\begin{equation}
G_{0}=M_{0}-T_{0}S_{0}-\Phi_{0}Q_{0}-\Omega_{a,0}J_{a,0}-\Omega_{b,0}J_{b,0}=\frac{\pi}{4G_{N}}(m-\frac{q^{2}r_{0}^{2}}{(r_{0}^{2}+a^{2})(r_{0}^{2}+b^{2})+abq})\,,
\end{equation}
The on-shell action at fixed potential $\beta_{0},\Phi_{0},\Omega_{a,0},\Omega_{b,0}$
is given by the background subtraction method
\begin{align}
I_{0}(\beta_{0},\Phi_{0},\Omega_{a,0},\Omega_{b,0}) & =\beta_{0}G_{0}=\frac{\pi^{2}r_{0}}{2G_{N}}(\frac{m\left(abq+(r_{0}^{2}+a^{2})(r_{0}^{2}+b^{2})\right)-q^{2}r_{0}^{2}}{r_{0}^{4}-(ab+q)^{2}})\,.
\end{align}

We define a new couple of angular velocities $\omega_{l,r}^{x(y)},\ J^{x(y)}$
\cite{Hristov:2023cuo}
\begin{align}
\omega_{l,r}^{x} & =\frac{1}{2}(\omega_{l,r}^{a}+\omega_{l,r}^{b})\,,\ \ \omega_{l,r}^{y}=\frac{1}{2}(\omega_{l,r}^{a}-\omega_{l,r}^{b})\,,\nonumber \\
J^{x} & =\frac{1}{2}(J^{a}+J^{b})\,,\ \ J^{y}=\frac{1}{2}(J^{a}-J^{b})\,.
\end{align}
where $\omega_{l,r}^{\alpha}:=\frac{1}{2}(\beta_{+}\Omega_{+}^{\alpha}\pm\beta_{-}\Omega_{-}^{\alpha}),\, \alpha=a,b$, and we find $\omega_{l}^{x}=\omega_{r}^{y}=0$ at 2-derivative level.
In this case we find that the parameters $m,\,q,\,a,\,b$ can
be inverted in terms of the potentials $\beta_{l},\,\varphi_{l},\,\omega_{l}^{y}\,(\omega_{r}^{x})$
\begin{equation}
m=\frac{3\beta_{l}^{2}-\frac{\varphi_{l}^{2}}{\sigma^{2}}}{6(\omega_{l,y}^{2}+\pi^{2})},\ \ q=\frac{\varphi_{l}\left(\sqrt{3}\beta_{l}+\frac{\varphi_{l}}{\sigma}\right)}{3\sigma(\omega_{l,y}^{2}+\pi^{2})},\ \ a-b=\frac{\omega_{l,y}\left(\sqrt{3}\beta_{l}+\frac{\varphi_{l}}{\sigma}\right)}{\sqrt{3}(\omega_{l,y}^{2}+\pi^{2})}\,,
\end{equation}
and 
\begin{equation}
m=\frac{3\beta_{r}^{2}-\frac{\varphi_{r}^{2}}{\sigma^{2}}}{6(\omega_{r,x}^{2}+\pi^{2})},\ \ q=\frac{\varphi_{r}\left(\sqrt{3}\beta_{r}-\frac{\varphi_{r}}{\sigma}\right)}{3\sigma(\omega_{r,x}^{2}+\pi^{2})},\ \ a+b=\frac{\omega_{r,x}\left(\sqrt{3}\beta_{r}-\frac{\varphi_{r}}{\sigma}\right)}{\sqrt{3}(\omega_{r,x}^{2}+\pi^{2})}\ ,
\end{equation}
where $\omega_{r,x}=\omega^x_r,\, \omega_{l,y}=\omega^y_l$. The  left- and right-moving thermodynamics is \cite{Hristov:2023cuo}
\begin{align}
M & =\frac{\pi\left(3\sigma^{2}\beta_{l}^{2}-\varphi_{l}^{2}\right)}{8\sigma^{2}G_{N}(\omega_{l,y}^{2}+\pi^{2})}=\frac{\pi\left(3\sigma^{2}\beta_{r}^{2}-\varphi_{r}^{2}\right)}{8\sigma^{2}G_{N}(\omega_{r,x}^{2}+\pi^{2})},\ Q=\frac{\pi\varphi_{l}\left(3\sigma\beta_{l}+\sqrt{3}\varphi_{l}\right)}{12\sigma^{3}G_{N}(\omega_{l,y}^{2}+\pi^{2})}=\frac{\pi\varphi_{r}\left(3\sigma\beta_{r}-\sqrt{3}\varphi_{r}\right)}{12\sigma^{3}G_{N}(\omega_{r,x}^{2}+\pi^{2})},\nonumber \\
J^{x} & =\frac{\pi\omega_{r,x}\left(\sqrt{3}\varphi_{r}-3\sigma\beta_{r}\right){}^{2}\left(3\sigma\beta_{r}+2\sqrt{3}\varphi_{r}\right)}{216\sigma^{3}G_{N}(\omega_{r,x}^{2}+\pi^{2})^{2}}\,,\ \ J^{y}=\frac{\pi\omega_{l,y}\left(9\sigma^{3}\beta_{l}^{3}-9\sigma\beta_{l}\varphi_{l}^{2}-2\sqrt{3}\varphi_{l}^{3}\right)}{72\sigma^{3}G_{N}(\omega_{l,y}^{2}+\pi^{2})^{2}}\,,\nonumber \\
I_{l} & =\frac{\omega_{l,y}^{2}+\pi^{2}}{2\pi^{2}}S_{l}=\frac{\pi(\sqrt{3}\beta_{l}+\frac{\varphi_{l}}{\sigma})^{2}(\sqrt{3}\beta_{l}-\frac{2\varphi_{l}}{\sigma})}{24\sqrt{3}G_{N}(\omega_{l,y}^{2}+\pi^{2})}\,,\nonumber \\
I_{r} & =\frac{\omega_{r,x}^{2}+\pi^{2}}{2\pi^{2}}S_{r}=\frac{\pi(\sqrt{3}\beta_{r}-\frac{\varphi_{r}}{\sigma})^{2}(\sqrt{3}\beta_{r}+\frac{2\varphi_{r}}{\sigma})}{24\sqrt{3}G_{N}(\omega_{r,x}^{2}+\pi^{2})}\,.
\end{align}
As before, these on-shell actions satisfy
 \begin{equation}
		\frac{\partial I_{l,r}}{\partial \beta_{l,r}} = M \ ,  \qquad \frac{\partial I_{l,r}}{\partial \varphi_{l,r}} = - Q\ , \qquad \frac{\partial I_{l}}{\partial \omega^y_{l}} = -2\, J^y\ , \qquad \frac{\partial I_{r}}{\partial \omega^x_{r}} = -2\, J^x\ .
\end{equation}

\subsection{$4$-derivative supergravity}
Next, let us apply the RS method \cite{Reall:2019sah} to obtain the
action of the charged rotating black hole in 5D supergravity (\ref{5DAction},
\ref{5DSugra}). After some calculations, we obtain the corrected
on-shell action $I(\beta_{0},\Phi_{0},\Omega_{a,0},\Omega_{b,0})$
at fixed potential $\beta_{0},\Phi_{0},\Omega_{a,0},\Omega_{b,0}$
\begin{equation}
I=\frac{\pi^{2}r_{0}}{2G_{N}}(\frac{m\left(abq+(r_{0}^{2}+a^{2})(r_{0}^{2}+b^{2})\right)-q^{2}r_{0}^{2}}{r_{0}^{4}-(ab+q)^{2}})+\Delta I\,,
\end{equation}
where
\begin{align}
\Delta I= & -\frac{\pi^{2}d_{1}}{2r_{0}^{3}G_{N}(r_{0}^{2}+a^{2})(r_{0}^{2}+b^{2})((ab+q)^{2}-r_{0}^{4})}\Big(r_{0}^{2}\left(a^{2}+b^{2}\right)(ab+q)^{3}(5ab+3q)\nonumber \\
 & -11r_{0}^{10}\left(a^{2}+b^{2}\right)+r_{0}^{8}\left(-3a^{4}+7a^{2}b^{2}+3abq-3b^{4}+6q^{2}\right)-r_{0}^{6}\left(a^{2}+b^{2}\right)\big(a^{4}\nonumber \\
 & -24a^{2}b^{2}-26abq+b^{4}-8q^{2}\big)+r_{0}^{4}(ab+q)(5a^{5}b+2a^{4}q+31a^{3}b^{3}+45a^{2}b^{2}q\nonumber \\
 & +5ab^{5}+19abq^{2}+2b^{4}q+3q^{3})-ab(ab+q)^{5}-9r_{0}^{12}\Big)\,.
\end{align}
To derive the left- and right-moving thermodynamic quantities, we
perform the redefinitions of the integral parameters $r_{0},\,q,\,a,\,b$,
then we will obtain the thermodynamics at fixed $M_{0},Q_{0},J_{a,0},J_{b,0}$.
For computational convenience, analogous to the method employed in
preceding sections, thermodynamics can be expressed in terms of the
parameter $(r_{+},r_{-},a,b)$ via
\begin{equation}
m=\frac{1}{2}(a^{2}+b^{2}+r_{+}^{2}+r_{-}^{2})\,,\ \ q=\pm r_{+}r_{-}-ab\,.\label{5Dmq}
\end{equation}
we choose the sign ``$+$'' in (\ref{5Dmq}). The other sign choice
``$-$'' corresponds to sending $\omega_{l}^{x}\leftrightarrow\omega_{l}^{y},\,J^{x}\leftrightarrow J^{y},\,b\rightarrow-b,\,\sigma\rightarrow-\sigma$
in the results obtained using the sign choice ``$+$''. 
They satisfy the first law and quantum statistical relation
\begin{align}
S_{l,r} & =-I_{l,r}+\beta_{l,r}M-\varphi_{l,r}Q-2\omega_{l,r}^{x}J^{x}-2\omega_{l,r}^{y}J^{y}\,,\nonumber \\
\delta I_{l,r} & =M\delta\beta_{l,r}-Q\delta\varphi_{l,r}-2J^{x}\delta\omega_{l,r}^{x}-2J^{y}\delta\omega_{l,r}^{y}\,,\nonumber \\
\delta S_{l,r} & =\beta_{l,r}\delta M-\varphi_{l,r}\delta Q-2\omega_{l,r}^{x}\delta J^{x}-2\omega_{l,r}^{y}\delta J^{y}\,,
\end{align}
For 2-derivative gravity, both $\omega_{l}^{x}$ and $\omega_{r}^{y}$ vanish identically. In contrast, 4-derivative gravity introduces corrections to these values $\omega_{l}^{x}=\mathcal{O}(d_{1})\,,\ \omega_{r}^{y}=\mathcal{O}(d_{1})$, providing a more refined understanding of  gravitational effects. In this case we find that $r_{+},\, r_{-},\,a,\,b$ can no longer be inverted in terms of the potentials $\beta_l,\varphi_l,\omega_l^x,\omega_l^y$, precisely because $\omega^x_l$ are zero at the leading order $\mathcal{O}(d_1^0)$, and we have the analogous problem in the right-moving sector. We can instead look at the entropy in the microcanonical ensemble of fixed $M, Q, J^x , J^y$.
The parameters that characterize the black hole solution can be explicitly expressed as
\begin{align}
a+b & =\frac{6J_{x}}{\sqrt{3}\sigma Q+2M}\,,\ \ (r_{+}-r_{-})^{2}=-\frac{36J_{x}^{2}}{(\sqrt{3}\sigma  Q+2M)^{2}}+\frac{8G_{N}(M-\sqrt{3}\sigma Q)}{3\pi}\,,\nonumber \\
a-b & =\frac{6J_{y}}{2M-\sqrt{3}\sigma Q}\,,\ \ (r_{+}+r_{-})^{2}=-\frac{36J_{y}^{2}}{(\sqrt{3}\sigma Q-2M)^{2}}+\frac{8G_{N}(\sqrt{3}\sigma Q+M)}{3\pi}\,,
\end{align}
and we find the entropy
\begin{align}
S_{l} & =S_{l,0}+\Delta S_{l}\,,\nonumber \\
S_{l,0} & =\frac{2\sqrt{\frac{2}{3}\pi G_{N}(\sqrt{3}\sigma Q+M)(4M^{2}-3\sigma ^{2}Q^{2})^{2}-9\pi^{2}J_{y}^{2}(3\sigma ^{2}Q^{2}+4\sqrt{3}\sigma MQ+4M^{2})}}{3\sqrt{3}\sigma Q+6M}\nonumber \\
\Delta S_{l} & =\frac{8\pi^{2}d_{1}(4M^{2}-3\sigma ^{2}Q^{2})}{3G_{N}S_{l,0}(3\sqrt{3}\sigma Q+6M)}\Big(2\sigma ^{3}Q^{3}G_{N}^{3}(4M^{2}-3\sigma ^{2}Q^{2})^{4}(3\sqrt{3}\sigma ^{2}Q^{2}+6\sigma MQ+\sqrt{3}M^{2})\nonumber \\
 & +162\pi^{2}G_{N}\big(-J_{x}^{2}J_{y}^{2}(4M^{2}-3\sigma ^{2}Q^{2})^{2}(3\sqrt{3}\sigma ^{3}Q^{3}+24\sigma ^{2}MQ^{2}+9\sqrt{3}\sigma M^{2}Q+2M^{3})\nonumber \\
 & -J_{y}^{4}(M^{2}-3\sigma ^{2}Q^{2})(9\sqrt{3}\sigma ^{5}Q^{5}+168\sigma ^{2}M^{3}Q^{2}+96\sqrt{3}\sigma ^{3}M^{2}Q^{3}+48\sqrt{3}\sigma M^{4}Q+16M^{5}\nonumber \\
 & +81\sigma ^{4}MQ^{4})+3MJ_{x}^{4}(27\sigma ^{6}Q^{6}-54\sqrt{3}\sigma ^{5}MQ^{5}+81\sigma ^{4}M^{2}Q^{4}+24\sqrt{3}\sigma ^{3}M^{3}Q^{3}+16M^{6}\nonumber \\
 & -72\sigma ^{2}M^{4}Q^{2})\big)-8748\pi^{3}MJ_{x}^{2}J_{y}^{2}\big(3\sqrt{3}\sigma ^{3}Q^{3}(J_{x}^{2}-J_{y}^{2})-8\sqrt{3}\sigma M^{2}QJ_{y}^{2}+4M^{3}(J_{x}^{2}-J_{y}^{2})\nonumber \\
 & -3\sigma ^{2}MQ^{2}(3J_{x}^{2}+5J_{y}^{2})\big)-3\pi G_{N}^{2}(4M^{2}-3\sigma ^{2}Q^{2})^{2}\big(27\sigma ^{6}Q^{6}(7J_{y}^{2}-3J_{x}^{2})+8M^{6}(J_{x}^{2}-J_{y}^{2})\nonumber \\
 & +9\sqrt{3}\sigma ^{5}MQ^{5}(39J_{y}^{2}-7J_{x}^{2})+9\sigma ^{4}M^{2}Q^{4}(41J_{x}^{2}+71J_{y}^{2})+33\sqrt{3}\sigma ^{3}M^{3}Q^{3}(J_{x}^{2}+3J_{y}^{2})\nonumber \\
 & -6\sigma ^{2}M^{4}Q^{2}(35J_{x}^{2}+17J_{y}^{2})-4\sqrt{3}\sigma M^{5}Q(7J_{x}^{2}+9J_{y}^{2})\big)\Big)\Big(36\pi \sigma ^{2}Q^{2}G_{N}J_{y}^{2}(4M^{3}-3\sqrt{3}\sigma ^{3}Q^{3}\nonumber \\
 & -9\sigma ^{2}MQ^{2})(4M^{2}-3\sigma ^{2}Q^{2})^{2}+4G_{N}^{2}(4\sigma M^{2}Q-3\sigma ^{3}Q^{3})^{4}+81\pi^{2}J_{x}^{4}(81\sigma ^{4}M^{2}Q^{4}+27\sigma ^{6}Q^{6}\nonumber \\
 & -54\sqrt{3}\sigma ^{5}MQ^{5}+24\sqrt{3}\sigma ^{3}M^{3}Q^{3}-72\sigma ^{2}M^{4}Q^{2}+16M^{6})+81\pi^{2}J_{y}^{4}(27\sigma ^{6}Q^{6}+81\sigma ^{4}M^{2}Q^{4}\nonumber \\
 & +54\sqrt{3}\sigma ^{5}MQ^{5}-24\sqrt{3}\sigma ^{3}M^{3}Q^{3}-72\sigma ^{2}M^{4}Q^{2}+16M^{6})-18\pi J_{x}^{2}\big(9\pi J_{y}^{2}(3\sigma ^{2}Q^{2}+M^{2})\nonumber \\
 & -2\sigma ^{2}Q^{2}G_{N}(3\sqrt{3}\sigma ^{3}Q^{3}-9\sigma ^{2}MQ^{2}+4M^{3})\big)(4M^{2}-3\sigma ^{2}Q^{2})^{2}\Big)^{-1}\,, \nonumber\\
 S_{r}&=S_{l}|_{J_{x}\leftrightarrow J_{y},\ \sigma \rightarrow-\sigma }\,,
\end{align}
where $J_x=J^x,\ J_y=J^y$. We can find that $S_{l,0}$ depends only on $M,\,Q,\,J_y$, while $\Delta S_l$ on all charges, $M,\,Q,\,J_x,\,J_y$. We refrain from giving the results the grand-canonical ensemble due to their considerable length. Due to the fact that $\omega_{l}^{x} \neq 0\,,\ \omega_{r}^{y} \neq 0$, we no longer find that the on-shell actions can be explicitly found in terms of the left- and right-moving chemical potentials.

\subsection{BPS limit and a puzzle from dimensional reduction}
\label{sec:5dBPS}
In the BPS limit \cite{Hristov:2023cuo}
\begin{equation}
M=\sqrt{3}\sigma Q\,,\ \ \ Q>0\,,
\end{equation}
it leads $a=-b\pm i(r_{+}-r_{-})$. For simplicity, we will focus
on the branch of the solution of the ``$+$'' sign. The other sign
choice corresponds to sending $i\rightarrow-i$ in the expression.
Imposing the BPS condition $a=-b+i(r_{+}-r_{-})$, we get the BPS
thermodynamics
\begin{align}
M & =\sqrt{3}\sigma Q=\frac{3\pi(b+ir_{-})(b-ir_{+})}{4G_{N}}\,,\nonumber \\
J^{x} & =\frac{3\pi(r_{-}-r_{+})(r_{-}-ib)(b-ir_{+})}{8G_{N}}\,,\ \ J^{y}=-\frac{\pi(b+ir_{-})(b-ir_{+})\big(2b+i(r_{-}-r_{+})\big)}{8G_{N}}\,,\nonumber \\
\beta_{l} & =\frac{\sqrt{3}}{\sigma}\varphi_{l}=\frac{3\pi(b+ir_{-})(b-ir_{+})}{r_{-}+r_{+}}+\frac{8\pi d_{1}}{(r_{-}+r_{+})^{3}(b+ir_{-})(b-ir_{+})}\big(6b^{4}+12ib^{3}(r_{-}-r_{+})\nonumber \\
 & -2b^{2}(r_{-}^{2}-16r_{+}r_{-}+r_{+}^{2})+4ib(r_{-}^{3}+4r_{+}r_{-}^{2}-4r_{+}^{2}r_{-}-r_{+}^{3})-r_{-}^{4}-r_{+}^{4}+8r_{-}^{2}r_{+}^{2}\big)\,,\nonumber \\
\omega_{l}^{x} & =0\,,\ \ \omega_{l}^{y}=\frac{\pi\left(i(r_{+}-r_{-})-2b\right)}{r_{-}+r_{+}}-\frac{16\pi d_{1}}{(r_{-}+r_{+})^{3}(b+ir_{-})(b-ir_{+})}\big(2b^{3}+3ib^{2}(r_{-}-r_{+})\nonumber \\
 & +b(r_{-}^{2}+8r_{+}r_{-}+r_{+}^{2})+i(r_{-}^{3}+2r_{+}r_{-}^{2}-2r_{+}^{2}r_{-}-r_{+}^{3})\big)\,,\nonumber \\
S_{l}^\text{BPS} & =\frac{\pi^{2}\left(r_{-}+r_{+}\right)(b+ir_{-})(b-ir_{+})}{4G_{N}}+\frac{4\pi^{2}d_{1}\left(-b^{2}-ib(r_{-}-r_{+})+r_{-}^{2}+r_{+}^{2}+r_{-}r_{+}\right)}{\left(r_{-}+r_{+}\right)G_{N}}\,,\nonumber \\
I_{l}^{{\rm BPS}} & =\frac{\pi^{2}(b+ir_{-})^{2}(b-ir_{+})^{2}}{2\left(r_{-}+r_{+}\right)G_{N}}-\frac{4\pi^{2}d_{1}}{(r_{-}+r_{+})^{3}G_{N}}\big(-2b^{4}-4ib^{3}(r_{-}-r_{+})-12b^{2}r_{-}r_{+}\nonumber \\
 & -2ib(r_{-}^{3}+3r_{+}r_{-}^{2}-3r_{+}^{2}r_{-}-r_{+}^{3})+r_{-}^{4}+r_{+}^{4}+2r_{-}r_{+}^{3}+2r_{-}^{3}r_{+}\big)\,,
\end{align}
and
\begin{align}
\beta_{r} & =\frac{\varphi_{r}}{\sqrt{3}\sigma}=-\frac{\pi(b+ir_{-})(b-ir_{+})}{r_{-}-r_{+}}-\frac{8\pi d_{1}\left(2b+i(r_{-}-r_{+})\right){}^{2}}{3(r_{-}-r_{+})(b+ir_{-})(b-ir_{+})}\,,\nonumber \\
S^{\rm BPS}_{r} & =-2\omega_{r}^{x}J^{x}=-\frac{3\pi^{2}(r_{-}-r_{+})(b+ir_{-})(b-ir_{+})}{4G_{N}}\,,\nonumber \\ 
\omega_{r}^{x} & =i\pi\,,\ \ \omega_{r}^{y}=I_{r}^{{\rm BPS}}=0\,,\ \ 
\end{align}
They satisfy the first law and quantum statistical relation
\begin{align}
 I_{l}^\text{BPS}& = \beta_l\, M -S_l^\text{BPS} - \varphi_{l}\, Q -2 \omega_l^y\, J^y = - S_l^\text{BPS} +2  \varphi_{l}\,Q- 2\omega_l^y\, J^y\,, \nonumber\\
 \delta I_{l}^{{\rm BPS}}&=2Q\delta\varphi_{l}-2J^{y}\delta\omega_{l}^{y}\,,\ \ \nonumber\\
I_{r}^\text{BPS}& =\beta_r\, M - S_r^\text{BPS}-\varphi_{r}\, Q - 2\omega^x_{r}\, J^x = -S_r^\text{BPS} -2 \pi i\, J^x  \ , \qquad \delta I_r^{\rm BPS} = 0\,.
\end{align}
Since the choice of integration constants does not affect the thermodynamic
relations $I_{l}(\beta_{l},\varphi_{l,}\omega_{l}^{y})$, we choose
new integration constants making the potential energy fixed at the
leading order $\beta_{l}\rightarrow\beta_{l,0}\,,\ \ \varphi_{l}\rightarrow\varphi_{l,0}\,,\ \ \omega_{l}^{y}\rightarrow\omega_{l,0}^{y}$
\begin{align}
r_{+} & \rightarrow r_{+}+\delta r_{+}\,,\ \ r_{-}\rightarrow r_{-}+\delta r_{-}\,,\nonumber \\
\delta r_{+} & =\frac{8d_{1}\big(3b^{3}+ib^{2}(5r_{-}-4r_{+})+b(r_{-}^{2}+12r_{+}r_{-}+2r_{+}^{2})+i(r_{-}^{3}+2r_{+}r_{-}^{2}-4r_{+}^{2}r_{-}-2r_{+}^{3})\big)}{3(r_{-}+r_{+})(b+ir_{-})^{2}(b-ir_{+})}\,,\nonumber \\
\delta r_{-} & =\frac{8d_{1}\big(3b^{3}+ib^{2}(4r_{-}-5r_{+})+b(2r_{-}^{2}+12r_{+}r_{-}+r_{+}^{2})+i(2r_{-}^{3}+4r_{+}r_{-}^{2}-2r_{+}^{2}r_{-}-r_{+}^{3})\big)}{3(r_{-}+r_{+})(b+ir_{-})(b-ir_{+})^{2}}\,.
\end{align}
Then, the left-moving thermodynamics at fixed potential $\beta_{l,0},\varphi_{l,0},\omega_{l,0}^{y}$
are
\begin{align}
\beta_{l} & =\beta_{l,0}=\frac{\sqrt{3}}{\sigma}\varphi_{l}=\frac{\sqrt{3}}{\sigma}\varphi_{l,0}=\frac{3\pi(b+ir_{-})(b-ir_{+})}{r_{-}+r_{+}}\,,\nonumber \\
\omega_{l}^{x} & =0\,,\ \ \ \ \omega_{l}^{y}=\frac{\pi\left(i(r_{+}-r_{-})-2b\right)}{r_{-}+r_{+}}\,,\nonumber \\
M & =\sqrt{3}\sigma Q=\frac{3\pi(b+ir_{-})(b-ir_{+})}{4G_{N}}+\frac{2\pi d_{1}(b^{2}+ib(r_{-}-r_{+})-r_{-}^{2}-r_{+}^{2}-r_{-}r_{+})}{G_{N}(b+ir_{-})(b-ir_{+})}\,,\nonumber \\
J^{x} & =\frac{3\pi(r_{-}-r_{+})(r_{-}-ib)(b-ir_{+})}{8G_{N}}+\frac{\pi d_{1}}{G_{N}(b+ir_{-})(b-ir_{+})}\big(2b^{3}+2ib^{2}(r_{-}-r_{+})\nonumber \\
 & +2b(2r_{-}^{2}+5r_{+}r_{-}+2r_{+}^{2})+3i(r_{-}-r_{+})(r_{-}+r_{+})^{2}\big)\,,\nonumber \\
J^{y} & =-\frac{\pi(b+ir_{-})(b-ir_{+})\big(2b+i(r_{-}-r_{+})\big)}{8G_{N}}+\frac{\pi d_{1}(r_{-}+r_{+})^{2}\left(2b+i(r_{-}-r_{+})\right)}{G_{N}(b+ir_{-})(b-ir_{+})}\,,\nonumber \\
S_{l}^\text{BPS} & =\frac{2\pi^{2}d_{1}(r_{-}+r_{+})\big(2b+i(r_{-}-r_{+})\big)^{2}}{G_{N}(b+ir_{-})(b-ir_{+})}+\frac{\pi^{2}(r_{-}+r_{+})(b+ir_{-})(b-ir_{+})}{4G_{N}}\,,\nonumber \\
I_{l}^\text{BPS} & =\frac{\pi^{2}(b+ir_{-})^{2}(b-ir_{+})^{2}}{2\left(r_{-}+r_{+}\right)G_{N}}-\frac{4\pi^{2}d_{1}(-b^{2}-ib(r_{-}-r_{+})+r_{-}^{2}+r_{+}^{2}+r_{-}r_{+})}{(r_{-}+r_{+})G_{N}}\,,\label{5DLeftBPS}
\end{align}
The parameters characterizing the black hole solution can be explicitly
inverted in terms of the $\varphi_{l},\omega_{l}^{y}$
\begin{equation}
r_{-}=-r_{+}+\frac{4\pi\varphi_{l}}{\sqrt{3}\sigma(\pi^{2}+\omega_{l,y}^{2})}\,,\ \ \ b=ir_{+}-\frac{2i\varphi_{l}}{\sqrt{3}\sigma (\pi+i\omega_{l,y})}\,.\label{rfb}
\end{equation}
Substituting eq.(\ref{rfb})
into the left-moving thermodynamics eq.(\ref{5DLeftBPS}), we obtain
the following results
\begin{align}
\label{eq:BPSentropy}
M & =\sqrt{3}\sigma Q=\frac{\pi\big(\varphi_{l}^{2}+2d_{1}\sigma^{2}(\omega_{l,y}^{2}-3\pi^{2})\big)}{\sigma^{2}G_{N}(\pi^{2}+\omega_{l,y}^{2})}\,,\quad J^{y}=\frac{2\pi\varphi_{l}\omega_{l,y}(\varphi_{l}^{2}-24\pi^{2}d_{1}\sigma^{2})}{3\sqrt{3}\sigma^{3}G_{N}(\omega_{l,y}^{2}+\pi^{2})^{2}}\,,\nonumber \\
I_{l}^\text{BPS} & = \frac{2\pi\varphi_{l}\big(\varphi_{l}^{2}+6d_{1}\sigma^{2}(\omega_{l,y}^{2}-3\pi^{2})\big)}{3\sqrt{3}\sigma^{3}G_{N}(\pi^{2}+\omega_{l,y}^{2})}\,,\ \quad S_{l}^\text{BPS}=\frac{4\pi^{3}(\varphi_{l}^{3}+24d_{1}\sigma^{2}\varphi_{l}\omega_{l,y}^{2})}{3\sqrt{3}\sigma^{3}G_{N}(\pi^{2}+\omega_{l,y}^{2})^{2}}\,.
\end{align}

For conventional purposes and simpler comparison with the following discussion, we can set $G_N = 1$ and $\varphi_\text{BPS} := - 2\varphi_l$, $\omega_{y, \text{BPS}}:= 2\omega^y_l$, such that
\be
\label{eq:5dbps}
\begin{split}
 I_{l}^\text{BPS} & = - S_l^\text{BPS} -  \varphi_\text{BPS}\,Q- \omega_{y, \text{BPS}}\, J^y\,, \\ 
 I_{l}^\text{BPS} & =- \frac{\pi\varphi_\text{BPS} \big(\varphi_\text{BPS}^{2}+6\, d_{1}\, \sigma^2 (\omega_{y, \text{BPS}}^{2}-12\pi^{2})\big)}{12\sqrt{3} \sigma^3\, (\pi^{2}+\frac{1}{4}\omega_{y, \text{BPS}}^{2})}\,.
\end{split}
\ee

This answer is in agreement with the recent result in \cite{Cassani:2024tvk},~\footnote{Due to the different supergravity conventions we use, the precise map between \eqref{eq:5dbps} and (7.9) in \cite{Cassani:2024tvk} involves setting $\sigma = 1/\sqrt{3}$ and relating $\omega^\text{here}_y = \omega_-^\text{there}$, as well as $d_1^\text{here} = (\alpha \lambda)^\text{there}$ upon taking the minimal supergravity limit.} derived as a limit from asymptotically AdS thermodynamics, see also \cite{Bobev:2022bjm,Cassani:2022lrk}. This result is also in agreement with the earlier calculation of \cite{deWit:2009de}, as can be most directly demonstrated by comparing the expression for the entropy in \eqref{eq:BPSentropy} with eq. (7.12) in \cite{deWit:2009de}. Notably, it is in disagreement with other calculations, see \cite{Castro:2007ci,Gupta:2021roy}, which relate more directly to a four-dimensional point of view that we discuss next. We suggest that the reason for the mismatch lies in the understanding of the $D=4$/$D=5$ connection at a higher-derivative level, but leave the full resolution for future study.

\subsection*{The $D=4$ point of view}

Adopting four-dimensional point of view focusing on the near-horizon geometry AdS$_2 \times$S$^2$ coming from the reduction of the three-sphere (here we set $\sigma=1$),~\footnote{We can think of the Kaluza-Klein (KK) compactification of 5d supergravity on a circle. The explicit map can be found in \cite{Gaiotto:2005gf,Behrndt:2005he} at two-derivative order, and in \cite{Butter:2014iwa} in the presence of four-derivative invariants.} the higher-derivative theory including the Weyl$^2$ invariant with coefficient $d_1$ can be defined by the so called prepotential, see \cite{Andrianopoli:1996cm}, 
\be
\label{eq:5dprepot}
	F_{5 d} (X; A_\mathbb{W}, A_\mathbb{T}) = - \frac{(X^1)^3}{X^0} +  d_1\, \frac{(3 A_\mathbb{W}+A_\mathbb{T})}{8}\, \frac{X^1}{X^0}\ ,
\ee
with two holomorphic sections $X^{0,1}$, and auxiliary higher-derivative fields $A_\mathbb{W}, A_\mathbb{T}$ governing the Weyl$^2$ and $\log$-invariants, respectively, as discussed earlier. Due to the fact that upon compactification, there appears an additional Kaluza-Klein vector multiplet in 4d, such that the theory contains an additional complex scalar and abelian vector.

In order to apply the four-dimensional fixed-point, or localization formula, that reproduces the on-shell action of the black holes (also known as OSV formula, see e.g. \cite{LopesCardoso:1998tkj,Ooguri:2004zv,Butter:2014iwa,Hristov:2021qsw}), we further need to give the conserved charges from $D=4$ perspective. The electric charge $Q$ simply becomes (upto a conventional $\sqrt{3}$ prefactor) the electric charge $Q^1$ carried by the gauge field $A^1$, conjugate to the chemical potential $\varphi^1_\text{BPS}$. Instead, the KK gauge field $A^0$ carries both an electric charge, which is $D=5$ angular momentum $J^y$, and magnetic charge $P^0 = 1$ responsible for the fibration of the additional circle (the fifth dimension) over S$^2$. The OSV formula therefore produces
\bea
\label{eq:4dOSV}
\begin{split}
	I^\text{OSV}_{5d} &= i\, (F_{5d} - \bar F_{5d}) \Big|_{(X^0 = \varphi^0_\text{BPS} + i \pi P^0, X^1 = \varphi^1_\text{BPS}; A_\mathbb{W} = - 16 \pi^2, A_\mathbb{T} = 0)} \\ 
& = - \frac{2 \pi\, (\varphi^1_\text{BPS})^3}{ (\varphi^0_\text{BPS})^2+\pi^2 } + d_1 \frac{12 \pi^3\, \varphi^1_\text{BPS}}{ (\varphi^0_\text{BPS})^2+\pi^2 }\ ,
\end{split}
\eea
in disagreement with \eqref{eq:5dbps} upon the identification $\varphi^0 = \omega_y/2$ and $\varphi^1 = \varphi/ (2 \sqrt{3})$ in the 4d and 5d pictures of the same physical system. Note that the disagreement is specifically in the missing term proportional to $d_1\, \omega_y^2$, which is exactly the missing term in \cite{Castro:2007ci,Gupta:2021roy}.

It is interesting to mention that this discrepancy was originally attributed in \cite{deWit:2009de}[Sec. 7] to the lack of clear understanding of the higher-derivative relation between the four- and five-dimensional supergravity theories. This gap has since been closed in \cite{Banerjee:2011ts} and \cite{Butter:2014iwa} and was already taken into account in the prepotential \eqref{eq:5dprepot}. The fact that the mismatch still persists is an indication that we need to revisit the application of fixed-point/localizaion formulas across dimensional reduction examples, particularly in the presence of topology change.~\footnote{The mismatch appears to hold specifically for black holes with spherical topology, going from S$^3$ horizon in $D=5$ to S$^2$ horizon in $D=4$. In contract, the dimensional reduction of black rings, \cite{deWit:2009de}, or black strings, \cite{Bobev:2021qxx}, in presence of HD terms is fully consistent and presents no such discrepancy. We also note that the mismatch is independent of the asymptotics and remains true also for BPS black holes in AdS$_5$ in presence of four-derivative corrections, whose four-dimensional reduction has been utilized in \cite{Hosseini:2017mds}.} It would be thus interesting to carefully repeat the steps in \cite{Cassani:2024kjn,Colombo:2025ihp} including HD corrections, and in turn relate via dimensional reduction the answer to the OSV formula and more general fixed-point results in $D=4$ HD supergravity, \cite{Ooguri:2004zv,Hristov:2021qsw,Hristov:2024cgj}.

\subsection{Almost BPS limit}
Let us now focus on the almost BPS limit
\begin{equation}
M=-\sqrt{3}\sigma Q\,,\ \ \ Q<0\,,\label{AlmostBPS}
\end{equation}
We choose the branch of the solution of the ``$+$'' sign, eq.(\ref{AlmostBPS})
leads $a=b+i(r_{+}+r_{-})$. Imposing the almost BPS condition, we
get the almost BPS thermodynamics 
\begin{align}
M & =-\sqrt{3}\sigma Q=\frac{3\pi(b+ir_{-})(b+ir_{+})}{4G_{N}}\,, \ \ 
J_{x}=\frac{\pi(b+ir_{-})(b+ir_{+})\big(2b+i(r_{-}+r_{+})\big)}{8G_{N}}\,,\nonumber \\
J_{y}&=\frac{3i\pi(r_{-}+r_{+})(b+ir_{-})(b+ir_{+})}{8G_{N}}\,,
\end{align}
the left-moving sector
\begin{align}
\beta_{l} & =-\frac{1}{\sqrt{3}\sigma}\varphi_{l}=\frac{\pi(b+ir_{-})(b+ir_{+})}{r_{-}+r_{+}}+\frac{8\pi d_{1}\big(2b+i(r_{-}+r_{+})\big)^{2}}{3(r_{-}+r_{+})(b+ir_{-})(b+ir_{+})}\,,\nonumber \\
S_{l}^\text{aBPS} & =-2\omega_{l}^{y}J^{y}=\frac{3\pi^{2}(r_{-}+r_{+})(b+ir_{-})(b+ir_{+})}{4G_{N}}\,,\nonumber \\
\omega_{l}^{x} & =I_{l}^\text{aBPS}=0\,,\ \ \ \omega_{l}^{y}=i\pi\,,\ \ 
\end{align}
and the right-moving sector
\begin{align}
\beta_{r} & =-\frac{\sqrt{3}}{\sigma}\varphi_{r}=-\frac{3\pi(b+ir_{-})(b+ir_{+})}{r_{-}-r_{+}}-\frac{8\pi d_{1}}{(r_{-}-r_{+})^{3}(b+ir_{-})(b+ir_{+})}\big(6b^{4}\nonumber \\
 & +12ib^{3}(r_{-}+r_{+})-2b^{2}(r_{-}^{2}+16r_{+}r_{-}+r_{+}^{2})+4ib(r_{-}^{3}-4r_{+}r_{-}^{2}-4r_{+}^{2}r_{-}+r_{+}^{3})\nonumber \\
 & -r_{-}^{4}-r_{+}^{4}+8r_{-}^{2}r_{+}^{2}\big)\,,\nonumber \\
\omega_{r}^{y} & =0\,,\ \ \ \omega_{r}^{x}=-\frac{i\pi(-2ib+r_{-}+r_{+})}{r_{-}-r_{+}}-\frac{16\pi d_{1}}{(r_{-}+r_{+}){}^{3}(b+ir_{-})(b+ir_{+})}\big(2b^{3}\nonumber \\
 & +3ib^{2}(r_{-}+r_{+})+b(r_{-}^{2}-8r_{+}r_{-}+r_{+}^{2})+i(r_{-}^{3}-2r_{+}r_{-}^{2}-2r_{+}^{2}r_{-}+r_{+}^{3})\big)\,,\nonumber \\
S_{r}^\text{aBPS} & =-\frac{\pi^{2}(r_{-}-r_{+})(b+ir_{-})(b+ir_{+})}{4G_{N}}+\frac{4\pi^{2}d_{1}\big(b^{2}+ib(r_{-}+r_{+})-r_{-}^{2}-r_{+}^{2}+r_{-}r_{+}\big)}{(r_{-}-r_{+})G_{N}}\,,\nonumber \\
I_{r}^\text{aBPS} & =-\frac{\pi^{2}(b+ir_{-})^{2}(b+ir_{+})^{2}}{2(r_{-}-r_{+})G_{N}}+\frac{4\pi^{2}d_{1}}{(r_{-}-r_{+}){}^{3}G_{N}}\big(-2b^{4}-4ib^{3}(r_{-}+r_{+})+12b^{2}r_{-}r_{+}\nonumber \\
 & -2ib(r_{-}^{3}-3r_{+}r_{-}^{2}-3r_{+}^{2}r_{-}+r_{+}^{3})+r_{-}^{4}+r_{+}^{4}-2r_{-}r_{+}^{3}-2r_{-}^{3}r_{+}\big)\,.
\end{align}
They satisfy the first law and quantum statistical relation
\be
\begin{split}
 I_{l}^\text{aBPS}& = \beta_l\, M -S_l^\text{aBPS} - \varphi_{l}\, Q - 2\omega_l^y\, J^y = - S_l^\text{aBPS} -2\pi i\, J^y\,, \qquad \delta I_l^{\rm aBPS} = 0\ ,\\ 
I_{r}^\text{aBPS}& =\beta_r\, M - S_r^\text{aBPS}-\varphi_{r}\, Q - 2\omega^x_{r}\, J^x = -S_r^\text{aBPS} +2 \varphi_r\, Q - 2\omega^x_r\, J^x  \ , \\ \delta I_r^{\rm aBPS}&=2Q\delta \varphi_r-2J^x\delta \omega^x_r\,.
\end{split}
\ee
We can express everything in terms of $\varphi_{r},\omega_{r,x}$:
\begin{align}
M & =-\sqrt{3}\sigma Q=\frac{\pi\big(\varphi_{r}^{2}+2d_{1}\sigma^{2}(\omega_{r,x}^{2}-3\pi^{2})\big)}{\sigma^{2}G_{N}(\pi^{2}+\omega_{r,x}^{2})}\,,\quad  J^{x}=-\frac{2\pi\varphi_{r}\omega_{r,x}\big(\varphi_{r}^{2}-24\pi^{2}d_{1}\sigma^{2}\big)}{3\sqrt{3}\sigma^{3}G_{N}(\omega_{r,x}^{2}+\pi^{2})^{2}}\,,\nonumber \\
I_{r}^\text{aBPS} & =-\frac{2\pi\varphi_{r}(\varphi_{r}^{2}+6d_{1}\sigma^{2}(\omega_{r,x}^{2}-3\pi^{2}))}{3\sqrt{3}\sigma^{3}G_{N}(\pi^{2}+\omega_{r,x}^{2})}\,,\quad S_{r}^\text{aBPS}=-\frac{4\pi^{3}(\varphi_{r}^{3}+24d_{1}\sigma^{2}\varphi_{r}\omega_{r,x}^{2})}{3\sqrt{3}\sigma^{3}G_{N}(\omega_{r,x}^{2}+\pi^{2})^{2}}\,.
\end{align}

Similarly to the BPS limit, we can set $G_N = 1$ and $\varphi_\text{aBPS} := - 2\varphi_r$, $\omega_{x, \text{aBPS}}:= 2\omega_{r,x}$, such that
\be
\label{eq:5dabps}
\begin{split}
 I_{r}^\text{aBPS} & = - S_r^\text{aBPS} -  \varphi_\text{aBPS}\,Q- \omega_{x, \text{aBPS}}\, J^x\,, \\ 
 I_{r}^\text{aBPS} & = \frac{\pi\varphi_\text{aBPS} \big(\varphi_\text{aBPS}^{2}+6 d_{1}\sigma^{2}(\omega_{x, \text{aBPS}}^{2}-12\pi^{2})\big)}{12\sqrt{3} \sigma^3\, (\pi^{2}+\frac{1}{4}\omega_{x, \text{aBPS}}^{2})}\,,
\end{split}
\ee
in clear analogy with \eqref{eq:5dbps}. Similarly to the discussion there, the four-dimensional point of view towards these black holes offers an interesting perspective, see \cite{Goldstein:2008fq,Bena:2009ev,Hristov:2012nu}. We refrain from repeating the steps outlined in \cite{Hristov:2021qsw}, but note that an analogous mismatch at higher-derivative level as discussed around \eqref{eq:4dOSV} persists also in this case.

\subsection*{Acknowledgements}
\noindent We wish to thank Robert Saskowski for initial discussions and collaboration on related subjects. P.H. and Y.P. are supported by the National Key R\&D Program No.~2022YFE0134300 and the National Natural Science Foundation of China (NSFC) Grant No.~12175164, No.~12247103 and No.~12447138. K.H. is supported in part by the Bulgarian NSF grant KP-06-N68/3.

\appendix

\section{Corrected solutions and thermodynamics}
\label{AppendixA}

The corrected black hole solutions $\Delta h,\,\Delta f,\,\Delta\psi$
in eq.(\ref{RNsolution}) are 
\begin{align}
\Delta h & =\frac{2(D-3)}{(D-2)r^{2(D-2)}}\Bigg(2(D-4)(D-2)m^{2}c_{3}+q^{2}\Big(2(D-4)c_{1}-(D^{2}-6D+10)c_{2}\nonumber \\
 & -2(2D^{2}-11D+16)c_{3}+4(D-2)c_{4}-(D-4)(D-2)c_{5}-2(D-3)(D-2)c_{6}\Big)\Bigg)\nonumber \\
 & -\frac{4(D-3)mq^{2}}{(D-2)r^{3D-7}}\Big((D-4)(D-1)c_{1}-c_{2}-Dc_{3}+2(D-2)(D-1)c_{4}+(D-2)c_{5}\nonumber \\
 & -(D-3)(D-2)c_{6}\Big)+\frac{(D-3)q^{4}}{(D-2)(3D-7)r^{2(2D-5)}}\Big((D-4)(11D^{2}-45D+44)c_{1}\nonumber \\
 & +(4D^{3}-33D^{2}+83D-64)c_{2}+2(D-2)(D^{2}-D-4)c_{5}-4(D-2)(D-3)^{2}c_{6}\nonumber \\
 & +2(4D^{3}-34D^{2}+87D-68)c_{3}-16(D-2)^{2}(D-3)c_{7}-8(D-2)^{2}(D-3)c_{8}\nonumber \\
 & +4(D-2)(5D^{2}-19D+16)c_{4}\Big)\:,\nonumber \\
\Delta f & =\Delta h-\frac{4(D-3)q^{2}f_{0}}{(D-2)r^{2(D-2)}}\Big((D-4)(2D-3)c_{1}+(D-2)(D-1)(c_{5}+c_{6})\nonumber \\
 & +(D^{2}-5D+5)c_{2}+(2D^{2}-9D+8)c_{3}+2(D-2)(2D-3)c_{4}\Big),\nonumber \\
\Delta\psi & =-\frac{2\sqrt{2}(D-3)^{3/2}\sigma q^{3}}{\sqrt{D-2}(3D-7)r^{3D-7}}\Big(-(7D-19)(D-2)c_{6}-8(D-2)^{2}(2c_{7}+c_{8})\nonumber \\
 & +(D^{2}-5D+5)c_{2}+(2D^{2}-9D+8)c_{3}+(D-2)\big(2(D+1)c_{4}-(D-5)c_{5}\big)\nonumber \\
 & +(D-4)(2D-3)c_{1}\Big)-\frac{4\sqrt{2}(D-3)^{3/2}\sqrt{D-2}\, m\sigma q}{r^{2(D-2)}}c_{6}\,.
\end{align}

The corrected thermodynamics $\Delta r,\ \Delta\beta,\,\Delta S,\ \Delta\Phi,\,\Delta I$
of static charged black hole are
\begin{align}
\Delta r & =\frac{r_{0}^{5-2D}}{(D-2)(3D-7)(r_{0}^{2D-6}-q^{2})}\Big[ ((D-3)(3D-7) q^{2}\big(c_{1}(D-4)+c_{2}(D-3)\nonumber \\
 & +c_{3}(3D-8)+(2c_{4}+c_{5}+c_{6})(D-2)\big)r_{0}^{2D-6}-c_{3}(D-4)(D-2)(3D-7)r_{0}^{4(D-3)}\nonumber \\
 & -(D-3)(D-2)q^{4}\big(5c_{1}(D-4)+c_{2}(4D-13)+c_{3}(11D-32)+4c_{4}(2D-3)\nonumber \\
 & +2(c_{5}+c_{6})(D-1)-8(2c_{7}+c_{8})(D-2)\big)\Big]\ ,\nonumber \\
\Delta S & =\frac{r_{0}^{2-D}\, \omega_{D-2}}{4G_{N}(3D-7)(r_{0}^{2D-6}-q^{2})}\Big[-2(c_{3}+c_{6})(D-3)(D-2)(3D-7) q^{2}r_{0}^{2D-6}\nonumber \\
 & +(D-3)q^{4}\big(c_{3}(D(7D-36)+48)+2c_{6}(D-2)(5D-13)+c_{2}(D(2D-11)+16)\nonumber \\
 & +4(D-2)(c_{4}(D-4)+c_{5}(D-3))+8(2c_{7}+c_{8})(D-2)^{2}+c_{1}(D-4)^{2}\big)\nonumber \\
 & +c_{3}(D-2)^{2}(3D-7) r_{0}^{4(D-3)}\Big]\,,\nonumber
\end{align}
 \begin{align}
 \Delta\beta & =-\frac{4\pi r_{0}^{11-4D}}{(D-3)(D-2)(3D-7)(1-q^{2}r_{0}^{6-2D})^{2}(r_{0}^{2D-6}-q^{2})}\Big[\nonumber \\
 & -(D-3)(D-2)q^{6}\big(c_{1}(D-4)^{2}+c_{2}(D(2D-11)+16)+c_{3}(D(7D-36)+48)\nonumber \\
 & +2c_{6}(D-2)(5D-13)+4(D-2)(c_{4}(D-4)+c_{5}(D-3))+8(2c_{7}+c_{8})(D-2)^{2}\big)\nonumber \\
 & +(D-2)^{2}(3D-7)\big(q^{2}(c_{3}(D-4)-2c_{6}(D-3))r_{0}^{4(D-3)}-c_{3}(D-4) r_{0}^{6(D-3)}\big)\nonumber \\
 & +(D-3) q^{4}r_{0}^{2D-6}\big(c_{1}(3D-8)(D-4)^{2}+c_{3}(3D(D(5D-38)+100)-272)\nonumber \\
 & +c_{2}(3D-8)(D(2D-11)+16)+4(D-2)(3D-8)(c_{4}(D-4)+c_{5}(D-3))\nonumber \\
 & +8c_{6}(D-2)(3(D-5)D+19)+8(2c_{7}+c_{8})(D-2)^{2}(3D-8)\big)\Big]\,,\nonumber \\
\Delta\Phi & =\frac{\sqrt{2}\sigma qr_{0}^{7-3D}}{(D-3)(3D-7)\sqrt{D^{2}-5D+6}(r_{0}^{2D-6}-q^{2})}\Big[ (D-3)^{3}q^{2}(q^{2}-2 r_{0}^{2D-6})\nonumber \\
 & \times\big(c_{2}(2D^{2}-11D+16)+c_{3}(7D^{2}-36D+48)+2c_{6}(D-2)(5D-13)\nonumber \\
 & +c_{1}(D-4)^{2}+4(D-2)\big(c_{4}(D-4)+c_{5}(D-3)\big)+8(2c_{7}+c_{8})(D-2)^{2}\big)\nonumber \\
 & +(D-3)^{2}(D-2)(3D-7)\big(c_{3}(D-4)+2c_{6}(D-3)\big)r_{0}^{4(D-3)}\Big]\,.\nonumber \\
\Delta I & =\frac{r_{0}^{-D-4}\, \omega_{D-2}}{4(D-3)(3D-7)G_{N}\left(r_{0}^{2D}-q^{2}r_{0}^{6}\right)}\Big[-(D-2)(3D-7) r_{0}^{2D}\big(c_{3}(D-2) r_{0}^{2D}\nonumber \\
 & -2(c_{3}+c_{6})(D-3)q^{2}r_{0}^{6}\big)-(D-3)q^{4}r_{0}^{12}\big(4(4c_{1}+4c_{2}+12c_{3}+8c_{4}+6c_{5}\nonumber \\
 & +13c_{6}+16c_{7}+8c_{8})-(8c_{1}+11c_{2}+36c_{3}+24c_{4}+20c_{5}+46c_{6}+64c_{7}\nonumber \\
 & +32c_{8})D+(c_{1}+2c_{2}+7c_{3}+4c_{4}+4c_{5}+10c_{6}+16c_{7}+8c_{8})D^{2}\big)\Big]\,.
\end{align}

\section{$D=4$ Kerr-Newman thermodynamics}
\label{AppendixB}

The redefinitions of the integral parameters in eq.(\ref{drdadq})
are
\begin{align}
\delta r_{0} & =4a^{2}c_{6}q^{2}[5r_{0}\left(a^{2}+r_{0}^{2}\right)^{3}\left(r_{0}^{4}-(a^{2}+q^{2})^{2}\right)\left(a^{4}+a^{2}(3q^{2}+2r_{0}^{2})-q^{2}r_{0}^{2}+r_{0}^{4}\right)]^{-1}\nonumber \\
 & \times\big(r_{0}^{8}\left(11a^{2}-35q^{2}\right)-a^{4}\left(a^{2}+q^{2}\right)^{3}+r_{0}^{6}\left(11q^{4}-10a^{4}+24a^{2}q^{2}\right)+a^{2}r_{0}^{2}(a^{2}+q^{2})(11a^{4}\nonumber \\
 & +29a^{2}q^{2}+12q^{4})+r_{0}^{4}\left(6a^{6}-26a^{4}q^{2}-41a^{2}q^{4}-3q^{6}\right)+15r_{0}^{10}\big)\nonumber \\
 & -8a^{2}(4c_{7}+c_{8})q^{4}r_{0}[15(a^{2}+r_{0}^{2})^{5}\left((a^{2}+q^{2})^{2}-r_{0}^{4}\right)\left(a^{4}+a^{2}(3q^{2}+2r_{0}^{2})-q^{2}r_{0}^{2}+r_{0}^{4}\right)]^{-1}\nonumber \\
 & \times\big(6a^{10}+5a^{8}\left(3q^{2}-4r_{0}^{2}\right)+a^{6}\left(9q^{4}-82q^{2}r_{0}^{2}-20r_{0}^{4}\right)+a^{4}\left(24r_{0}^{6}-65q^{4}r_{0}^{2}+56q^{2}r_{0}^{4}\right)\nonumber \\
 & +a^{2}\left(51q^{4}r_{0}^{4}-62q^{2}r_{0}^{6}-2r_{0}^{8}\right)-3q^{4}r_{0}^{6}+41q^{2}r_{0}^{8}-20r_{0}^{10}\big)\nonumber \\
 & +q^{4}f_{5D,2}[16a^{5}r_{0}^{4}\left((a^{2}+q^{2})^{2}-r_{0}^{4}\right)\left(a^{4}+a^{2}\left(3q^{2}+2r_{0}^{2}\right)-q^{2}r_{0}^{2}+r_{0}^{4}\right)]^{-1}\Big(-6\pi a^{12}\nonumber \\
 & -12a^{11}r_{0}+15\pi a^{10}r_{0}^{2}+34a^{9}r_{0}^{3}+3\pi a^{8}\left(q^{2}-r_{0}^{2}\right)\left(2q^{2}+r_{0}^{2}\right)+6\pi a^{6}r_{0}^{4}(3q^{2}-5r_{0}^{2})\nonumber \\
 & +2a^{7}(6q^{4}r_{0}-3q^{2}r_{0}^{3}-22r_{0}^{5})-2a^{5}(2q^{4}r_{0}^{3}+13q^{2}r_{0}^{5}+16r_{0}^{7})+6\pi a^{4}r_{0}^{4}(q^{4}+2q^{2}r_{0}^{2}+4r_{0}^{4})\nonumber \\
 & -2a^{3}\left(12q^{4}r_{0}^{5}+q^{2}r_{0}^{7}+20r_{0}^{9}\right)+3\pi a^{2}r_{0}^{6}\left(4q^{4}+2q^{2}r_{0}^{2}+5r_{0}^{4}\right)+15r_{0}^{9}(2a-\pi r_{0})\left(r_{0}^{2}-q^{2}\right)\nonumber \\
 & +6\tan^{-1}\left(\frac{r_{0}}{a}\right)\left(a^{2}+r_{0}^{2}\right)\big(-5r_{0}^{8}\left(2a^{2}+q^{2}\right)+a^{2}r_{0}^{6}\left(2a^{2}+3q^{2}\right)+2a^{6}\left(a^{4}-q^{4}\right)\nonumber \\
 & +a^{2}r_{0}^{4}\left(8a^{4}-7a^{2}q^{2}-4q^{4}\right)+a^{4}r_{0}^{2}\left(-7a^{4}+a^{2}q^{2}+2q^{4}\right)+5r_{0}^{10}\big)\Big)\,,\nonumber 
 \end{align}
 \begin{align}
\delta a & =-8ac_{6}q^{2}[5\left(a^{2}+r_{0}^{2}\right){}^{3}\left(a^{2}+q^{2}+r_{0}^{2}\right)\left(a^{4}+a^{2}\left(3q^{2}+2r_{0}^{2}\right)-q^{2}r_{0}^{2}+r_{0}^{4}\right)]^{-1}\big(5r_{0}^{8}\nonumber \\
 & +r_{0}^{6}(4a^{2}-9q^{2})-a^{2}r_{0}^{2}(4a^{2}+5q^{2})(a^{2}+q^{2})+2r_{0}^{4}(q^{4}-3a^{4}+6a^{2}q^{2})+a^{4}(a^{2}+q^{2})^{2}\big)\nonumber \\
 & +8a\left(4c_{7}+c_{8}\right)q^{4}r_{0}^{2}[15\left(a^{2}+r_{0}^{2}\right){}^{5}\left(a^{2}+q^{2}+r_{0}^{2}\right)\left(a^{4}+a^{2}\left(3q^{2}+2r_{0}^{2}\right)-q^{2}r_{0}^{2}+r_{0}^{4}\right)]^{-1}\nonumber \\
 & \times\big(9a^{8}+a^{6}\left(9q^{2}-8r_{0}^{2}\right)-a^{4}\left(49q^{2}r_{0}^{2}+30r_{0}^{4}\right)+51a^{2}q^{2}r_{0}^{4}-19q^{2}r_{0}^{6}+13r_{0}^{8}\big)\nonumber \\
 & +q^{4}f_{5D,2}[32a^{6}r_{0}^{3}\left(a^{2}+q^{2}+r_{0}^{2}\right)\left(a^{4}+a^{2}(3q^{2}+2r_{0}^{2})-q^{2}r_{0}^{2}+r_{0}^{4}\right)]^{-1}\Big(33\pi a^{10}+66a^{9}r_{0}\nonumber \\
 & +3\pi a^{8}\left(q^{2}+3r_{0}^{2}\right)+a^{7}\left(6q^{2}r_{0}-4r_{0}^{3}\right)+6\pi a^{6}r_{0}^{2}\left(2q^{2}-7r_{0}^{2}\right)+6\pi a^{4}r_{0}^{4}\left(r_{0}^{2}-3q^{2}\right)\nonumber \\
 & +a^{3}\left(34q^{2}r_{0}^{5}-28r_{0}^{7}\right)+22a^{5}q^{2}r_{0}^{3}+\pi a^{2}\left(9r_{0}^{8}-12q^{2}r_{0}^{6}\right)+15r_{0}^{7}(2a-\pi r_{0})\left(r_{0}^{2}-q^{2}\right)\nonumber \\
 & -6\left(a^{4}-r_{0}^{4}\right)\left(11a^{6}+a^{4}\left(q^{2}+3r_{0}^{2}\right)+a^{2}\left(4q^{2}r_{0}^{2}-3r_{0}^{4}\right)-5q^{2}r_{0}^{4}+5r_{0}^{6}\right)\tan^{-1}\left(\frac{r_{0}}{a}\right)\Big)\nonumber \\
\delta q & =\frac{2c_{6}q\left(a^{4}(a^{2}+q^{2})^{2}+r_{0}^{4}(3q^{4}-14a^{4}+30a^{2}q^{2})-2a^{2}r_{0}^{2}(4a^{4}+13a^{2}q^{2}+6q^{4})-6q^{2}r_{0}^{6}+5r_{0}^{8}\right)}{5r_{0}^{2}\left(a^{2}+r_{0}^{2}\right){}^{2}\left(a^{4}+a^{2}\left(3q^{2}+2r_{0}^{2}\right)-q^{2}r_{0}^{2}+r_{0}^{4}\right)}\nonumber \\
 & -\frac{4(4c_{7}+c_{8})q^{3}\left(6a^{8}+a^{6}(9q^{2}-8r_{0}^{2})-a^{4}(65q^{2}r_{0}^{2}+28r_{0}^{4})+a^{2}(51q^{2}r_{0}^{4}-8r_{0}^{6})-3q^{2}r_{0}^{6}+6r_{0}^{8}\right)}{15\left(a^{2}+r_{0}^{2}\right){}^{4}\left(a^{4}+a^{2}\left(3q^{2}+2r_{0}^{2}\right)-q^{2}r_{0}^{2}+r_{0}^{4}\right)}\nonumber \\
 & -q^{3}\left(a^{2}+r_{0}^{2}\right)f_{5D,2}[16a^{5}r_{0}^{5}\left(a^{4}+a^{2}\left(3q^{2}+2r_{0}^{2}\right)-q^{2}r_{0}^{2}+r_{0}^{4}\right)]^{-1}\Big(3\pi a^{6}(a^{2}-q^{2})+6\pi a^{6}r_{0}^{2}\nonumber \\
 & +6a^{5}r_{0}(a^{2}-q^{2})-6\pi r_{0}^{6}\left(a^{2}+q^{2}\right)+2ar_{0}^{5}\left(5a^{2}+6q^{2}\right)-3\pi a^{2}q^{2}r_{0}^{4}+2a^{3}r_{0}^{3}\left(5a^{2}+q^{2}\right)\nonumber \\
 & +6\left(-a^{8}+a^{6}\left(q^{2}-2r_{0}^{2}\right)+a^{2}r_{0}^{4}\left(q^{2}+2r_{0}^{2}\right)+2q^{2}r_{0}^{6}+r_{0}^{8}\right)\tan^{-1}\left(\frac{r_{0}}{a}\right)+6ar_{0}^{7}-3\pi r_{0}^{8}\Big)\,.
\end{align}
where $f_{5D,2}:=c_{2}+4c_{3}+2c_{5}+4c_{6}+8c_{7}+6c_{8}$. The corrected
thermodynamics $\Delta\beta,\,\Delta S,\,\Delta\Omega,\ \Delta\Phi,\,\Delta I$
in eq.(\ref{5DKNmqj}) are
\begin{align}
\Delta\beta & =16\pi c_{6}q^{2}r_{0}[5(a^{2}+r_{0}^{2})^{2}\left(a^{2}+q^{2}-r_{0}^{2}\right){}^{3}\left(a^{2}+q^{2}+r_{0}^{2}\right)]^{-1}\big(r_{0}^{6}\left(30a^{2}+q^{2}\right)+a^{2}\left(a^{2}+q^{2}\right)^{3}\nonumber \\
 & +r_{0}^{2}\left(2a^{2}-q^{2}\right)\left(a^{2}+q^{2}\right)\left(5a^{2}+3q^{2}\right)+r_{0}^{4}\left(-20a^{4}-23a^{2}q^{2}+3q^{4}\right)-5r_{0}^{8}\big)\nonumber \\
 & -32\pi\left(4c_{7}+c_{8}\right)q^{4}r_{0}^{3}[15\left(a^{2}+r_{0}^{2}\right){}^{4}\left(a^{2}+q^{2}-r_{0}^{2}\right){}^{3}\left(a^{2}+q^{2}+r_{0}^{2}\right)]^{-1}\big(a^{6}\left(3q^{2}+26r_{0}^{2}\right)\nonumber \\
 & +a^{4}\left(3q^{4}+19q^{2}r_{0}^{2}-50r_{0}^{4}\right)+a^{2}\left(-10q^{4}r_{0}^{2}-19q^{2}r_{0}^{4}+46r_{0}^{6}\right)+3r_{0}^{4}\left(q^{2}+r_{0}^{2}\right)\left(q^{2}-2r_{0}^{2}\right)\big)\nonumber \\
 & +\pi q^{4}f_{5D,2}[4a^{5}r_{0}^{2}\left(a^{2}+q^{2}-r_{0}^{2}\right){}^{3}\left(a^{2}+q^{2}+r_{0}^{2}\right)]^{-1}\Big(-3\pi r_{0}^{8}\left(5a^{2}+6q^{2}\right)+36ar_{0}^{7}\left(a^{2}+q^{2}\right)\nonumber \\
 & -3\pi a^{6}\left(a^{2}+q^{2}\right)\left(5a^{2}+q^{2}\right)-6a^{5}r_{0}\left(a^{2}+q^{2}\right)\left(5a^{2}+q^{2}\right)+3\pi r_{0}^{6}\left(-6a^{4}-4a^{2}q^{2}+q^{4}\right)\nonumber \\
 & +6ar_{0}^{5}\left(8a^{4}+2a^{2}q^{2}-q^{4}\right)+3\pi a^{2}r_{0}^{4}\left(10a^{4}+4a^{2}q^{2}+q^{4}\right)+3\pi a^{4}r_{0}^{2}\left(3a^{4}-4a^{2}q^{2}-q^{4}\right)\nonumber \\
 & +6(a^{2}+r_{0}^{2})^{2}\big(r_{0}^{4}(11a^{2}+6q^{2})+a^{2}(a^{2}+q^{2})(5a^{2}+q^{2})-r_{0}^{2}(13a^{4}+8a^{2}q^{2}+q^{4})-3r_{0}^{6}\big)\tan^{-1}(\frac{r_{0}}{a})\nonumber \\
 & +4a^{3}r_{0}^{3}\left(7a^{4}-3a^{2}q^{2}-q^{4}\right)-18ar_{0}^{9}+9\pi r_{0}^{10}\Big)\,,\nonumber \\
\Delta S & =\frac{4\pi c_{3}}{G_{N}}+\frac{4\pi c_{6}q^{2}\big(a^{4}+a^{2}(q^{2}-10r_{0}^{2})-3q^{2}r_{0}^{2}+5r_{0}^{4}\big)}{5G_{N}\left(a^{2}+r_{0}^{2}\right){}^{2}\left(a^{2}+q^{2}-r_{0}^{2}\right)}-\frac{8\pi(4c_{7}+c_{8})q^{4}r_{0}^{2}(3a^{4}-10a^{2}r_{0}^{2}+3r_{0}^{4})}{15G_{N}\left(a^{2}+r_{0}^{2}\right){}^{4}\left(a^{2}+q^{2}-r_{0}^{2}\right)}\nonumber \\
 & -\pi q^{4}f_{5D,2}[16a^{5}r_{0}^{3}G_{N}\left(a^{2}+q^{2}-r_{0}^{2}\right)]^{-1}\big(3\pi a^{6}+6a^{5}r_{0}+3\pi a^{4}r_{0}^{2}+4a^{3}r_{0}^{3}-3\pi a^{2}r_{0}^{4}\nonumber \\
 & -6\left(a-r_{0}\right)\left(a+r_{0}\right)\left(a^{2}+r_{0}^{2}\right){}^{2}\tan^{-1}\left(\frac{r_{0}}{a}\right)+6ar_{0}^{5}-3\pi r_{0}^{6}\big)\,,\nonumber 
 \end{align}
 \begin{align}
\Delta\Omega & =\frac{8ac_{6}q^{2}r_{0}^{2}\left(5a^{4}+a^{2}\left(7q^{2}-6r_{0}^{2}\right)+2q^{4}-9q^{2}r_{0}^{2}+5r_{0}^{4}\right)}{5\left(a^{2}+r_{0}^{2}\right){}^{4}\left(\left(a^{2}+q^{2}\right)^{2}-r_{0}^{4}\right)}+[15\left(a^{2}+r_{0}^{2}\right)^{6}\left((a^{2}+q^{2})^{2}-r_{0}^{4}\right)]^{-1}\nonumber \\
 & \times8a\left(4c_{7}+c_{8}\right)q^{4}r_{0}^{2}\left(r_{0}^{4}\left(25a^{2}+19q^{2}\right)-a^{2}r_{0}^{2}\left(23a^{2}+26q^{2}\right)+3a^{4}\left(a^{2}+q^{2}\right)-13r_{0}^{6}\right)\nonumber \\
 & +q^{4}f_{5D,2}[32a^{6}r_{0}^{3}\left(a^{2}+r_{0}^{2}\right)^{2}\big(\left(a^{2}+q^{2}\right)^{2}-r_{0}^{4}\big)]^{-1}\Big(-9\pi a^{10}-18a^{9}r_{0}+3\pi a^{8}\left(5r_{0}^{2}-3q^{2}\right)\nonumber \\
 & -18a^{7}r_{0}\left(q^{2}-2r_{0}^{2}\right)+6\pi a^{6}r_{0}^{2}\left(3r_{0}^{2}-q^{2}\right)-6a^{5}r_{0}^{3}\left(q^{2}-8r_{0}^{2}\right)-30\pi a^{4}r_{0}^{6}-9\pi a^{2}r_{0}^{6}\left(2q^{2}+r_{0}^{2}\right)\nonumber \\
 & +6\left(a^{2}+r_{0}^{2}\right){}^{2}\left(r_{0}^{4}\left(13a^{2}+5q^{2}\right)-a^{2}r_{0}^{2}\left(11a^{2}+4q^{2}\right)+3a^{4}\left(a^{2}+q^{2}\right)-5r_{0}^{6}\right)\tan^{-1}\left(\frac{r_{0}}{a}\right)\nonumber \\
 & +a^{3}\left(26q^{2}r_{0}^{5}+28r_{0}^{7}\right)+30ar_{0}^{7}\left(q-r_{0}\right)\left(q+r_{0}\right)+15\pi r_{0}^{8}\left(r_{0}^{2}-q^{2}\right)\Big)\,,\nonumber \\
\Delta\Phi & =-4\sigma qc_{6}[5r_{0}(a^{2}+r_{0}^{2})^{4}\big((a^{2}+q^{2})^{2}-r_{0}^{4}\big)]^{-1}\big((6a^{6}-20q^{2}a^{4}-29q^{4}a^{2}+3q^{6})r_{0}^{4}-(5a^{2}+q^{2})r_{0}^{8}\nonumber \\
 & -a^{4}\left(a^{2}+q^{2}\right)^{3}+\left(-14a^{4}+6q^{2}a^{2}-3q^{4}\right)r_{0}^{6}+a^{2}\left(a^{2}+q^{2}\right)\left(9a^{4}+25q^{2}a^{2}+10q^{4}\right)r_{0}^{2}+5r_{0}^{10}\big)\nonumber \\
 & -8\sigma q^{3}(4c_{7}+c_{8})r_{0}[15(a^{2}+r_{0}^{2})^{6}\big((a^{2}+q^{2})^{2}-r_{0}^{4}\big)]^{-1}\big((15q^{2}-14r_{0}^{2})a^{8}+(9q^{4}-70r_{0}^{2}q^{2}-20r_{0}^{4})a^{6}\nonumber \\
 & +(20r_{0}^{6}+36q^{2}r_{0}^{4}-59q^{4}r_{0}^{2})a^{4}+(14r_{0}^{8}-10q^{2}r_{0}^{6}+31q^{4}r_{0}^{4})a^{2}+3r_{0}^{6}(q^{2}+r_{0}^{2})(q^{2}-2r_{0}^{2})+6a^{10}\big)\nonumber \\
 & +\sigma q^{3}f_{5D,2}[8a^{5}r_{0}^{4}\left(a^{2}+r_{0}^{2}\right)^{2}\big((a^{2}+q^{2})^{2}-r_{0}^{4}\big)]^{-1}\Big(-3\pi r_{0}^{12}+6ar_{0}^{11}-3\pi\left(2a^{2}+3q^{2}\right)r_{0}^{10}\nonumber \\
 & +2a\left(5a^{2}+9q^{2}\right)r_{0}^{9}+3\pi\left(a^{4}+2q^{2}a^{2}+3q^{4}\right)r_{0}^{8}+3a^{6}\pi\left(a^{2}-2q^{2}\right)r_{0}^{4}+12a^{2}\pi\left(a^{2}+q^{2}\right)^{2}r_{0}^{6}\nonumber \\
 & +2a\left(2a^{4}-9q^{2}a^{2}-9q^{4}\right)r_{0}^{7}-2a^{3}\left(2a^{4}+17q^{2}a^{2}+9q^{4}\right)r_{0}^{5}-2a^{5}\left(5a^{4}+15q^{2}a^{2}+q^{4}\right)r_{0}^{3}\nonumber \\
 & +6\tan^{-1}(\frac{r_{0}}{a})(a^{2}+r_{0}^{2})^{2}\big(a^{8}-(q^{4}-5r_{0}^{2}q^{2}+2r_{0}^{4})a^{4}+2q^{2}r_{0}^{2}(q^{2}-4r_{0}^{2})a^{2}+r_{0}^{8}+3q^{2}r_{0}^{6}-3q^{4}r_{0}^{4}\big)\nonumber \\
 & -3a^{8}\pi\left(2a^{2}+5q^{2}\right)r_{0}^{2}+6a^{7}\left(q^{4}-a^{4}\right)r_{0}+3a^{8}\pi\left(q^{4}-a^{4}\right)\Big)\,,\label{Potential} \nonumber\\
\Delta I & =-\frac{\pi\left(a^{4}+a^{2}\left(q^{2}+2r_{0}^{2}\right)-q^{2}r_{0}^{2}+r_{0}^{4}\right)}{G_{N}\left(a^{2}+q^{2}-r_{0}^{2}\right)}+4\pi c_{6}q^{2}[5G_{N}\left(a^{2}+r_{0}^{2}\right){}^{3}\left(a^{2}+q^{2}-r_{0}^{2}\right)^{3}]^{-1}\Big(\nonumber \\
 & +r_{0}^{8}\left(25a^{2}+13q^{2}\right)+r_{0}^{6}\left(10a^{4}-78a^{2}q^{2}-11q^{4}\right)+a^{2}r_{0}^{2}\left(a^{2}+q^{2}\right)\left(11a^{4}-13a^{2}q^{2}-12q^{4}\right)\nonumber \\
 & +a^{4}\left(a^{2}+q^{2}\right)^{3}+r_{0}^{4}\left(-10a^{6}+32a^{4}q^{2}+57a^{2}q^{4}+3q^{6}\right)-5r_{0}^{10}\Big)-\frac{4\pi c_{3}}{G_{N}}\nonumber \\
 & -8\pi(4c_{7}+c_{8})q^{4}r_{0}^{2}[15G_{N}(a^{2}+r_{0}^{2})^{5}(a^{2}+q^{2}-r_{0}^{2})^{3}]^{-1}\big(a^{6}(9q^{4}-46q^{2}r_{0}^{2}-14r_{0}^{4})-3r_{0}^{6}(q^{2}-r_{0}^{2})^{2}\nonumber \\
 & +a^{8}(12q^{2}+13r_{0}^{2})+a^{4}(-65q^{4}r_{0}^{2}+102q^{2}r_{0}^{4}+6r_{0}^{6})+3a^{2}(17q^{4}r_{0}^{4}-30q^{2}r_{0}^{6}+9r_{0}^{8})+3a^{10}\big)\nonumber \\
 & +\frac{\pi q^{4}f_{5D,2}}{8a^{5}r_{0}^{3}G_{N}\left(a^{2}+q^{2}-r_{0}^{2}\right)^{3}}\Big(-9\pi a^{10}-18a^{9}r_{0}+6\pi a^{8}(r_{0}^{2}-q^{2})+3\pi a^{6}(q^{4}-4q^{2}r_{0}^{2}+6r_{0}^{4})\nonumber \\
 & +6a^{7}r_{0}\left(3r_{0}^{2}-2q^{2}\right)+a^{5}\left(6q^{4}r_{0}-20q^{2}r_{0}^{3}+22r_{0}^{5}\right)+6\pi a^{4}r_{0}^{4}\left(q^{2}-2r_{0}^{2}\right)+3\pi a^{2}r_{0}^{4}\left(q^{4}-3r_{0}^{4}\right)\nonumber \\
 & -2a^{3}r_{0}^{3}\left(q^{4}+4q^{2}r_{0}^{2}-11r_{0}^{4}\right)-12ar_{0}^{5}\left(q^{2}-r_{0}^{2}\right){}^{2}+6\pi r_{0}^{6}\left(q^{2}-r_{0}^{2}\right){}^{2}+6\left(a^{2}+r_{0}^{2}\right)\tan^{-1}\left(\frac{r_{0}}{a}\right)\nonumber \\
 & \times\big(3a^{8}+a^{6}(2q^{2}-5r_{0}^{2})-a^{4}(q^{2}-r_{0}^{2})^{2}+a^{2}r_{0}^{2}\left(q^{4}-4q^{2}r_{0}^{2}+5r_{0}^{4}\right)-2r_{0}^{4}\left(q^{2}-r_{0}^{2}\right)^{2}\big)\Big)\,.
\end{align}

\bibliographystyle{JHEP}
\bibliography{refs.bib}

\end{document}